%% file: main.tex
\documentclass[11pt,oneside]{article}

\input{preamble}
\bodyproofsfalse

\title{The value of conceptual knowledge}
\author{%
Benjamin Davies and Anirudh Sankar%
\thanks{
Department of Economics, Stanford University; bldavies@stanford.edu and asankar@stanford.edu.
We thank
Steve Callander,
Arun Chandrasekhar,
Ben Golub,
Matt Jackson,
Annie Liang,
Jann Spiess,
and
seminar participants at Motu and Stanford
for helpful discussions and comments.
}
}
\date{Draft version: \today}

\begin{document}

\maketitle

\begin{abstract}
    \noindent
    We study the instrumental value of conceptual knowledge when making statistical decisions.
    Such knowledge tells agents how unknown, payoff-relevant states relate.
    It is distinct from the statistical knowledge gained from observing signals of those states.
    We formalize this distinction in a tractable framework used by economists and statisticians.
    Conceptual knowledge is valuable because it empowers agents to design more informative signals.
    It is more valuable when states are more ``reducible'': when they can be explained with fewer common concepts.
    Its value is non-monotone in the number of signals and vanishes when agents have infinitely many signals.
    Agents who know more concepts can attain the same payoffs with fewer signals.
    This is especially true when states are highly reducible.
    
    \vskip\baselineskip
\end{abstract}


\clearpage
\section{Introduction}

Humans use mental models to make sense of the world \citep{Johnson-Laird-1983-}.
The building blocks of these models are ``concepts'': mental representations we use to describe objects in our environment and how they relate \citep{Murphy-2002-}.
Understanding these relationships empowers us to use information about one object to draw inferences about another \citep{Mitchell-2021-NYAS}.

This paper studies the interaction between
(i)~the conceptual knowledge embedded in mental models and
(ii)~the statistical knowledge inferred from data.
We ask: when and why is conceptual knowledge instrumentally ``valuable''?

For example, suppose a farmer wants to learn which fertilizers to apply to his crops.
He views fertilizers as ``black boxes,'' knowing \emph{that} they help crops grow but not \emph{why}.
He does not know how fertilizers' effects relate and cannot extrapolate one from another's.
So when he tries different fertilizers to learn their effects, he has to try them separately.

Now suppose the farmer knows fertilizers supply nitrogen, a nutrient that helps crops grow.
This tells him how fertilizers' effects relate: they share a common ``nitrogen component.''
He can use this component to extrapolate one fertilizer's effect from another's.
Moreover, when he tries different fertilizers to learn their effects, he can combine them to isolate the nitrogen component.
This is better than trying each fertilizer separately because it allows him to learn their effects from one trial rather than many.

Nitrogen is a mental construct---the farmer cannot see it.
All he can see are the effects of trying different fertilizers.
But his conceptual knowledge of nitrogen allows him to learn more efficiently.
It empowers him to run trials that are more informative and instrumentally valuable.
\emph{How much more valuable} is a quantity we define and characterize in this paper.
We call this quantity the ``value of conceptual knowledge.''

Quantifying this value is important for designing interventions that give people information.
In empirical work motivating and guided by this paper \citep{Sankar-etal-2025-}, we experimentally study the effect of teaching concepts to farmers.
We compare farmers given concepts and data to farmers given data only, and find that farmers given concepts make more profitable decisions.
These results provide empirical support for the theoretical predictions made in this paper.
The results suggest a way to improve interventions that give people data: we should also give them conceptual tools to interpret data.
By explaining how these tools work and quantifying their value, this paper helps in designing interventions that improve people's lives more cost-effectively.

Quantifying the value of conceptual knowledge is also important for comparing human and artificial intelligence (AI).
Humans currently have an edge in using concepts to unite seemingly unrelated phenomena:
Newton isolated a concept (gravity) that unites apples falling on Earth with the orbits of other planets;
Watson and Crick isolated a concept (DNA) that unites crime scene investigation with the origins of domesticated rice;
Bernoulli isolated a concept (risk aversion) that unites choices in poker with choices between insurance policies.
These and other concepts allow humans to learn from limited data \citep{Tenenbaum-etal-2011-Science}.
In contrast, AI relies on recognizing patterns in large, rich sets of data \citep{Goodfellow-etal-2016-,Halevy-etal-2009-IEEE}.
By understanding what conceptual knowledge \emph{is}, and when and why it is valuable, we can better allocate inferential tasks between humans and AI.

\paragraph{Contributions}

This paper makes three major contributions.
First, we distinguish two types of knowledge: statistical and conceptual.
Statistical knowledge comes from observing patterns in data; conceptual knowledge tells us how data are generated and what patterns to expect \emph{ex ante}.
We formalize this distinction in a tractable framework used by economists and statisticians.
This framework gives us a transparent model of how statistical and conceptual knowledge interact.

Second, we use our framework to define and characterize the value of conceptual knowledge.
Our definition builds on that of the instrumental value of information \citep{Howard-1966-IEEE,Raiffa-Schlaifer-1961-}: whereas information is valuable because it leads to better decisions, conceptual knowledge is valuable because it leads to better information.
We quantify \emph{how much better} and study this quantity's comparative statics.
Our framework admits a closed-form expression for the value of optimally acquired information, allowing us to analyze and clarify how and why this value changes.
Moreover, our focus on a specific decision problem allows us to generate insights and intuitions that are not immediate from studying abstract problems (as in, e.g., \cite{Blackwell-1951-ProceedingsoftheSecondBerkeleySymposiumonMathematicalStatisticsandProbability,Blackwell-1953-AoMS} and \cite{Whitmeyer-2025-JPE}).%
\footnote{
For example, our non-monotonicity result (Theorem~\ref{thm:vck-tau}) could not be derived from \cite{Blackwell-1951-ProceedingsoftheSecondBerkeleySymposiumonMathematicalStatisticsandProbability,Blackwell-1953-AoMS} or \citeapos{Whitmeyer-2025-JPE} frameworks without imposing more structure on them.
}

Third, we use our framework to formalize what it means to have ``deeper'' conceptual knowledge.
This allows us to compare the marginal values of having deeper knowledge or more data.
It also identifies conceptual knowledge as an economic good one may acquire in the same way data are goods one acquires.
In this way, we advance the literature on learning and information acquisition that treats conceptual knowledge as fixed and minimally restrictive \cite[e.g.,][]{Bardhi-2024-ECTA,Callander-2011-AER,Schwartzstein-2014-JEEA}.

This paper also connects to the literatures on
the value of information,
model-based inference,
human cognition,
and
human-machine comparisons.
We discuss these literatures and our contributions to them after presenting our main results.

\paragraph{Overview}

Section~\ref{sec:illustrative} elaborates on our leading example of a farmer learning about fertilizers.

Section~\ref{sec:framework} extends the example to a more general setting.
We consider a Bayesian agent who makes a statistical decision.
His environment contains a collection of unknown, real-valued states.
He learns about these states from noisy signals.
Then he takes real-valued actions.
His loss equals the mean squared difference between the actions and states.
He takes the actions that minimize his posterior expected loss.

The signals give the agent statistical knowledge.
His prior on the states encodes his conceptual knowledge.
This knowledge tells him how states relate.
It allows him to mentally represent states as combinations of ``concepts,'' which we model as eigenvectors of the state vector's prior variance matrix.
The corresponding eigenvalues index concepts' explanatory power: an eigenvalue is larger when the corresponding concept explains more of the states' prior variances.

We say states are more ``reducible'' when their prior variances are explained by fewer common concepts.
This happens when the eigenvalues of the prior variance matrix are more spread out.
For example, if one eigenvalue is much larger than the others, then the state vector is likely to be close to a one-dimensional subspace of the many-dimensional state space.
The more spread out are the eigenvalues, the more the agent can ``reduce'' the state vector by representing it as a low-dimensional combination of high-dimensional concepts.
This dimension reduction is what makes conceptual knowledge valuable.

Section~\ref{sec:preliminaries} contains preliminary results that we draw upon in later sections.
We characterize the instrumental ``value'' of the agent's signals, derive sharp bounds on this value (see Proposition~\ref{prop:voi-bounds}), and formalize what it means for eigenvalues to be ``more spread out.''

We define and characterize the ``value of conceptual knowledge'' in Section~\ref{sec:vck}.
First, we suppose the agent designs an ``optimal sample'' containing the most valuable signals.
This sample focuses on the concepts with the most explanatory power (see Proposition~\ref{prop:optimal}).
Next, we consider a counterfactual ``na\"ive'' agent who does not know which concepts have more or less explanatory power.
This prevents him from focusing on the concepts with the most power.
As a result, his optimal sample is less valuable than the conceptually knowledgeable agent's.
We derive the values of optimal samples with and without conceptual knowledge, and call their difference the ``value'' of such knowledge.
This difference equals the payoff gain from having conceptual knowledge and using it to design a more valuable sample.

Our first main result (Theorem~\ref{thm:vck-eigenvalues}) says that conceptual knowledge is more valuable when states are more reducible.
If the states can be explained by a few common concepts, then the agent gains a lot from identifying those concepts and focusing on them when he designs signals (i.e., ``asking the right questions'').
In contrast, if every concept has the same explanatory power, then the agent gains nothing from identifying those concepts because he designs the same sample that he would if he was na\"ive.

Our second main result (Theorem~\ref{thm:vck-tau}) says that the value of conceptual knowledge
(i)~is non-monotone in the number of signals and
(ii)~vanishes when there are infinitely many signals.
If the agent can observe more signals, then he can learn more about the concepts he focuses on, raising the payoff gain from knowing which to focus on.
However, having more signals also prompts him to broaden his focus, lowering the gain from knowing which concepts to focus on.
The first effect dominates the second when the number of signals is sufficiently small.
As it becomes arbitrarily large, the agent's posterior becomes independent of his prior, and so the conceptual knowledge embedded in his prior becomes irrelevant and loses its instrumental value.

In Section~\ref{sec:deeper}, we extend our measure of the value of conceptual knowledge to one of ``deeper'' knowledge.
We suppose the agent knows some, but not all, of the relevant concepts, and refer to the ``depth'' of his knowledge as the number he knows.
Our third main result (Theorem~\ref{thm:deeper-vck}) says that deeper conceptual knowledge is weakly more valuable.
However, if the agent knows enough concepts, then knowing more yields no additional value because it does not change the optimal sample he designs.

Finally, in Section~\ref{sec:versus}, we study the trade-off between conceptual and statistical knowledge.
Our framework gives us a precise language for describing this trade-off: is the agent better off knowing more concepts or having more signals?
Our fourth main result (Theorem~\ref{thm:versus-size}) says that if the agent knows more concepts, then he can attain the same welfare with fewer signals, especially when states are highly reducible.
This is because he can design better samples and extract more value from each signal, lowering the number he needs to attain a given welfare target.

Section~\ref{sec:literature} discusses related literature.
Section~\ref{sec:conclusion} concludes.
Appendix~\ref{app:additional} contains additional discussions and results.
Appendix~\ref{app:proofs} contains proofs of our mathematical claims.

\section{An illustrative example}
\label{sec:illustrative}

This section elaborates on the example presented in our introduction.
The example is inspired by our empirical work in Uganda, where we study the role that conceptual knowledge plays when farmers learn about fertilizers \citep{Sankar-etal-2025-}.

\paragraph{Environment}

A Bayesian farmer wants to learn the effect~$\theta_k\in\R$ of applying fertilizer~$k\in\{1,2\}$ to his crops.
His prior on~$\theta\equiv(\theta_1,\theta_2)$ is a normal distribution with variance~$\Var(\theta)$.
He observes the outcome
\[ y=\theta_1w_1+\theta_2w_2+u \]
of using~$w_1\in\R$ more units of fertilizer~1 and~$w_2\in\R$ more units of fertilizer~2.%
\footnote{
We interpret negative values of~$w_k$ as using less of fertilizer~$k$ than the farmer uses currently.
}
The vector~$w=(w_1,w_2)$ has Euclidean length~$\norm{w}=1$ and the error~$u\in\R$ is independently normally distributed with variance~$\vu>0$.%
\footnote{
We normalize~$\norm{w}=1$ so that only the direction of~$w$ (and not its magnitude) affects the informativeness of~$y$.
}
It captures the randomness in~$y$ due to variation in unobserved factors.

The farmer's data~$\samp\equiv\{(w,y)\}$ comprise the vector~$w$ and outcome~$y$.
These data are valuable insofar as they make the farmer's beliefs about~$\theta$ more precise.
We measure the value of~$\samp$ via the mean difference
\[ \voi(\samp)\equiv\frac{1}{2}\sum_{k=1}^2\left(\Var(\theta_k)-\Var(\theta_k\mid\samp)\right) \]
between the prior and posterior variances of~$\theta_1$ and~$\theta_2$.

The farmer chooses the vector~$w$ that maximizes~$\voi(\samp)$ subject to the constraint~$\norm{w}=1$.
Intuitively, he chooses the combination of fertilizers that teaches him as much as possible about their effects.
That he can only choose \emph{one} combination reflects the scarcity and cost of relevant data: our Ugandan setting is one of many where humans must learn from limited data.

\paragraph{Conceptual knowledge}

The farmer knows the two fertilizers supply equal amounts of nitrogen, a nutrient that helps crops grow.
He cannot see or touch nitrogen; it is a mental construct.
But he can use his conceptual knowledge of nitrogen to express each fertilizer's effect~$\theta_k$ as the sum of a common ``nitrogen effect'' and an idiosyncratic effect.
He encodes these effects by the scalars
\[ \gamma_1\equiv\frac{\theta_1+\theta_2}{\sqrt{2}} \quad\text{and}\quad \gamma_2\equiv\frac{\theta_1-\theta_2}{\sqrt{2}}, \]
allowing him to express the effect vector
\[ \theta=\gamma_1\evec_1+\gamma_2\evec_2 \]
as a linear combination of two unit vectors
\[ \evec_1=\frac{1}{\sqrt{2}}\begin{bmatrix} 1 \\ 1 \end{bmatrix} \quad\text{and}\quad \evec_2=\frac{1}{\sqrt{2}}\begin{bmatrix} 1 \\ -1 \end{bmatrix}. \]
These vectors form an orthonormal basis for the Euclidean space~$\R^2$ containing~$\theta$.
The common and idiosyncratic effects~$\gamma_1$ and~$\gamma_2$ are the coordinates of~$\theta$ over this basis.

The farmer knows~$\evec_1$ and~$\evec_2$, but does not know~$\gamma_1$ or~$\gamma_2$.
Knowing~$\evec_1$ and~$\evec_2$ makes learning~$\theta$ equivalent to learning~$\gamma\equiv(\gamma_1,\gamma_2)$.
Moreover, since~$\evec_1$ and~$\evec_2$ are orthonormal, his data~$\samp$ have value%
\footnote{
We derive this expression for~$\voi(\samp)$ in Section~\ref{sec:voi}.
}
\[ \voi(\samp)=\frac{1}{2}\sum_{k=1}^2\left(\Var(\gamma_k)-\Var(\gamma_k\mid\samp)\right). \]
Thus, knowing about nitrogen allows the farmer to reframe his problem from learning about the effect vector~$\theta$ to learning about the coordinate vector~$\gamma$.
He does not have to learn each fertilizer's effect separately; instead, he can learn the nitrogen and idiosyncratic effects, and extrapolate the overall effects.
This is important because he only has one observation~$(w,y)$ from which to infer two unknowns~$\theta_1$ and~$\theta_2$.
Knowing how these unknowns relate (via~$\evec_1$ and~$\evec_2$) allows him to learn about both at the same time by choosing~$w$ appropriately.

The farmer's choice of~$w$ depends on the relative contributions of~$\gamma_1$ and~$\gamma_2$ to the prior variances of~$\theta_1$ and~$\theta_2$.
He knows~$\gamma_1$ contributes more: the fertilizers' effects are mostly determined by how much nitrogen they supply.
So he assumes~$\gamma_1$ and~$\gamma_2$ are independently distributed with variances~$\lambda_1=\vm(1+\rho)$ and~$\lambda_2=\vm(1-\rho)$.
The sum
\begin{align*}
    \lambda_1+\lambda_2
    &= \Var\left(\frac{\theta_1+\theta_2}{\sqrt{2}}\right)+\Var\left(\frac{\theta_1-\theta_2}{\sqrt{2}}\right) \\
    &= \Var(\theta_1)+\Var(\theta_2)
\end{align*}
of these variances equals the sum of the prior variances of~$\theta_1$ and~$\theta_2$.
The parameter~$\rho\in[0,1)$ determines the share
\[ \frac{\lambda_1}{\lambda_1+\lambda_2}=\frac{1+\rho}{2} \]
of this sum contributed by~$\gamma_1$.
This share equals~$1/2$ when~$\rho=0$, in which case~$\gamma_1$ and~$\gamma_2$ contribute equally.
It equals one in the limit as~$\rho\to1$, in which case only~$\gamma_1$ contributes.
The larger is~$\rho$, the more likely is~$\theta$ to be close to the one-dimensional subspace of~$\R^2$ spanned by~$\evec_1$.

The coordinate vector~$\gamma\equiv(\gamma_1,\gamma_2)$ has variance
\[ \Var(\gamma)=\vm\begin{bmatrix} 1+\rho & 0 \\ 0 & 1-\rho \end{bmatrix}, \]
and so the effect vector
\[ \theta=\frac{1}{\sqrt{2}}\begin{bmatrix} 1 & 1 \\ 1 & -1 \end{bmatrix}\gamma \]
has prior variance
\begin{align}
    \Var(\theta)
    &= \left(\frac{1}{\sqrt{2}}\begin{bmatrix} 1 & 1 \\ 1 & -1 \end{bmatrix}\right)\Var(\gamma)\left(\frac{1}{\sqrt{2}}\begin{bmatrix} 1 & 1 \\ 1 & -1 \end{bmatrix}\right)^T \notag \\
    &= \vm\begin{bmatrix} 1 & \rho \\ \rho & 1 \end{bmatrix}. \label{eq:illustrative-Sigma}
\end{align}
Thus~$\theta_1$ and~$\theta_2$ have equal prior variances~$\vm$ and correlation~$\rho$.
Intuitively, the more~$\theta_1$ and~$\theta_2$ are determined by the common effect~$\gamma_1$, the more likely they are to have similar values.

The prior variance matrix~\eqref{eq:illustrative-Sigma} has eigendecomposition
\[ \Var(\theta)=\lambda_1\evec_1\evec_1^T+\lambda_2\evec_2\evec_2^T. \]
Each eigenvalue~$\lambda_k$ equals the prior variance of~$\theta$ in the direction of the corresponding eigenvector~$\evec_k$.
So~$\theta$ has the most prior variance in the direction of~$\evec_1$ and the least in the direction of~$\evec_2$.

\paragraph{Value of information}

The value~$\voi(\samp)$ of the farmer's data is largest when~$w=\pm\evec_1$ and smallest~$w=\pm\evec_2$ (see Proposition~\ref{aprop:singleton-voi}).
For example, choosing~$w=\evec_1$ makes~$y=\gamma_1+u$ a ``pure signal'' of~$\gamma_1$.
This makes~$\samp$ maximally valuable because it provides information about the component of~$\theta$ with the most prior variance, leading to the largest difference between prior and posterior variances.
In contrast, choosing~$w=\evec_2$ makes~$y=\gamma_2+u$ a pure signal of~$\gamma_2$.
This makes~$\samp$ \emph{minimally} valuable because it provides information about the component of~$\theta$ with the \emph{least} prior variance, leading to the \emph{smallest} difference between prior and posterior variances.%
\footnote{
In general, the data~$\samp$ are most valuable when they contain information about the components of~$\theta$ with the most prior variance.
We formalize and prove this claim in Sections~\ref{sec:voi}--\ref{sec:optimal}.
}

\paragraph{Value of conceptual knowledge}

The data~$\samp$ have maximal value
\[ \voi^*\equiv\max_{\norm{w}=1}\voi(\samp). \]
The farmer attains~$\voi^*$ by choosing~$w=\pm\evec_1$.
This choice relies on his conceptual knowledge: he must know~$\theta_1$ and~$\theta_2$ share a common nitrogen effect.
Without this knowledge, the farmer would have no way to represent~$\theta_k$ as the sum of components with differential contributions to its prior variance.
So he would assume equal contributions (i.e., $\rho=0$) and his data would have maximal value
\[ \voi\z\equiv\max_{\norm{w}=1}\left[\voi(\samp)\big\rvert_{\rho=0}\right]. \]
The difference
\[ \vck\equiv\voi^*-\voi\z \]
between~$\voi^*$ and~$\voi\z$ captures the value of the farmer's conceptual knowledge: the value of knowing about nitrogen and using this knowledge to make his data more valuable.

The value~$\vck$ of the farmer's conceptual knowledge is larger when~$\rho$ is larger.%
\footnote{
We have
\[ \voi^*=\frac{(1+\rho)^2\sm^4}{2\left((1+\rho)\vm+\vu\right)}\quad\text{and}\quad\voi\z=\frac{\sm^4}{2(\vm+\vu)} \]
by Proposition~\ref{aprop:singleton-voi} and the definition of~$\voi\z$.
So~$\partial\voi^*/\partial\rho>0$ and~$\partial\voi\z/\partial\rho=0$, from which it follows that~$\partial\vck/\partial\rho>0$.
}\textsuperscript{,}%
\footnote{
For example, the correlation~$\rho$ will be close to one when the fertilizers supply nitrogen only, and close to zero when their nutrient profiles are very different.
}
Intuitively, if most of fertilizers' effects come from supplying nitrogen, then the farmer can refine his prior a lot by isolating the nitrogen effect when he tries fertilizers.
We formalize this intuition in Section~\ref{sec:vck}, and generalize it to a setting in which~$\samp$ has arbitrary size and~$\theta$ has arbitrary length.
In this setting, conceptual knowledge is more valuable when the eigenvalues of the prior variance matrix~$\Var(\theta)$ are more spread out (see Theorem~\ref{thm:vck-eigenvalues}).
This is why raising~$\rho$ raises~$\vck$: it raises~$\lambda_1=(1+\rho)\vm$ and lowers~$\lambda_2=(1-\rho)\vm$ without changing their mean~$(\lambda_1+\lambda_2)/2=\vm$.

\section{Framework}
\label{sec:framework}

We consider a Bayesian agent who collects data before making a statistical decision.
This section describes the agent's environment, formalizes his conceptual knowledge about that environment, explains our modeling assumptions, and presents some specific examples.

\subsection{Environment}

\paragraph{Prior}

There is a true but unknown vector~$\theta\equiv(\theta_1,\ldots,\theta_\dimF)$ of real-valued ``states.''
The agent's prior on~$\theta$ is a probability distribution~$\prior$ over the~$\dimF$-dimensional Euclidean space~$\R^\dimF$.
This distribution is normal with mean~$\mu\in\R^\dimF$ and variance~$\Sigma\in\R^{\dimF\times\dimF}$:
\[ \prior=\Ncal(\mu,\Sigma). \]
We assume~$\dimF\ge2$ is finite and~$\Sigma$ is invertible.

The agent derives~$\prior$ from his conceptual knowledge about~$\theta$.
We explain this derivation in Section~\ref{sec:knowledge}.

\paragraph{Data}

The agent observes a sample~$\samp\equiv\{(w^{(i)},y^{(i)})\}_{i=1}^n$ of size~$n$.
Each ``observation''~$(w^{(i)},y^{(i)})$ comprises
a ``covariate''~$w^{(i)}\in\R^\dimF$ with Euclidean length~$\norm{w^{(i)}}=1$,%
\footnote{
\label{fn:covariate-length}%
Assuming the covariates have unit length normalizes the scales of the signals~$y^{(1)},\ldots,y^{(n)}$ so that only the directions of~$w^{(1)},\ldots,w^{(n)}$ (and not their magnitudes) affect signals' informativeness.
It also ensures the Gram matrix~\eqref{eq:gram} always has trace~$n$ (see Footnote~\ref{fn:gram-trace}).
This allows us to identify optimal samples of size~$n$ via a specific Gram matrix~\eqref{eq:optimal-gram}.
}
and an ``outcome''
\begin{equation}
    \label{eq:outcome}
    y^{(i)}=\theta^Tw^{(i)}+u^{(i)}
\end{equation}
equal to the sum of
\[ \theta^Tw^{(i)}=\sum_{k=1}^\dimF\theta_kw_k^{(i)} \]
and an independently normally distributed error~$u^{(i)}$ with mean zero and variance~$\vu>0$.
Thus, the outcome~$y^{(i)}$ provides a noisy signal of a weighted combination of states, where the weights are determined by the covariate~$w^{(i)}\equiv(w_1^{(i)},\ldots,w_\dimF^{(i)})$.

\paragraph{Actions and losses}

The agent uses his prior~$\prior$, the sample~$\samp$, and Bayes' rule to form posterior beliefs about~$\theta$.
Then he chooses a~$\dimF$-vector~$a\equiv(a_1,\ldots,a_\dimF)$ of real-valued actions.
These actions induce a loss
\[ \loss(\theta,a)\equiv\frac{1}{\dimF}\sum_{k=1}^\dimF(a_k-\theta_k)^2 \]
equal to the mean squared difference between them and the corresponding states.%
\footnote{
Suppose~$p_1,\ldots,p_\dimF$ are strictly positive and sum to one.
Let~$D$ be the~$\dimF\times\dimF$ diagonal matrix with~${kk}^\text{th}$ entry~$\dimF p_k$, and let~$a'\equiv Da$ and~$\theta'\equiv D\theta$.
Then
\[ \loss(a',\theta')
    =\sum_{k=1}^\dimF p_k(a_k-\theta_k)^2 \]
is a weighted average of the squared differences between the actions and corresponding states.
The weights~$p_1,\ldots,p_\dimF$ encode the agent's preferences: the larger is~$p_k$, the larger is the loss from taking an action~$a_k$ different than the state~$\theta_k$.
We focus on the case with~$p_k=1/\dimF$ for each~$k$, which makes~$D$ equal the identity matrix and~$\theta'$ equal~$\theta$.
However, we can easily generalize our analysis to a setting with non-equal weights by replacing~$\theta$ with~$\theta'$.
Then what matters are the eigenvalues and eigenvectors of~$\Var(\theta')=D\Sigma D^T$, rather than those of~$\Sigma$.
This does not change our results or insights substantively.
}

Let~$\E$ take expectations with respect to the prior distribution~$\prior$.
The agent chooses the action vector that minimizes his posterior expected loss:%
\footnote{
In Appendix Section~\ref{asec:statistical-learning}, we explain how the choice problem~\eqref{eq:action-vector} is equivalent to a prediction problem that arises in the machine and statistical learning literatures.
This equivalence comes from interpreting~$\theta_1,\ldots,\theta_\dimF$ as values of an unknown function.
}
\begin{equation}
    \label{eq:action-vector}
    a\in\argmin_{a'\in\R^\dimF}\E[\loss(\theta,a')\mid\samp].
\end{equation}
Intuitively, he wants to estimate~$\theta_1,\ldots,\theta_\dimF$ accurately, and the accuracy of his estimates~$a_1,\ldots,a_\dimF$ is determined by their squared errors~$(\theta_k-a_k)^2$.
Thus, our framework aligns with least squares estimation, a tool used throughout empirical economics and statistics.%
\footnote{
Indeed, if the prior distribution~$\prior$ is diffuse, then~\eqref{eq:action-vector} equals the Ordinary Least Squares estimate of~$\theta$ given~$\samp$.
}

\paragraph{Value of~$\samp$}

If the agent did not observe the sample~$\samp$, then his minimized prior and posterior expected losses would be equal.
The information in~$\samp$ is instrumentally valuable because it helps the agent take actions with lower expected losses.
Accordingly, we define the ``value of~$\samp$'' to be the difference between his minimized prior and posterior expected losses:%
\footnote{
\citet[p.\! 90]{Raiffa-Schlaifer-1961-} define a similar object and call it the ``(expected) value of sample information.''
}
\begin{equation}
    \label{eq:voi-definition}
    \voi(\samp)\equiv\min_{a'\in\R^\dimF}\E[\loss(\theta,a')]-\min_{a'\in\R^\dimF}\E[\loss(\theta,a')\mid\samp].
\end{equation}

\paragraph{Covariate selection}

The agent chooses the covariates~$w^{(1)},\ldots,w^{(n)}$ that maximize~\eqref{eq:voi-definition} subject to the length constraints~$\norm{w^{(i)}}=1$.
Intuitively, he wants to design the signals~$y^{(1)},\ldots,y^{(n)}$ so that they provide as much payoff-relevant information as possible about the state vector~$\theta$.

\paragraph{Timing}

First, nature draws~$\theta$ from the prior distribution~$\prior$.
Second, the agent chooses the covariates~$w^{(1)},\ldots,w^{(n)}$ and observes the outcomes~$y^{(1)},\ldots,y^{(n)}$.
Third, he combines~$\prior$ and the sample~$\samp\equiv\{(w^{(i)},y^{(i)})\}_{i=1}^n$ to form posterior beliefs about~$\theta$.
Finally, he chooses the action vector~\eqref{eq:action-vector} that minimizes his posterior expected loss.

\subsection{Conceptual knowledge}
\label{sec:knowledge}

\paragraph{Mental model and concepts}

Our definition of conceptual knowledge draws upon psychologists' and cognitive scientists':
concepts are the building blocks of mental models \citep{Johnson-Laird-1983-}, are used to describe objects and how they relate \citep{Murphy-2002-}, and allow humans to generalize across objects \citep{Mitchell-2021-NYAS}.

Accordingly, our agent's conceptual knowledge allows him to describe the states~$\theta_1,\ldots,\theta_\dimF$ and how they relate.
It gives him a mental model of
\begin{equation}
    \label{eq:theta-decomposition}
    \theta=\sum_{k=1}^\dimF\gamma_k\evec_k
\end{equation}
as an unknown combination of known vectors~$\evec_1,\ldots,\evec_\dimF\in\R^\dimF$.
We call these vectors ``concepts.''%
\footnote{
The vectors~$\evec_1,\ldots,\evec_\dimF$ may not correspond to physical features of the agent's environment.
Instead they are mental constructs he uses to make sense of his environment.
For example, nutrients like nitrogen are mental constructs: no farmer sees them.
All farmers see are the effects of applying fertilizers.
This is why we call~$\evec_1,\ldots,\evec_\dimF$ ``concepts.''
}
They are the building blocks of the agent's mental model.
They capture his environment's generalizable structure: each state~$\theta_j$ depends on the~$j^\text{th}$ component of~$\evec_k$ via a coefficient~$\gamma_k\in\R$ that is independent of~$j$.%
\footnote{
The coefficients~$\gamma_1,\ldots,\gamma_\dimF$ are akin to ``deep parameters'' that determine the ``reduced-form'' states~$\theta_1,\ldots,\theta_\dimF$ via the structural relationships encoded by~$\evec_1,\ldots,\evec_\dimF$ \citep{Lucas-1976-CRCSPP}.
}
This allows the agent to generalize across states: signals of~$\theta_1$ provide information about~$\gamma_1,\ldots,\gamma_\dimF$, from which he can extrapolate~$\theta_2,\ldots,\theta_\dimF$.

For example, suppose~$\theta_1,\ldots,\theta_\dimF$ are the effects of applying different fertilizers.
Then~$\evec_1,\ldots,\evec_\dimF$ could encode nutrient quantities and~$\gamma_1,\ldots,\gamma_\dimF$ the fertilizer-invariant effects of supplying different nutrients.%
\footnote{
Alternatively, if~$\theta_1,\ldots,\theta_\dimF$ are the prices of financial assets, then~$\evec_1,\ldots,\evec_\dimF$ could encode payoffs in different states of nature and~$\gamma_1,\ldots,\gamma_\dimF$ the prices of Arrow-Debreu securities \citep[see, e.g.,][]{Varian-1987-JEP}.
}
Moreover, if a farmer learns one fertilizer's overall effect, then he can extrapolate the others' via their joint dependence on~$\gamma_1,\ldots,\gamma_\dimF$.

For convenience and without loss of generality, we assume~$\evec_1,\ldots,\evec_\dimF$ are orthonormal for the remainder of the paper.

\paragraph{Eigendecomposition}

The agent knows the concepts~$\evec_1,\ldots,\evec_\dimF$ included in his mental model~\eqref{eq:theta-decomposition}.
He does not know the coefficients~$\gamma_1,\ldots,\gamma_\dimF$, but he knows some contribute more to the states' prior variances than others.
Specifically, he knows each coefficient~$\gamma_k$ is independently distributed with variance~$\lambda_k>0$ non-decreasing in~$k$.%
\footnote{
\label{fn:gamma-wlog}
It is without loss of generality to assume~$\gamma_1,\ldots,\gamma_\dimF$ are independently distributed.
This is because~$\Lambda$ is positive-semidefinite, and so, by the spectral theorem, there is an orthogonal matrix~$A\in\R^{\dimF\times\dimF}$ and diagonal matrix~$\Lambda'\in\R^{\dimF\times\dimF}$ such that~$\Lambda=A\Lambda'A^T$.
Then~$\emat'\equiv\emat A$ is orthogonal and~$\Sigma$ has eigendecomposition~$\emat'\Lambda'(\emat')^T$, so we can carry out our analysis by replacing~$\emat$ with~$\emat'$ and~$\Lambda$ with~$\Lambda'$.
Likewise, it is without loss to assume~$\lambda_1\ge\cdots\ge\lambda_\dimF$ because we can permute the indices of the eigenpairs~$(\lambda_k,\evec_k)$ without changing~$\Sigma$.
}
Then~$\theta$ has prior variance
\begin{align}
    \Sigma
    &= \emat\Lambda\emat^T \notag \\
    &= \sum_{k=1}^\dimF\lambda_k\evec_k\evec_k^T, \label{eq:Sigma-eigendecomposition}
\end{align}
where
\[ \Lambda\equiv\begin{bmatrix} \lambda_1 \\ & \ddots \\ & & \lambda_\dimF \end{bmatrix} \]
is the~$\dimF\times\dimF$ diagonal matrix with entries~$\lambda_1\ge\cdots\ge\lambda_\dimF\ge0$ and
\[ \emat\equiv\begin{bmatrix} \evec_1 & \cdots & \evec_\dimF \end{bmatrix} \]
is the~$\dimF\times\dimF$ orthogonal matrix with columns~$\evec_1,\ldots,\evec_\dimF$.

Equation~\eqref{eq:Sigma-eigendecomposition} is an eigendecomposition of~$\Sigma$.
The~$k^\text{th}$ largest eigenvalue~$\lambda_k=\Var(\gamma_k)$ of~$\Sigma$ equals the prior variance of~$\theta$ in the direction of the corresponding unit eigenvector~$\evec_k$.
The trace
\[ \trace(\Sigma)=\sum_{k=1}^\dimF\lambda_k \]
of~$\Sigma$ equals the sum of the eigenvalues~$\lambda_1,\ldots,\lambda_\dimF$.
So these eigenvalues' mean
\begin{align*}
    \overline\lambda
    &\equiv \frac{1}{\dimF}\sum_{k=1}^\dimF\lambda_k \\
    &= \frac{1}{\dimF}\sum_{k=1}^\dimF\Var(\theta_k)
\end{align*}
equals the mean of the states' prior variances.
The ratio~$\lambda_k/\trace(\Sigma)$ equals the share of these variances contributed by~$\gamma_k$.
If the shares contributed by~$\gamma_1,\ldots,\gamma_\dimF$ are equal, then~$\lambda_k=\trace(\Sigma)/\dimF=\overline\lambda$ is constant in~$k$ and so~$\Sigma=\emat\Lambda\emat^T$ is proportional to~$\dimF\times\dimF$ identity matrix~$I_\dimF$:
\[ \emat\left(\overline\lambda I_\dimF\right)\emat^T=\overline\lambda I_\dimF. \]
In contrast, if~$\lambda_1/\trace(\Sigma)\approx1$, then~$\gamma_1$ contributes most of the states' prior variances.

\paragraph{Reducibility}

The distribution of~$\lambda_1,\ldots,\lambda_\dimF$ around their mean~$\overline\lambda=\trace(\Sigma)/\dimF$ captures the states' ``reducibility.''
They are more ``reducible'' when their prior variances are explained by fewer common concepts: when~$\lambda_1,\ldots,\lambda_\dimF$ are more spread out around~$\overline\lambda$.%
\footnote{
We formalize what it means for~$\lambda_1,\ldots,\lambda_\dimF$ to be ``more spread out'' in Section~\ref{sec:mps}.
}\textsuperscript{,}%
\footnote{
If~$\lambda_2=\cdots=\lambda_\dimF$ (as in Example~\ref{eg:pairwise}), then the distribution of~$\lambda_1,\ldots,\lambda_\dimF$ is fully determined by the leading eigenvalue~$\lambda_1$ and the ``spectral gap''~$(\lambda_1-\lambda_2)$.
This gap appears elsewhere in the statistics literature; for example, spectral gaps determine Markov chains' mixing times \citep{Levin-etal-2008-} and whether principal components can be estimated consistently \citep{Yu-etal-2015-Biometrika}.
}
The agent's conceptual knowledge allows him to ``reduce'' the state vector~$\theta$ by representing it as a low-dimensional combination of higher-dimensional concepts.

If the agent had no conceptual knowledge---i.e., if he did not have a mental model of~$\theta$ as a combination of concepts with different explanatory powers---then he would not be able to reduce states in the manner described above.
His prior variance matrix
\[ \Sigma\z\equiv\overline\lambda I_\dimF \]
would equal the prior variance matrix in the case when~$\lambda_k=\overline\lambda$ for each~$k\in\{1,\ldots,\dimF\}$.
We refer to~$\prior\z\equiv\Ncal(\mu,\Sigma\z)$ as a ``na\"ive'' prior because it ignores the covariances among states stemming from their dependence on common concepts.

We can use the true prior~$\prior\equiv\Ncal(\mu,\Sigma)$ and na\"ive prior~$\prior\z$ to measure how much the agent's conceptual knowledge allows him to reduce states.
Since~$\prior$ and~$\prior\z$ are normal distributions with equal means, the Kullback-Leibler (hereafter ``KL'') divergence from~$\prior$ and~$\prior\z$ equals%
\footnote{
See \citet[Section~A.5]{Rasmussen-Williams-2006-} for a derivation of~\eqref{eq:divergence}.
}
\begin{align}
    \KLdiv
    &= \frac{1}{2}\left(\trace((\Sigma\z)^{-1}\Sigma)-\dimF+\ln\left(\frac{\det(\Sigma\z)}{\det(\Sigma)}\right)\right) \notag \\
    &= -\frac{1}{2}\sum_{k=1}^\dimF\ln\left(\frac{\lambda_k}{\overline\lambda}\right). \label{eq:divergence}
\end{align}
The KL divergence~\eqref{eq:divergence} measures the information gain from using~$\prior$ as a prior rather than~$\prior\z$.
This information is purely conceptual: it does not depend on the sample~$\samp$.
It comes from knowing how to represent states as low-dimensional combinations of high-dimensional concepts.
We study the instrumental value of this dimension reduction in Section~\ref{sec:vck}.

The KL divergence~\eqref{eq:divergence} equals zero when the eigenvalues~$\lambda_1,\ldots,\lambda_\dimF$ of~$\Sigma$ are equal and is larger when they are more spread out (see Proposition~\ref{aprop:divergence}).%
\footnote{
If~$\prior\z$ has mean~$\mu\z\in\R^\dimF$, then~\eqref{eq:divergence} becomes
\[ \KLdiv=\frac{1}{2}\left(\frac{1}{\overline\lambda}\norm{\mu-\mu\z}-\sum_{k=1}^\dimF\ln\left(\frac{\lambda_k}{\overline\lambda}\right)\right). \]
So even if~$\mu\z\not=\mu$, the KL divergence from~$\prior$ to~$\prior\z$ is non-negative and does not fall when~$\lambda_1,\ldots,\lambda_\dimF$ undergo a MPS (see Proposition~\ref{aprop:divergence}).
But it is strictly larger than zero when~$\mu\z\not=\mu$, even if~$\lambda_1=\cdots=\lambda_\dimF$.
}
This is because spreading~$\lambda_1,\ldots,\lambda_\dimF$ narrows the distribution~$\prior$ to a lower-dimensional subspace of~$\R^\dimF$.

For example, consider the prior variance matrix~\eqref{eq:illustrative-Sigma} derived in Section~\ref{sec:illustrative}.
This matrix has eigenvalues~$\lambda_1=\vm(1+\rho)$ and~$\lambda_2=\vm(1-\rho)$, which have mean~$\overline\lambda=\vm$ and become more spread out as~$\rho\in[0,1)$ grows.
The KL divergence
\[ \KLdiv=-\frac{1}{2}\ln(1-\rho^2) \]
from~$\prior$ to~$\prior\z$ equals zero when~$\rho=0$ and grows as~$\rho$ grows.
If~$\rho=0$, then the true prior~$\prior$ has equal variance in all directions of~$\R^2$, and so the agent gains nothing from knowing the eigenvectors~$\evec_1$ and~$\evec_2$ of~\eqref{eq:illustrative-Sigma}.
The larger is~$\rho$, the more concentrated is~$\prior$ around the subspace spanned by~$\evec_1$, and so the more the agent gains from knowing~$\evec_1$ and~$\evec_2$.

\paragraph{Relationship to PCA}

The example above illustrates the connection between our ideas and principal component analysis (hereafter ``PCA'').
PCA is a dimension reduction technique that projects a distribution onto its highest variance dimensions.
Traditional PCA estimates these dimensions from data.
In contrast, our agent derives them from his conceptual knowledge: he knows which dimensions have the highest variance \emph{before} observing any data.
This ``pre-data PCA'' allows him to collect more instrumentally valuable data; we call this benefit the ``value of conceptual knowledge'' and quantify it in Sections~\ref{sec:vck}--\ref{sec:versus}.

\subsection{Modeling assumptions}

We assume states and outcomes are jointly normally distributed under the agent's prior, and his actions are real-valued and induce quadratic losses.
This setup is common in the literature on statistical decisions \citep{Hastie-etal-2009-}.
It is also implicit in empirical economics papers that estimate linear models via Ordinary Least Squares.
Our agent has a linear model~\eqref{eq:theta-decomposition} of the unknown state vector.
If his prior is diffuse, then his optimal actions equal the estimates obtained via OLS.

We also assume the agent knows how the states covary \emph{a priori}.
This separates the conceptual knowledge embedded in his prior from the statistical knowledge he infers from his sample.
The assumption allows us to measure the agent's conceptual and statistical knowledge on independent scales, and to study their relative contributions to his welfare (see Sections~\ref{sec:deeper} and~\ref{sec:versus}).

Finally, we assume there is a correct model of the agent's environment (i.e., a true prior variance matrix) that he can know at different ``depths'' (see Sections~\ref{sec:deeper} and~\ref{sec:versus}).
This separates our paper from the literatures on model uncertainty \citep{Chatfield-1995-JRSS,Marinacci-2015-JEEA} and mis-specification \citep{Esponda-Pouzo-2016-ECTA,Spiegler-2016-QJE}, which study agents who do not know the correct model or use an incorrect model.
Our analysis complements those literatures: rather than asking ``what if the agent does not know the correct model?'' we ask ``what does he gain from knowing the correct model?''%
\footnote{
However, our analysis connects to robust approaches \citep{Gilboa-Schmeidler-1989-JME,Hansen-Sargent-2001-AER,Klibanoff-etal-2005-ECTA}.
The na\"ive prior is robust in that it commits minimally to any particular covariance structure; it spreads variance evenly across all dimensions.
Theorem~\ref{thm:vck-eigenvalues} can be interpreted as saying the cost of this robust approach is larger when the correct model has more structure (i.e., the eigenvalues~$\lambda_1,\ldots,\lambda_\dimF$ are more spread out).
In this way, we quantify the value of imposing correct structural restrictions versus maintaining robustness to mis-specification.
}

\subsection{Examples}

Below are two examples of how the prior variance matrix~$\Sigma$ encodes conceptual knowledge about the states.
The first example generalizes the setting described in Section~\ref{sec:illustrative}.
It builds~$\Sigma$ from first principles, starting with the eigenvalues and eigenvectors.
The second example builds~$\Sigma$ from knowledge of how the states are generated, then derives the eigenvalues and eigenvectors.

\begin{example}[Pairwise correlated states]
    \label{eg:pairwise}
    Suppose the agent knows each state~$\theta_k$ has two components: a common component that is proportional to the states' mean and an idiosyncratic component that is independent across states.
    He encodes the common component by the unit vector
    \[ \evec_1=\frac{1}{\sqrt{\dimF}}\ones{\dimF}, \]
    where~$\ones{\dimF}\equiv(1,\ldots,1)$ is the~$\dimF$-vector of ones.
    He encodes the idiosyncratic components by unit vectors~$\evec_2,\ldots,\evec_\dimF$ that are orthogonal to~$\evec_1$ and each other.
    The~$k^\text{th}$ coefficient~$\gamma_k$ in~\eqref{eq:theta-decomposition} has prior variance
    \[ \lambda_k=\vm\begin{cases}
        1+\rho(\dimF-1) & \text{if}\ k=1 \\
        1-\rho & \text{if}\ k>1,
    \end{cases} \]
    where~$\vm>0$ is the mean of~$\lambda_1,\ldots,\lambda_\dimF$ and where~$\rho\in[0,1)$ determines the share
    \[ \frac{\lambda_1}{\lambda_1+\cdots+\lambda_\dimF}=\frac{1}{\dimF}+\rho\left(1-\frac{1}{\dimF}\right) \]
    of the prior variances of~$\theta_1,\ldots,\theta_\dimF$ contributed by the coefficient~$\gamma_1$ on~$\evec_1$.
    This share equals~$1/\dimF$ when~$\rho=0$, in which case~$\lambda_k$ is constant in~$k$ and so~$\gamma_1,\ldots,\gamma_\dimF$ contribute to the prior variances of~$\theta_1,\ldots,\theta_\dimF$ equally.
    It equals one in the limit as~$\rho\to1$, in which case only~$\gamma_1$ contributes.

    Since~$\evec_1,\ldots,\evec_\dimF$ are orthonormal, the sum
    \[ \sum_{k=1}^\dimF\evec_k\evec_k^T=I_\dimF \]
    of their outer products equals the~$\dimF\times\dimF$ identity matrix.
    Therefore, the prior variance matrix
    \begin{align}
        \Sigma
        &= \lambda_1\evec_1\evec_1^T+\lambda_\dimF\left(I_\dimF-\evec_1\evec_1^T\right) \notag \\
        &= \rho\vm\ones{\dimF}\ones{\dimF}^T+(1-\rho)\vm I_\dimF \notag \\
        &= \vm\begin{bmatrix}
            1 & \rho & \cdots \\
            \rho & 1 & \\
            \vdots & & \ddots
        \end{bmatrix} \label{eq:pairwise-Sigma}
    \end{align}
    is the~$\dimF\times\dimF$ matrix with diagonal entries equal to~$\vm$ and off-diagonal entries equal to~$\rho\vm$.
    Thus, under the agent's prior, the states have equal variances~$\vm$ and pairwise correlations~$\rho$.
\end{example}

\begin{example}[Random walk]
    \label{eg:brownian}
    Let~$\nu>0$.
    Suppose the agent knows~$\theta_1,\ldots,\theta_\dimF$ are values of a random walk with known initial value~$\theta_0\in\R$ and unknown, independently distributed increments
    \[ \theta_k-\theta_{k-1}\sim\Ncal(0,\nu^2). \]
    Then the prior variance matrix
    \begin{equation}
        \label{eq:brownian-Sigma}
        \Sigma=\nu^2\begin{bmatrix} 1 & 1 & \cdots & 1 \\ 1 & 2 & \cdots & 2 \\ \vdots & \vdots & & \vdots \\ 1 & 2 & \cdots & \dimF \end{bmatrix}
    \end{equation}
    has~${jk}^\text{th}$ entry~$\Sigma_{jk}=\nu^2\min\{j,k\}$.
    \cite{Fortiana-Cuadras-1997-LinAlgApp} show that~\eqref{eq:brownian-Sigma} has~$k^\text{th}$ largest eigenvalue
    \[ \lambda_k=\frac{\nu^2}{4}\csc^2\left(\frac{(2k-1)\pi}{2\dimF+1}\right) \]
    and that the corresponding unit eigenvector~$\evec_k$ has~$j^\text{th}$ component
    \[ [\evec_k]_j=\frac{2}{\sqrt{2\dimF+1}}\sin\left(\frac{j(2k-1)\pi}{2\dimF+1}\right). \]
\end{example}

\begin{figure}[!t]
    \centering
    \includegraphics[width=0.8\linewidth]{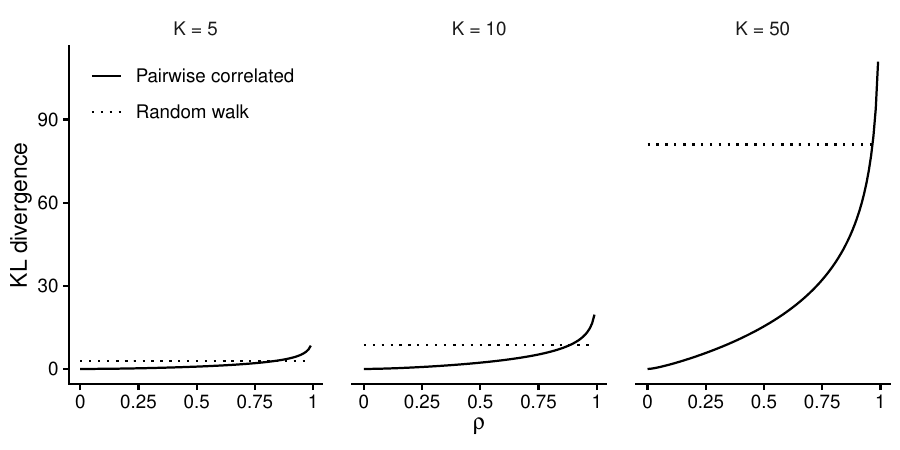}
    \caption{KL divergences~$\KLdiv$ when states are pairwise correlated (i.e., when~$\theta$ has prior variance~\eqref{eq:pairwise-Sigma}) and when they follow a random walk (i.e., when~$\theta$ has prior variance~\eqref{eq:brownian-Sigma} with~$\nu^2=2\vm/(\dimF+1)$)}
    \label{fig:brownian}
\end{figure}
The eigenvalues of~\eqref{eq:pairwise-Sigma} have mean~$\vm$, whereas the eigenvalues of~\eqref{eq:brownian-Sigma} have mean~$\nu^2(\dimF+1)/2$.
Choosing~$\nu^2=2\vm/(\dimF+1)$ equates these two means but does not equate the eigenvalues' distributions, nor the KL divergences~$\KLdiv$ those distributions imply.
We illustrate this fact in Figure~\ref{fig:brownian}.
It shows that assuming states follow a random walk is equivalent, in terms of how much prior structure it imposes, to assuming a large pairwise correlation.%
\footnote{
\cite{Callander-2011-AER} and others use Brownian motions (the continuous-time analogues of random walks) as tools for modeling ``complexity.''
They define ``complex'' environments as those in which only local learning is possible: learning a state provides some information about nearby states but little about distant states (see also \cite{Bardhi-2024-ECTA}).
The limiting case is when learning a state provides \emph{no} information about others; in our framework, this happens when the states are uncorrelated.
Yet Figure~\ref{fig:brownian} suggests that Brownian motions are as structurally restrictive as assuming states are highly correlated.
}
This is especially true when there are many states: if~$\dimF=5$, then the equivalent correlation is about~$0.82$; if~$\dimF=50$, then it is about~$0.97$.

\section{Preliminaries}
\label{sec:preliminaries}

This section contains preliminary results that we draw upon in later sections.
We characterize the optimal action vector~\eqref{eq:action-vector}, the posterior expected loss it induces, and the value~\eqref{eq:voi-definition} of the sample~$\samp$.
We establish sharp lower and upper bounds on this value, and explain how the agent constructs ``optimal'' samples.
Finally, we formalize what it means for the eigenvalues of the prior variance matrix to be ``more spread out.''

\subsection{Value of~\texorpdfstring{$\samp$}{S}}
\label{sec:voi}

Let~$\Var$ take variances with respect to the prior distribution~$\prior$.
Lemma~\ref{lem:expected-loss} characterizes the optimal action vector~\eqref{eq:action-vector} and the posterior expected loss it induces.%
\footnote{
Lemma~\ref{lem:expected-loss} holds even when~$\prior$ is not normal.
We assume~$\prior$ is normal so that we can derive closed-form expressions for the posterior variances of~$\theta_1,\ldots,\theta_\dimF$ and, thus, the value~\eqref{eq:voi} of~$\samp$.
}
This vector equals the posterior mean of~$\theta$.
It induces a posterior expected loss equal to the mean of the posterior variances of~$\theta_1,\ldots,\theta_\dimF$.

\begin{lemma}
    \label{lem:expected-loss}
    The optimal action vector~$a=\E[\theta\mid\samp]$ induces posterior expected loss
    \begin{equation}
        \label{eq:expected-loss}
        \E[\loss(\theta,a)\mid\samp]=\frac{1}{\dimF}\sum_{k=1}^\dimF\Var(\theta_k\mid\samp).
    \end{equation}
\end{lemma}
\ifbodyproofs\input{proofs/expected-loss}\fi

If the sample~$\samp$ was empty, then~\eqref{eq:expected-loss} would equal the minimized prior expected loss
\begin{equation}
    \label{eq:prior-expected-loss}
    \min_{a'\in\dimF}\E[\loss(\theta,a')]=\frac{1}{\dimF}\sum_{k=1}^\dimF\Var(\theta_k).
\end{equation}
Substituting~\eqref{eq:expected-loss} and~\eqref{eq:prior-expected-loss} into~\eqref{eq:voi-definition} yields and expression for the sample's value~$\voi(\samp)$ in terms of the states' prior and posterior variances:
\begin{equation}
    \label{eq:voi}
    \voi(\samp)=\frac{1}{\dimF}\sum_{k=1}^\dimF\left(\Var(\theta_k)-\Var(\theta_k\mid\samp)\right).
\end{equation}
Intuitively, the sample is valuable insofar as it lowers states' variances, allowing the agent to estimate them more accurately.
Moreover, since~$\evec_1,\ldots,\evec_\dimF$ are known and orthonormal, we have%
\footnote{
We have~$\theta=\emat\gamma$, where~$\emat$ is the orthogonal matrix with known columns~$\evec_1,\ldots,\evec_\dimF$ and~$\gamma$ is the vector of unknown coefficients~$\gamma_1,\ldots,\gamma_\dimF$.
So the prior variances of~$\theta_1,\ldots,\theta_\dimF$ and~$\gamma_1,\ldots,\gamma_\dimF$ have equal sums:
\begin{align*}
    \sum_{k=1}^\dimF\Var(\theta_k)
    = \trace\left(\Var(\theta)\right)
    = \trace\left(\Var(\emat\gamma)\right)
    = \trace\left(\emat\Var(\gamma)\emat^T\right)
    \overset{\star}{=} \trace\left(\Var(\gamma)\right)
    = \sum_{k=1}^\dimF\Var(\gamma_k),
\end{align*}
where~$\star$ uses the cyclic property of matrix traces and the orthogonality of~$\emat$.
Similarly, the posterior variances of~$\theta_1,\ldots,\theta_\dimF$ and~$\gamma_1,\ldots,\gamma_\dimF$ have equal sums.
Substituting these sums into~\eqref{eq:voi} yields~\eqref{eq:voi-gamma}.
}
\begin{equation}
    \label{eq:voi-gamma}
    \voi(\samp)=\frac{1}{\dimF}\sum_{k=1}^\dimF\left(\Var(\gamma_k)-\Var(\gamma_k\mid\samp)\right).
\end{equation}
Thus, equivalently, the sample is valuable insofar as it lowers the variances of~$\gamma_1,\ldots,\gamma_\dimF$.

\subsection{Bounds on~\texorpdfstring{$\voi(\samp)$}{π(S)}}

We can express~\eqref{eq:voi} in terms of the traces of the prior and posterior variance matrices:
\[ \voi(\samp)=\frac{1}{\dimF}\left(\trace(\Sigma)-\trace\left(\Var(\theta\mid\samp)\right)\right). \]
Lemma~\ref{lem:posterior-variance} characterizes~$\Var(\theta\mid\samp)$ in terms of the prior variance matrix~$\Sigma$ and ``Gram matrix''
\begin{equation}
    \label{eq:gram}
    \gram\equiv\sum_{i=1}^nw^{(i)}(w^{(i)})^T.
\end{equation}

\begin{lemma}
    \label{lem:posterior-variance}
    The state vector has posterior variance
    \begin{equation}
        \label{eq:posterior-variance}
        \Var(\theta\mid\samp)=\left(\Sigma^{-1}+\frac{1}{\vu}\gram\right)^{-1}.
    \end{equation}
\end{lemma}
\ifbodyproofs\input{proofs/posterior-variance}\fi

The Gram matrix~\eqref{eq:gram} is symmetric and positive semi-definite.
So, by the spectral theorem, there is a~$\dimF\times\dimF$ diagonal matrix
\[ \Delta\equiv\begin{bmatrix} \delta_1 \\ & \ddots \\ & & \delta_\dimF \end{bmatrix} \]
with entries~$\delta_1\ge\cdots\ge\delta_\dimF\ge0$ and a~$\dimF\times\dimF$ orthogonal matrix
\[ \Omega=\begin{bmatrix} \omega_1 & \cdots & \omega_\dimF \end{bmatrix} \]
such that
\begin{align}
    \gram
    &= \Omega\Delta\Omega^T \notag \\
    &= \sum_{k=1}^\dimF\delta_k\omega_k\omega_k^T. \label{eq:gram-eigendecomposition}
\end{align}
Then~$\delta_1,\ldots,\delta_\dimF$ are the eigenvalues of~$\gram$ and~$\omega_1,\ldots,\omega_\dimF\in\R^\dimF$ are the corresponding unit eigenvectors.
Proposition~\ref{prop:voi-bounds} uses the eigendecompositions~\eqref{eq:Sigma-eigendecomposition} and~\eqref{eq:gram-eigendecomposition} of the prior variance and Gram matrices to provide sharp bounds on~$\voi(\samp)$.

\begin{proposition}
    \label{prop:voi-bounds}
    The value~$\voi(\samp)$ of~$\samp$ satisfies
    \begin{equation}
        \label{eq:voi-bounds}
        \frac{1}{\dimF}\sum_{k=1}^\dimF\left(\lambda_k-\left(\frac{1}{\lambda_k}+\frac{\delta_{\dimF-k+1}}{\vu}\right)^{-1}\right)
        \,\overset{\star}{\le}\,
        \voi(\samp)
        \,\overset{\star\star}{\le}\,
        \frac{1}{\dimF}\sum_{k=1}^\dimF\left(\lambda_k-\left(\frac{1}{\lambda_k}+\frac{\delta_k}{\vu}\right)^{-1}\right),
    \end{equation}
    where~$\star$ holds with equality if~$\omega_k=\evec_{\dimF-k+1}$ for each~$k\in\{1,\ldots,\dimF\}$ and~$\star\star$ holds with equality if~$\omega_k=\evec_k$ for each~$k\in\{1,\ldots,\dimF\}$.
\end{proposition}
\ifbodyproofs\input{proofs/voi-bounds}\fi

Proposition~\ref{prop:voi-bounds} says that the sample~$\samp$ is most valuable when the eigenvectors of~$\Sigma$ and~$\gram$ are maximally ``aligned'': when~$\evec_k=\omega_k$ for each~$k\in\{1,\ldots,\dimF\}$ and hence~$\emat=\Omega$.
Then~$\samp$ contains more information about components of~$\theta$ with larger prior variances.
In contrast, the sample is \emph{least} valuable when the eigenvectors of~$\Sigma$ and~$\gram$ are maximally ``\emph{mis}-aligned'': when~$\evec_k=\omega_{\dimF-k+1}$ for each~$k\in\{1,\ldots,\dimF\}$.
Then~$\samp$ contains \emph{less} information about components of~$\theta$ with larger prior variances.

\subsection{Optimal samples}
\label{sec:optimal}

Suppose the eigenvectors of~$\Sigma$ and~$\gram$ are maximally aligned (and hence~$\emat=\Omega$).
Then, by Proposition~\ref{prop:voi-bounds}, the value~$\voi(\samp)$ of~$\samp$ rises when the trace
\[ \trace(\Var(\theta\mid\samp))=\sum_{k=1}^\dimF\left(\frac{1}{\lambda_k}+\frac{\delta_k}{\vu}\right)^{-1} \]
of the posterior variance matrix falls.
This trace depends on the eigenvalues~$\delta_1,\ldots,\delta_\dimF$ of~$\gram$, which are non-negative, non-increasing, and sum to~$n$.%
\footnote{
\label{fn:gram-trace}
Indeed
\[ \sum_{k=1}^\dimF\delta_k=\trace(\gram)=\trace\left(\sum_{i=1}^nw^{(i)}(w^{(1)})^T\right)\overset{\star}{=}\sum_{i=1}^n\trace\left((w^{(i)})^Tw^{(i)}\right)\overset{\star\star}{=}n, \]
where~$\star$ uses the linearity and cyclic property of matrix traces, and~$\star\star$ uses the fact that~$\norm{w^{(i)}}=1$ for each~$i$.
}
So~$\voi(\samp)$ is maximized when~$\delta_1,\ldots,\delta_\dimF$ solve
\begin{equation}
    \label{eq:optimal-problem}
    \min_{\delta_1,\ldots,\delta_\dimF\in\R}\ \ 
    \sum_{k=1}^\dimF\left(\frac{1}{\lambda_k}+\frac{\delta_k}{\vu}\right)^{-1}
    \ \ \text{subject to}\ \ 
    \delta_1\ge\ldots\ge\delta_\dimF\ge0\ \ \text{and}\ \sum_{k=1}^\dimF \delta_k=n.
\end{equation}
Proposition~\ref{prop:optimal} describes a solution to~\eqref{eq:optimal-problem}.
It uses the integer
\begin{equation}
    \label{eq:dimSopt}
    \dimSopt\equiv\max\left\{k\in\{1,\ldots,\dimF\}:\sum_{j=1}^k\frac{1}{\lambda_j}+\frac{n}{\vu}\ge\frac{k}{\lambda_k}\right\}
\end{equation}
to provide a sharp upper bound
\begin{equation}
    \label{eq:optimal-voi}
    \voi^*\equiv\frac{1}{\dimF}\left(\sum_{k=1}^\dimSopt\lambda_k-(\dimSopt)^2\left(\sum_{k=1}^\dimSopt\frac{1}{\lambda_k}+\frac{n}{\vu}\right)^{-1}\right)
\end{equation}
on the value of~$\samp$.%
\footnote{
Proposition~\ref{prop:optimal} echoes \citeapos{Liang-etal-2022-ECTA} Theorem~1, which says that if there are two unknown states (which \citeauthor{Liang-etal-2022-ECTA} call ``attributes''), then one should prioritize learning about the state with more prior variance.
}

\begin{proposition}
    \label{prop:optimal}
    Define
    \begin{equation}
        \label{eq:optimal-delta}
        \delta_k^*\equiv\begin{cases}
            \frac{n}{\dimSopt}+\vu\left(\frac{1}{\dimSopt}\sum_{j=1}^{\dimSopt}\frac{1}{\lambda_j}-\frac{1}{\lambda_k}\right) & \text{if}\ k\le\dimSopt \\
            0 & \text{if}\ k>\dimSopt
        \end{cases}
    \end{equation}
    for each~$k\in\{1,\ldots,\dimF\}$.
    Then~$\voi(\samp)\le\voi^*$ with equality if~$\samp$ induces Gram matrix
    \begin{equation}
        \label{eq:optimal-gram}
        \gram=\sum_{k=1}^\dimF\delta_k^*\evec_k\evec_k^T.
    \end{equation}
\end{proposition}
\ifbodyproofs\input{proofs/optimal}\fi

We call the sample ``optimal'' if it induces the Gram matrix~\eqref{eq:optimal-gram}.
The agent can construct such a sample as follows: for each~$k\in\{1,\ldots,\dimF\}$, collect~$\delta_k^*$ observations with covariate~$\evec_k$.%
\footnote{
This may be infeasible for two reasons:
(i)~the eigenvalues~$d_1^*,\ldots,d_\dimF^*$ may not be integers;
(ii)~the agent may not be able to choose~$\evec_1,\ldots,\evec_\dimF$ as covariates (since, e.g., it would require him to combine negative quantities of fertilizers).
We abstract from these issues for convenience and expositional clarity. 
}
Then the outcomes~$y^{(1)},\ldots,y^{(n)}$ are pure signals of the coefficients~$\gamma_1,\ldots,\gamma_\dimF$.
For example, suppose~$w^{(1)}=\evec_1$.
Then, since~$\evec_1,\ldots,\evec_\dimF$ are orthonormal, we have
\begin{align*}
    y^{(1)}
    &= \theta^Tw^{(1)}+u^{(1)} \\
    &= \left(\sum_{k=1}^\dimF\gamma_k\evec_k\right)^T\evec_1+u^{(1)} \\
    &= \gamma_1+u^{(1)}
\end{align*}
An optimal sample contains pure signals of~$\gamma_1,\ldots,\gamma_\dimSopt$, the coefficients in~\eqref{eq:theta-decomposition} that contribute most to the states' prior variances.
However, it provides no information about~$\gamma_{\dimSopt+1},\ldots,\gamma_\dimF$, the coefficients that contribute least to the states' prior variances.
Thus, the agent optimally ``regularizes'' by focusing on the coefficients that ``matter'' and ignoring those that do not.
The number~$\dimSopt$ that ``matter'' grows as the sample size~$n$ grows.
We call~$\dimSopt$ the ``rank'' of an optimal sample because it is the rank of the Gram matrix~\eqref{eq:optimal-gram}.

If~$\samp$ is optimal, then the posterior variance matrix~$\Var(\theta\mid\samp)$ has~$k^\text{th}$ largest eigenvalue
\[ \left(\frac{1}{\lambda_k}+\frac{\delta_k^*}{\vu}\right)^{-1}=\begin{cases}
    \dimSopt\left(\sum_{j=1}^{\dimSopt}\frac{1}{\lambda_j}+\frac{n}{\vu}\right)^{-1} & \text{if}\ k\le\dimSopt \\
    \lambda_k & \text{if}\ k>\dimSopt
\end{cases} \]
and trace%
\footnote{
The two terms on the RHS of~\eqref{eq:optimal-trace} correspond to the ``sampling'' and ``extrapolation'' errors discussed in Appendix Section~\ref{asec:oos}.
}
\begin{equation}
    \label{eq:optimal-trace}
    \sum_{k=1}^\dimF\left(\frac{1}{\lambda_k}+\frac{\delta_k^*}{\vu}\right)^{-1}=(\dimSopt)^2\left(\sum_{k=1}^{\dimSopt}\frac{1}{\lambda_k}+\frac{n}{\vu}\right)^{-1}+\sum_{k>\dimSopt}\lambda_k.
\end{equation}
The eigenvalues of~$\Var(\theta\mid\samp)$ are the posterior variances of the unknown coefficients~$\gamma_1,\ldots,\gamma_\dimF$.
So if~$\samp$ is optimal, then it equates the posterior variances of~$\gamma_1,\ldots,\gamma_\dimSopt$ to each other and the posterior variances of~$\gamma_{\dimSopt+1},\ldots,\gamma_\dimF$ to their prior variances.%
\footnote{
This equality of large eigenvalues and ignorance of small eigenvalues is reminiscent of \citeapos{Arrow-1963-AER} theorem on the optimality of deductible insurance contracts.
Such contracts second-degree stochastically dominate all other contracts with the same premia \citep{Gollier-Schlesinger-1996-ET}.
They provide full coverage against risks above a minimum threshold.
Similarly, optimal samples provide ``full coverage against posterior variance'' above a minimum threshold.
}
Intuitively, the agent has a target variance and designs~$\samp$ so as to bring the posterior variances of~$\gamma_1,\ldots,\gamma_\dimF$ below that target.%
\footnote{
This strategy is called ``reverse water-filling'' in rate-distortion theory---see \citet[Chapter~10]{Cover-Thomas-2006-}.
It also appears in \citeapos{Ilut-Valchev-2025-} model of abstract reasoning.
}
This minimizes the trace~\eqref{eq:optimal-trace} given the sample size~$n$.

\subsection{Mean-preserving spreads}
\label{sec:mps}

Finally, consider the eigenvalues~$\lambda_1,\ldots,\lambda_\dimF$ of the prior variance matrix~$\Sigma$.
Let~$\ecdf:(0,\infty)\to[0,1]$ be their (empirical) cumulative distribution function (hereafter ``CDF''):
\begin{align}
    \ecdf(z)
    &= \frac{\abs{\{k\in\{1,\ldots,\dimF\}:\lambda_k\le z\}}}{\dimF} \label{eq:eigenvalue-cdf}
\end{align}
for all~$z>0$.
A ``mean-preserving spread'' (hereafter ``MPS'') of~$\ecdf$ is a CDF~$\ecdf':(0,\infty)\to[0,1]$ such that
\begin{enumerate}

    \item[(i)]
    The distributions described by~$\ecdf$ and~$\ecdf'$ have the same mean:
    \[ \int_0^\infty z\,\der\ecdf(z)=\int_0^\infty z\,\der\ecdf'(z). \]

    \item[(ii)]
    For all~$z>0$, the area under~$\ecdf'$ from~0 to~$z$ is at least the area under~$\ecdf$ from~0 to~$z$:
    \[ \int_0^z\left(\ecdf'(t)-\ecdf(t)\right)\,\der t\ge0. \]

\end{enumerate}
These are the ``integral conditions'' from \cite{Rothschild-Stiglitz-1970-JET}.
Condition~(ii) says that~$\ecdf'$ has more weight in its tails than~$\ecdf$, capturing the idea of eigenvalues being more spread out.

We say~$\lambda_1,\ldots,\lambda_\dimF$ ``undergo a MPS'' when their CDF~\eqref{eq:eigenvalue-cdf} undergoes a MPS.
This changes the trace of the posterior variance matrix without changing the trace of~$\Sigma$.
So if~$\lambda_1,\ldots,\lambda_\dimF$ undergo a MPS, then the agent's posterior expected loss changes but his prior expected loss does not.
This makes MPSs useful for analyzing how the value of~$\samp$ depends on the distribution of~$\lambda_1,\ldots,\lambda_\dimF$.
We discuss this dependence in Section~\ref{sec:vck} and Appendix Section~\ref{asec:voi}, in which we state results that depend on the following lemma:

\begin{lemma}
    \label{lem:mps}
    Let~$\lambda_k>0$ and~$\lambda_k'>0$ be non-increasing in~$k\in\{1,\ldots,\dimF\}$, and let~$\ecdf$ and~$\ecdf'$ be their CDFs defined as in~\eqref{eq:eigenvalue-cdf}.
    The following are equivalent:
    \begin{itemize}

        \item[(i)]
        $\ecdf'$ is a mean-preserving spread of~$\ecdf$.

        \item[(ii)]
        $\sum_{k=1}^\dimF g(\lambda_k')\ge\sum_{k=1}^\dimF g(\lambda_k)$ for all convex functions~$g:(0,\infty)\to\R$.

        \item[(iii)]
        $\sum_{j=1}^k\lambda_j'\ge\sum_{j=1}^k\lambda_j$ for each~$k\in\{1,\ldots,\dimF\}$, with equality when~$k=\dimF$.

        \item[(iv)]
        $\sum_{j=k}^\dimF\lambda_j'\le\sum_{j=k}^\dimF\lambda_j$ for each~$k\in\{1,\ldots,\dimF\}$, with equality when~$k=1$.

    \end{itemize}
\end{lemma}
\ifbodyproofs\input{proofs/mps}\fi

For example, consider the prior variance matrix~\eqref{eq:pairwise-Sigma} constructed in Example~\ref{eg:pairwise}.
This matrix has eigenvalues~$\lambda_1=(1+\rho(\dimF-1))\vm$ and~$\lambda_2=\cdots=\lambda_\dimF=(1-\rho)\vm$.
Their~$k^\text{th}$ partial sum
\[ \sum_{j=1}^k\lambda_j=\left(k+\rho(\dimF-k)\right)\vm \]
is increasing in~$\rho$ when~$k<\dimF$ and constant in~$\rho$ when~$k=\dimF$.
Thus, by Lemma~\ref{lem:mps}, the eigenvalues of~\eqref{eq:pairwise-Sigma} undergo a MPS when~$\rho$ rises.

\section{Value of conceptual knowledge}
\label{sec:vck}

Whereas information is valuable insofar as it helps the agent make better decisions (i.e., take actions~$a_1,\ldots,a_\dimF$ that estimate the states~$\theta_1,\ldots,\theta_\dimF$ more accurately), conceptual knowledge is valuable insofar as it helps him obtain better information.

We formalize this idea as follows.
Suppose the agent collects an optimal sample with value~$\voi^*$ (see Section~\ref{sec:optimal}).
Designing this sample relies on his conceptual knowledge: his mental model~\eqref{eq:theta-decomposition} of~$\theta$ as a combination of concepts with different explanatory powers.
If the agent did not have this knowledge, then he would use the ``na\"ive'' prior~$\prior\z=\Ncal(\mu,\Sigma\z)$ described in Section~\ref{sec:knowledge}.
He would assume~$\theta$ has prior variance~$\Sigma\z=\overline\lambda I_\dimF$, a matrix with eigenvalues~$\lambda_1\z=\cdots=\lambda_\dimF\z=\overline\lambda$.
So, by analogy to~\eqref{eq:dimSopt} and~\eqref{eq:optimal-voi}, the agent's optimal sample would have rank
\begin{align*}
    \dimS\z
    &\equiv \max\left\{k\in\{1,\ldots,\dimF\}:\sum_{j=1}^k\frac{1}{\lambda_j\z}+\frac{n}{\vu}\ge\frac{k}{\lambda_k\z}\right\} \\
    &= \dimF
\end{align*}
and value
\begin{align*}
    \voi\z
    &\equiv \frac{1}{\dimF}\left(\sum_{k=1}^{\dimS\z}\lambda_k\z-(\dimS\z)^2\left(\sum_{k=1}^{\dimS\z}\frac{1}{\lambda_k\z}+\frac{n}{\vu}\right)^{-1}\right) \\
    &= \frac{\overline\lambda\tau}{\dimF+\tau},
\end{align*}
where
\[ \tau\equiv\frac{n/\vu}{1/\overline\lambda} \]
indexes the precision of the data relative to the agent's prior.
This sample would provide equal information about each state: the optimal Gram matrix~\eqref{eq:optimal-gram} would have equal eigenvalues.%
\footnote{
Indeed, if~$\lambda_1,\ldots,\lambda_\dimF$ are equal (to~$\overline\lambda$), then~$\delta_1^*,\ldots,\delta_\dimF^*$ are also equal (to~$n/\dimF$).
}
Intuitively, if the agent did not know which concepts had more explanatory power, then he would have no reason to prioritize some components of~$\theta$ over others when collecting data, and so he would collect the same amount on every component.

The true and na\"ive prior variance matrices~$\Sigma$ and~$\Sigma\z$ have equal traces, and so replacing the true prior~$\prior=\Ncal(\mu,\Sigma)$ with the na\"ive prior~$\prior\z=\Ncal(\mu,\Sigma\z)$ does not change the agent's prior expected loss (see Section~\ref{sec:voi}).
But the two priors imply different optimal samples and different posterior expected losses~\eqref{eq:expected-loss}.
The difference
\begin{align*}
    \vck
    &\equiv \voi^*-\voi\z \notag \\
    &= \frac{\overline\lambda}{\dimF}\left(\sum_{k=1}^\dimSopt\frac{\lambda_k}{\overline\lambda}-(\dimSopt)^2\left(\sum_{k=1}^\dimSopt\frac{\overline\lambda}{\lambda_k}+\tau\right)^{-1}-\frac{\dimF\tau}{\dimF+\tau}\right)
\end{align*}
between~$\voi^*$ and~$\voi\z$ equals the decline in the agent's posterior expected loss from having conceptual knowledge and using it to design an optimal sample.

Accordingly, we call~$\vck$ the ``value of conceptual knowledge.''
This value depends on the eigenvalues~$\lambda_1,\ldots,\lambda_\dimF$ of~$\Sigma$ and the precision index~$\tau$ (which jointly determine the rank~$\dimSopt$).
We characterize this dependence in Theorems~\ref{thm:vck-eigenvalues} and~\ref{thm:vck-tau}.

\begin{theorem}
    \label{thm:vck-eigenvalues}
    The value~$\vck$ of conceptual knowledge
    \begin{enumerate}

        \item[(i)]
        is non-negative,

        \item[(ii)]
        equals zero when the eigenvalues~$\lambda_1,\ldots,\lambda_\dimF$ are equal, and

        \item[(iii)]
        does not fall when~$\lambda_1,\ldots,\lambda_\dimF$ undergo a mean-preserving spread.

    \end{enumerate}
\end{theorem}
\ifbodyproofs\input{proofs/vck-eigenvalues}\fi

Theorem~\ref{thm:vck-eigenvalues} says that conceptual knowledge is more valuable when states are more reducible.
If a few common concepts explain most of states' prior variances, then the agent gains a lot from identifying those concepts and learning about the corresponding coefficients~$\gamma_1,\ldots,\gamma_\dimSopt$ (i.e., ``asking the right questions'').
In contrast, if every concept has the same explanatory power, then the agent gains nothing from identifying those concepts because he designs the same sample as he would if he was na\"ive.

\begin{theorem}
    \label{thm:vck-tau}
    There is a finite threshold~$\tau'\ge0$ such that the value~$\vck$ of conceptual knowledge is increasing in the precision index~$\tau$ if and only if~$\tau<\tau'$.
    This threshold equals zero if and only if~$\lambda_1,\ldots,\lambda_\dimF$ are equal.
    Moreover,
    \[ \lim_{\tau\to\infty}\vck=0. \]
\end{theorem}
\ifbodyproofs\input{proofs/vck-tau}\fi

Theorem~\ref{thm:vck-tau} says that the value of conceptual knowledge is non-monotone in the precision index~$\tau$ (and, thus, the sample size~$n\equiv\tau\vu/\overline\lambda$).
This is because raising~$\tau$ has two effects: %
%
\begin{enumerate}

    \item[(i)]
    it gives the agent more information about the unknown coefficients~$\gamma_1,\ldots,\gamma_\dimSopt$, raising the gain from knowing which concepts to focus on;

    \item[(ii)]
    it leads the agent to learn about more coefficients (i.e., it raises~$\dimSopt$), lowering the gain from knowing which concepts to focus on.

\end{enumerate}
The first effect dominates the second precisely when~$\tau<\tau'$.
The threshold~$\tau'$ equals zero if and only if~$\lambda_1,\ldots,\lambda_\dimF$ are equal, in which case conceptual knowledge has no value because it does not change the optimal sample from what a na\"ive agent would design.

Theorem~\ref{thm:vck-tau} also says that the value of conceptual knowledge vanishes as~$\tau$ (and thus~$n$) grows without bound.
This is because the agent's posterior becomes less dependent on his prior as~$\tau$ grows and is independent in the limit as~$\tau\to\infty$.
Intuitively, if the agent has infinite data, then he does not benefit from doing ``pre-data PCA'' (see Section~\ref{sec:knowledge}) because he can do traditional (post-data) PCA.
Having access to unlimited data washes out the benefit of knowing what data to collect.
But this washout relies on having \emph{unrestricted} access: the agent must be able to choose covariates~$w^{(1)},\ldots,w^{(n)}$ that span the~$\dimF$-dimensional Euclidean space containing the state vector~$\theta$.
If the covariates do not span~$\R^\dimF$, then~$\samp$ may contain no information about some components of~$\theta$, the agent's posterior expected loss may be arbitrarily large, and the value of~$\samp$ may be arbitrarily small.
We illustrate this possibility in Appendix Section~\ref{asec:oos}.

As an illustration of Theorems~\ref{thm:vck-eigenvalues} and~\ref{thm:vck-tau}, consider the prior variance matrix~\eqref{eq:pairwise-Sigma} constructed in Example~\ref{eg:pairwise}.
Its eigenvalues are equal when~$\rho=0$ and undergo a MPS when~$\rho\in[0,1)$ rises.
So, by Theorem~\ref{thm:vck-eigenvalues}, the value~$\vck$ of conceptual knowledge equals zero when~$\rho=0$ and is non-decreasing in~$\rho$.
Moreover, by Theorem~\ref{thm:vck-tau}, there is a threshold~$\tau'\ge0$ such that~$\vck$ is increasing in the precision index~$\tau$ if and only if~$\tau<\tau'$.
We characterize this threshold below.

\begin{proposition}
    \label{prop:vck-pairwise}
    Suppose the states~$\theta_1,\ldots,\theta_\dimF$ have equal prior variances~$\vm>0$ and pairwise correlation~$\rho\in[0,1)$.
    Then the value~$\vck$ of conceptual knowledge
    \begin{enumerate}

        \item[(i)]
        equals zero when~$\rho=0$,

        \item[(ii)]
        is increasing in~$\rho$, and

        \item[(iii)]
        is increasing in the precision index~$\tau$ if and only if
        \begin{equation}
            \tau<\frac{\rho\dimF}{1+\rho(\dimF-1)}. \label{eq:vck-pairwise-condition}
        \end{equation}

    \end{enumerate}
\end{proposition}
\ifbodyproofs\input{proofs/vck-pairwise}\fi

Whereas Theorem~\ref{thm:vck-eigenvalues} implies~$\vck$ is non-decreasing in~$\rho$, Proposition~\ref{prop:vck-pairwise} says~$\vck$ is increasing in~$\rho$.
This is because Theorem~\ref{thm:vck-eigenvalues} holds for an arbitrary MPS, which may not affect the largest~$\dimSopt$ eigenvalues and thus may not change the value~\eqref{eq:optimal-voi} of an optimal sample.
But this is impossible for the MPS induced by raising~$\rho$, which raises the largest eigenvalue~$\lambda_1=\left(1+\rho(\dimF-1)\right)\vm$ of~\eqref{eq:pairwise-Sigma}.

\begin{figure}[!t]
    \centering
    \includegraphics[width=0.8\linewidth]{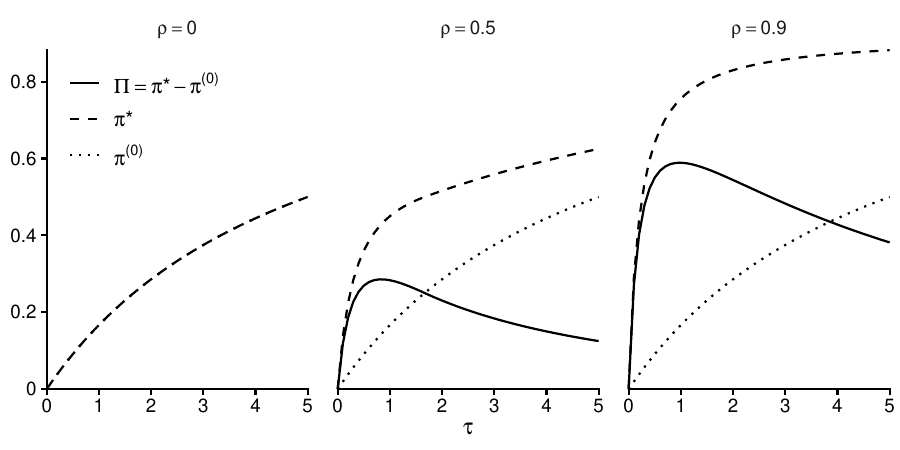}
    \caption{Values~$\vck$, $\voi^*$, and~$\voi\z$ when~$\theta$ has prior variance~\eqref{eq:pairwise-Sigma} and~$(\dimF,\vm,\vu)=(5,1,1)$}
    \label{fig:pairwise}
\end{figure}
Figure~\ref{fig:pairwise} shows how~$\vck$ depends on~$\rho$ and~$\tau$.
If~$\rho=0$, then the values~$\voi^*$ and~$\voi\z$ of the optimal samples collected by agents with and without conceptual knowledge are equal, and so~$\vck=0$ for all~$\tau>0$.
In contrast, if~$\rho>0$, then~$\voi^*>\voi\z$ for all~$\tau>0$ and hence~$\vck>0$ for all~$\tau>0$.
Both~$\voi^*$ and~$\voi\z$ grow as~$\tau$ grows, and~$\voi^*$ grows faster if and only if~\eqref{eq:vck-pairwise-condition} holds.
Thus~$\vck=\voi^*-\voi\z$ is increasing in~$\tau$ if and only if~\eqref{eq:vck-pairwise-condition} holds.

\section{Deeper conceptual knowledge}
\label{sec:deeper}

Next, we study the value of ``deepening'' the agent's conceptual knowledge.
We model this process as follows.
Suppose the agent knows the trace
\[ \trace(\Sigma)=\sum_{k=1}^\dimF\lambda_k \]
of the true prior variance matrix~$\Sigma$ and its first~$\dimM\in\{0,1,\ldots,\dimF\}$ eigenpairs~$(\lambda_k,\evec_k)$, but does not know the last~$(\dimF-\dimM)$ eigenpairs.
Intuitively, he knows the~$\dimM$ concepts with the most explanatory power, but does not know the~$(\dimF-\dimM)$ concepts with the least explanatory power.
So he assumes the components of~$\theta$ orthogonal to~$\evec_1,\ldots,\evec_\dimM$ have equal prior variances; specifically, he assumes~$\theta$ has prior variance
\begin{equation}
    \label{eq:SigmaM}
    \Sigma\M=\sum_{k\le\dimM}\lambda_k\evec_k\evec_k^T+\lambda_\dimF\M\left(I_\dimF-\sum_{k\le\dimM}\evec_k\evec_k^T\right),
\end{equation}
where
\[ \lambda_\dimF\M\equiv\frac{1}{\dimF-\dimM}\sum_{k>\dimM}\lambda_k \]
is the mean of the smallest~$(\dimF-\dimM)$ eigenvalues of~$\Sigma$.%
\footnote{
If~$\dimM=\dimF$, then the bracketed term in~\eqref{eq:SigmaM} equals zero and so~$\lambda_\dimF^{(\dimF)}$ can be defined arbitrarily.
We define~$\lambda_\dimF^{(\dimF)}\equiv\lambda_\dimF$.
}
The matrix~\eqref{eq:SigmaM} has the same trace as~$\Sigma$ but (possibly) different eigenvalues; its~$k^\text{th}$ largest eigenvalue
\[ \lambda_k\M\equiv\begin{cases}
    \lambda_k & \text{if}\ k\le\dimM \\
    \lambda_\dimF\M & \text{if}\ k>\dimM
\end{cases} \]
equals that of~$\Sigma$ if and only if~$k\le\dimM$.
The eigenvalues of~$\Sigma\M$ have mean
\[ \frac{1}{\dimF}\sum_{k=1}^\dimF\lambda_k\M=\overline\lambda \]
independently of~$\dimM$.
Likewise~$\lambda_\dimF\z=\overline\lambda$ by definition.
Thus~$\Sigma\z=\overline\lambda I_\dimF$ is the na\"ive prior variance matrix discussed in Sections~\ref{sec:knowledge} and~\ref{sec:vck}.
The parameter~$\dimM$ interpolates between~$\Sigma\z$ and~$\Sigma^{(\dimF)}=\Sigma$.
It captures the ``depth'' of the agent's conceptual knowledge: the larger is~$\dimM$, the more concepts he knows and the richer is his mental model of~$\theta$.%
\footnote{
Since~$\lambda_1\ge\cdots\ge\lambda_\dimF$ (by assumption), there are non-increasing returns to knowing more concepts (i.e., increasing~$\dimM$), since each additional concept contributes a non-increasing share of states' prior variances.
Intuitively, when the agent acquires conceptual knowledge, he prioritizes concepts with more explanatory power.
For example, he takes classes or reads textbooks that provide ``high-level summaries'' before ``digging into the details.''
}\textsuperscript{,}%
\footnote{
For example, a farmer could deepen his conceptual knowledge by learning about different nutrients that crops need and fertilizers supply (e.g., nitrogen, phosphorus, and potassium).
He could also learn about the nuances of nutrients he already knows about.
Nutrients can take different chemical forms (e.g., nitrogen can take the form of ammonium and nitrate) that have different effects when applied to different crops.
Learning about these differences gives the farmer new concepts that explain fertilizers' overall effects.
}

We say the agent has ``$\dimM$-deep conceptual knowledge'' if his prior on~$\theta$ has variance~$\Sigma\M$.
Suppose he has such knowledge and designs an optimal sample.
Then, by analogy to~\eqref{eq:dimSopt} and~\eqref{eq:optimal-voi}, this sample has rank
\[ \dimSM\equiv\max\left\{k\in\{1,\ldots,\dimF\}:\overline\lambda\left(\frac{k}{\lambda_k\M}-\sum_{j=1}^k\frac{1}{\lambda_j\M}\right)\le\tau\right\} \]
and value
\[ \voi\M\equiv\frac{\overline\lambda}{\dimF}\left(\sum_{k=1}^\dimSM\frac{\lambda_k\M}{\overline\lambda}-\left(\dimSM\right)^2\left(\sum_{k=1}^\dimSM\frac{\overline\lambda}{\lambda_k\M}+\tau\right)^{-1}\right), \]
where~$\tau\equiv n\overline\lambda/\vu$ is the precision index defined in Section~\ref{sec:vck}.
For example, letting~$\dimM=0$ yields the rank~$\dimS\z$ and value~$\voi\z$ of an optimal sample collected by a na\"ive agent.
We refer to the difference
\[ \vck\M\equiv\voi\M-\voi\z \]
between~$\voi\M$ and~$\voi\z$ as the ``value of~$\dimM$-deep conceptual knowledge.''
We characterize the relationship between~$\dimSM$ and~$\dimM$ in Lemma~\ref{lem:deeper-dimS}, and the relationship between~$\vck\M$ and~$\dimM$ in Theorem~\ref{thm:deeper-vck}.

\begin{lemma}
    \label{lem:deeper-dimS}
    There is a threshold~$\dimM'\in\{0,\ldots,\dimF\}$ such that
    \[ \dimSM=\begin{cases}
        \dimF & \text{if}\ \dimM\le\dimM' \\
        \dimM & \text{if}\ \dimM'<\dimM<\dimSopt \\
        \dimSopt & \text{if}\ \dimM\ge\dimSopt
    \end{cases} \]
    for each~$\dimM\in\{0,\ldots,\dimF\}$.
    This threshold is non-decreasing in the precision index~$\tau$.
\end{lemma}
\ifbodyproofs\input{proofs/deeper-dimS}\fi

\begin{theorem}
    \label{thm:deeper-vck}
    The value~$\vck\M$ of $\dimM$-deep conceptual knowledge
    \begin{enumerate}

        \item[(i)]
        is non-negative,

        \item[(ii)]
        equals zero when~$\dimM=0$,

        \item[(iii)]
        is non-decreasing in~$\dimM$, and

        \item[(iv)]
        equals the value~$\vck$ of full knowledge when~$\dimM\ge\dimSopt$.

    \end{enumerate}
\end{theorem}
\ifbodyproofs\input{proofs/deeper-vck}\fi

Theorem~\ref{thm:deeper-vck} says that deeper knowledge is (weakly) more valuable.
Intuitively, knowing more concepts allows the agent to design samples that provide more payoff-relevant information.

The value of~$\dimM$-deep conceptual knowledge is bounded above by the value~$\vck^{(\dimF)}=\vck$ of ``full'' knowledge, and attains this bound when~$\dimM\ge\dimSopt$.
Thus, the agent gains no additional value from knowing more than the~$\dimSopt$ concepts with the most explanatory power.
This is because he ignores the other~$(\dimF-\dimSopt)$ concepts when he designs samples (since~$\delta_k^*=0$ for each~$k>\dimSopt$), so knowing those concepts does not change his optimal sample.

For example, suppose the true prior variance matrix~$\Sigma$ has~$k^\text{th}$ largest eigenvalue
\[ \lambda_k=\frac{\dimF\decay(1-\decay)^{k-1}}{1-(1-\decay)^\dimF} \]
with~$0<\decay<1$.
Then~$\lambda_1,\ldots,\lambda_\dimF$ are strictly positive, have mean~$\overline\lambda=1$,
are constant in the limit as~$\decay\to0$, and undergo a MPS as~$\decay$ rises.%
\footnote{
For each~$k\in\{1,\ldots,\dimF\}$ we have~$\lambda_k\to1$ as~$\decay\to0$ by L'H\^opital's rule.
Moreover, the partial sum
\begin{align*}
    \sum_{j=1}^k\lambda_k
    &= \frac{\dimF\left(1-(1-\decay)^k\right)}{1-(1-\decay)^\dimF}
\end{align*}
is non-decreasing in~$\decay$ and is constant in~$\decay$ when~$k=\dimF$.
Thus, by Lemma~\ref{lem:mps}, the eigenvalues~$\lambda_1,\ldots,\lambda_\dimF$ undergo a MPS when~$\decay$ rises.
}
This parameter determines the rate
\[ \frac{\lambda_{k+1}-\lambda_k}{\lambda_k}=-\decay \]
at which~$\lambda_k$ decays as~$k$ grows.
Intuitively, the larger is~$\decay$, the faster concepts' marginal explanatory power falls.
Thus, if~$\decay$ is larger, then states are more reducible.

\begin{figure}[!t]
    \centering
    \includegraphics[width=0.8\linewidth]{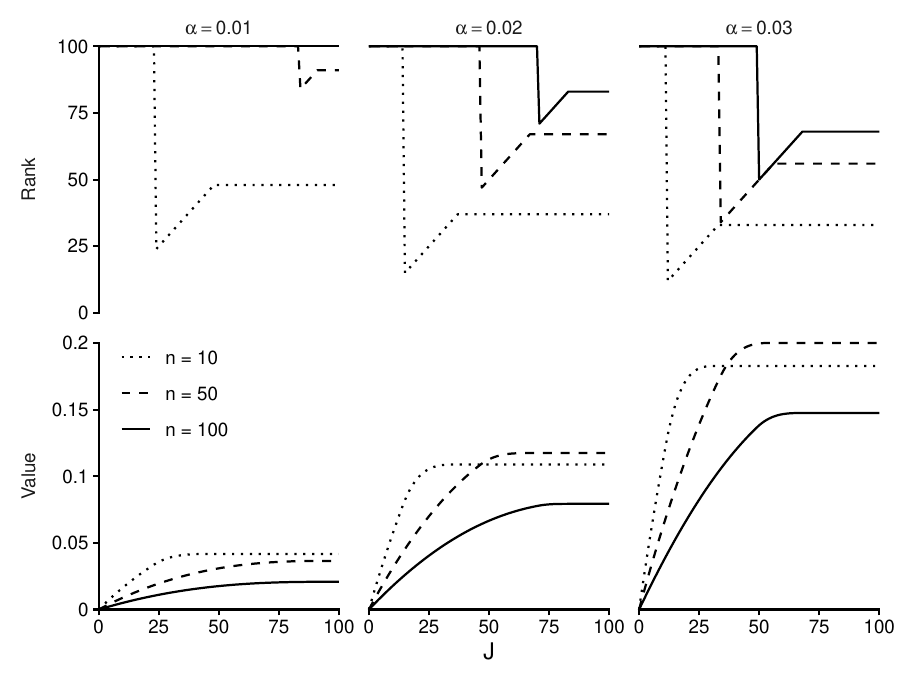}
    \caption{Rank~$\dimSM$ and value~$\vck\M$ when~$\lambda_{k+1}=(1-\decay)\lambda_k$ and~$(\dimF,\overline\lambda,\vu)=(100,1,1)$}
    \label{fig:deeper}
\end{figure}
Figure~\ref{fig:deeper} shows how~$\dimSM$ and~$\vck\M$ depend on~$\dimM$ when~$(\dimF,\overline\lambda,\vu)=(100,1,1)$ and~$\lambda_k$ decays at rate~$\decay\in\{0.01,0.02,0.03\}$.
If~$\dimM$ is sufficiently small, then the agent designs an optimal sample with rank~$\dimS\M=100$;
otherwise, he designs a sample with rank~$\dimSM=\min\{\dimM,\dimSopt\}$.
The threshold depth at which he switches from~100 to~$\min\{\dimM,\dimSopt\}$ rises as the sample size~$n$ rises, consistent with Lemma~\ref{lem:deeper-dimS}.%
\footnote{
Here~$\tau=n$ because~$\overline\lambda=\vu=1$.
}
The value~$\vck\M$ of~$\dimM$-deep conceptual knowledge is increasing in~$\dimM$ when~$\dimM<\dimSopt$ and constant in~$\dimM$ when~$\dimM\ge\dimSopt$.
This value is increasing in~$\decay$, consistent with Theorem~\ref{thm:vck-eigenvalues}: conceptual knowledge is more valuable when~$\lambda_1,\ldots,\lambda_\dimF$ are more spread out.
Likewise~$\vck\M$ is non-monotone in~$n$, consistent with Theorem~\ref{thm:vck-tau}: raising~$n$ allows the agent to learn more about the ``in-sample'' coefficients~$\gamma_1,\ldots,\gamma_\dimSM$ (raising~$\vck\M$), but also prompts him to expand his sample and learn about more coefficients (lowering~$\vck\M$).

\section{More concepts or more data?}
\label{sec:versus}

Finally, we study the trade-off between conceptual and statistical knowledge.
Our model in Section~\ref{sec:deeper} provides a formal language for describing this trade-off: would the agent rather know more concepts (i.e., increase the depth~$\dimM$) or have more data (i.e., increase the sample size~$n$)?

Suppose the agent has~$\dimM$-deep conceptual knowledge and designs an optimal sample of size~$n$.
The value~$\voi\M$ of this sample indexes the agent's welfare: it is larger when his minimized posterior expected loss is smaller.
Lemma~\ref{lem:versus-voi} says the agent is better off with deeper knowledge or more data.
Intuitively, if he knows more concepts, then he can ``ask better questions.''
If he has more data, then he can obtain better answers to his questions, making his posterior beliefs more precise and his expected loss lower.
\begin{lemma}
    \label{lem:versus-voi}
    The value~$\voi\M$ of an optimal sample designed by an agent with~$\dimM$-deep conceptual knowledge is
    \begin{enumerate}

        \item[(i)]
        non-increasing in~$\dimM$ and

        \item[(ii)]
        increasing in the sample size~$n$.

    \end{enumerate}
\end{lemma}
\ifbodyproofs\input{proofs/versus-voi}\fi

Now suppose the agent has a target value~$\voi_0\ge0$.
Let
\[ n_{\pi_0}\M\equiv\min\{n\ge0:\pi\M\ge\pi_0\} \]
be the minimum sample size necessary to attain this value.
This size is smaller when the agent knows more concepts and when states are more reducible:

\begin{theorem}
    \label{thm:versus-size}
    Fix~$\voi_0\ge0$.
    The minimum sample size~$n_{\pi_0}\M$ necessary to design a sample with value~$\voi_0$
    \begin{enumerate}
        
        \item[(i)]
        is non-increasing in the depth~$\dimM$ of the agent's conceptual knowledge and

        \item[(ii)]
        does not rise when the eigenvalues~$\lambda_1,\ldots,\lambda_\dimF$ undergo a mean-preserving spread.

    \end{enumerate}
\end{theorem}
\ifbodyproofs\input{proofs/versus-size}\fi

Theorem~\ref{thm:versus-size} says that if the agent knows more concepts, then he can attain the same welfare with fewer observations, especially when states are highly reducible.
This is because he can design better samples and extract more value from each observation, lowering the number he needs to attain the target~$\voi_0$.

\begin{figure}[!t]
    \centering
    \includegraphics[width=0.8\linewidth]{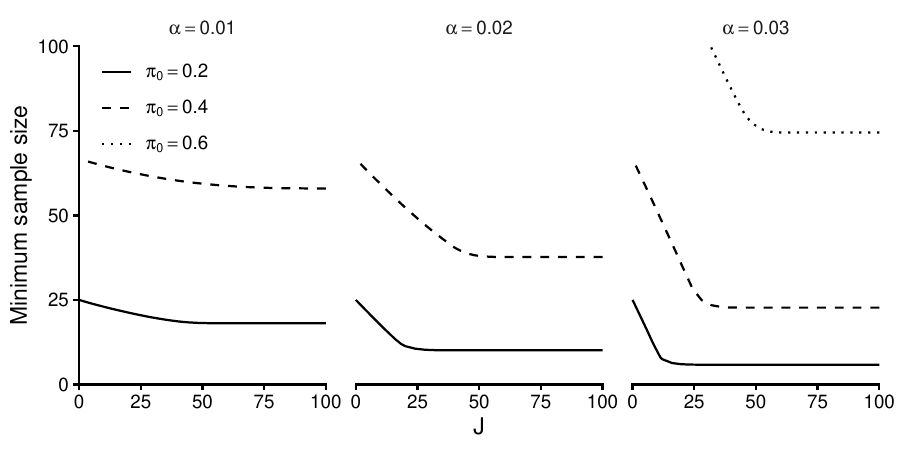}
    \caption{Minimum sample sizes~$n_{\voi_0}\M$ when~$\lambda_{k+1}=(1-\decay)\lambda_k$ and~$(\dimF,\overline\lambda,\vu)=(100,1,1)$}
    \label{fig:versus}
\end{figure}
As in illustration of Theorem~\ref{thm:versus-size}, suppose the true eigenvalues~$\lambda_1,\ldots,\lambda_\dimF$ have mean~$\overline\lambda=1$ and decay at rate~$\decay$ as in Section~\ref{sec:deeper}.
Figure~\ref{fig:versus} shows how the minimum sample size~$n_{\pi_0}\M$ depends on the welfare target~$\voi_0$ and depth~$\dimM$ when~$(\dimF,\vu)=(100,1)$ and~$\decay\in\{0.01,0.02,0.03\}$.
Given~$\voi_0$, the size~$n_{\pi_0}\M$ is decreasing in~$\dimM$ when~$\dimM<\dimSopt$ and constant in~$\dimM$ when~$\dimM\ge\dimSopt$.
Intuitively, if the agent knows too few concepts, then he cannot design samples that focus on all of the ``right'' concepts.
Giving him more concepts empowers him to design better samples, extract more value from each observation, and require fewer observations to attain~$\voi_0$.
However, once he knows all the ``right'' concepts, giving him more does not change how he designs samples or the marginal value of each observation.
Then the only way to for him to obtain \emph{more valuable} data is to obtain \emph{more} data, thus making~$n_{\pi_0}\M$ constant in~$\dimM\ge\dimSopt$.

The curves in Figure~\ref{fig:versus} are indifference curves: they trace out sets of depth-size pairs~$(\dimM,n)$ that allow the agent to attain different welfare targets~$\voi_0$.
The slope of each curve equals the marginal rate of substitution (hereafter ``MRS'') between concepts and data.
Intuitively, this MRS captures the number of observations the agent would give up to know another concept.
It depends on the depth-size pair~$(\dimM,n)$, the target~$\voi_0$, and the parameter~$\decay$ indexing states' reducibility.
Raising this parameter raises the rank~$\dimSopt$ of the optimal sample he would design if he had full knowledge (see Figure~\ref{fig:deeper}).
So raising~$\decay$ can have three effects on the MRS between concepts and data:
\begin{enumerate}

    \item
    If~$\dimM<\dimSopt$ before and after~$\decay$ rises, then the MRS rises in absolute value;

    \item
    If~$\dimM<\dimSopt$ before~$\decay$ rises but~$\dimM\ge\dimSopt$ after, then the MRS falls in absolute value (to zero);

    \item
    If~$\dimM\ge\dimSopt$ before and after~$\decay$ rises, then the MRS remains unchanged (at zero).

\end{enumerate}
So if states are more reducible, then the MRS between concepts and data may be higher or lower, depending on how many concepts the agent already knows.
We defer analyzing this dependence to future research.

\section{Related literature}
\label{sec:literature}

\paragraph{Value of information}

A large literature studies the instrumental value of information.
Seminal work by \cite{Blackwell-1951-ProceedingsoftheSecondBerkeleySymposiumonMathematicalStatisticsandProbability,Blackwell-1953-AoMS} shows that information sources can be ordered: one is more informative than another if it produces a sufficient statistic for the other's information.
\cite{Howard-1966-IEEE} and \cite{Raiffa-Schlaifer-1961-} link informativeness to instrumental value, defining the ``value of information'' as the difference in expected payoffs with and without it.
More recently, \cite{Brooks-etal-2024-AER} extend Blackwell's order to settings with multiple sources, 
\cite{Frankel-Kamenica-2019-AER} characterize measures of information's value, and \cite{Whitmeyer-2025-JPE} studies value-increasing transformations in abstract decision problems.%
\footnote{
Others study information's value in specific decision problems.
For example, \cite{Lehmann-1988-AnnStat} and \cite{Athey-Levin-2018-ResEcon} study its value in monotone decision problems, \cite{Persico-2000-ECTA} studies its value in auctions, and \cite{Cabrales-etal-2013-AER} study its value in investment decisions.
}

Our contribution is to illustrate \emph{why} some information is more valuable than others, focusing on a specific decision problem relevant to economists and statisticians.
We show that the value of a sample depends on its alignment with the agent's conceptual structure: the prior variance matrix.
Proposition~\ref{prop:voi-bounds} provides sharp bounds: a sample is most (least) valuable when the eigenvectors of the induced Gram and prior variance matrices are maximally aligned (mis-aligned).
Proposition~\ref{prop:optimal} characterizes the design of maximally valuable samples, showing that they focus on dimensions with the most prior variance.
Theorem~\ref{thm:vck-eigenvalues} quantifies how much this focus raises samples' value.

We separate samples' \emph{design} (i.e., the covariates they contain) from their \emph{quality} (i.e., signals' precision).
\citeauthor{Blackwell-1951-ProceedingsoftheSecondBerkeleySymposiumonMathematicalStatisticsandProbability} and his successors take samples as given, and ask how valuable they are.
In contrast, we show how one can use prior knowledge to make samples more valuable.
Thus, our framework allows us to transform the abstract notion of ``informativeness'' into concrete experimental design guidance.

\paragraph{Model-based inference}

Another literature emphasizes the importance of models for interpreting data.
This literature includes arguments by statisticians \citep[e.g.,][]{Cox-1990-StatSci} and computer scientists \cite[e.g.,][]{Wolpert-1996-NeuralComp} that models are essential for inference.
In economics, \cite{Lucas-1976-CRCSPP} argues that models are necessary to make counterfactual predictions; \cite{Koopmans-1947-REStat} and \cite{Wolpin-2013-} make similar arguments.
\cite{Manski-2003-} shows formally that data alone are insufficient for inference: one must also make assumptions about the data-generating process (i.e., impose a model).

Recognizing that models are important, many authors study the consequences of not knowing the true model or using the wrong model.
\cite{Gilboa-Schmeidler-1989-JME}, \cite{Klibanoff-etal-2005-ECTA}, and \cite{Marinacci-2015-JEEA} develop decision theories under model uncertainty, while \cite{Cerreia-Vioglio-etal-2025-REStud} and \cite{Spiegler-2016-QJE} develop decision theories under potential mis-specification.

Others study the competitive and political forces that shape the models agents use \citep{Dasaratha-etal-2025-,Izzo-etal-2023-AJPS,MontielOlea-Prat-2025-}.
These forces arise in the literature that treats models as ``narratives'': stories that explain observed events \citep{Aina-2025-,Eliaz-etal-2025-JEEA,Levy-etal-2022-AER,Schwartzstein-Sunderam-2021-AER}.%
\footnote{
Papers in the ``narratives'' literature focus on settings where many models are plausibly correct (e.g., financial markets and political campaigns).
In contrast, many real-world decisions are made in settings where there is a single, objectively correct model that can be learned through education and introspection.
For example,  there is an objectively correct set of mechanisms through which fertilizers help crops grow, which a farmer could discover experimentally or be taught as in \cite{Sankar-etal-2025-}.
Such settings are our focus in this paper.
}

In contrast, we treat models as dimension-reduction devices.
In our framework, a model specifies
(i)~which components of the states vary together (i.e., the eigenvectors~$\evec_1,\ldots,\evec_\dimF$) and
(ii)~how much variance each component contributes (i.e., the eigenvalues~$\lambda_1,\ldots,\lambda_\dimF$).
The distribution of these variances determines how much the agent's model reduces the~$\dimF$-dimensional state space to a lower-dimensional subspace.
Theorem~\ref{thm:vck-eigenvalues} characterizes the instrumental value of this reduction.

We also formalize what it means to have a ``richer'' model: knowing more eigenpairs of the prior variance matrix (see Section~\ref{sec:deeper}).
We introduce a parameter~$\dimM$ that interpolates between having a diffuse model ($\dimM=0$), a partial model ($\dimM<\dimF$), and a complete model ($\dimM=\dimF$).%
\footnote{
Thus~$\dimM$ offers an alternative to the ``completeness'' measure proposed by \cite{Fudenberg-etal-2022-JPE}.
}\textsuperscript{,}%
\footnote{
\cite{Mailath-Samuelson-2020-AER} argue that ``in practice, people work with models that are deliberately incomplete, including the most salient variables and excluding others.''
Indeed, Theorem~\ref{thm:deeper-vck} implies that the agent does not benefit from using models embedding depths greater than the rank~$\dimSopt$ of an optimal sample.
}
Introducing this parameter allows us to study the trade-off between having a richer model (i.e., increasing~$\dimM$) or more data (i.e., increasing the sample size~$n$).%
\footnote{
In contrast, \cite{Dominitz-Manski-2017-REStud} study the trade-off between having more data or ``better'' data.
}
Theorem~\ref{thm:versus-size} characterizes this trade-off.

Thus, our analysis contributes by describing \emph{why} models are important and quantifying how important they are.
Our approach formalizes \citeapos{Jackson-2019-TheFutureofEconomicDesign} claim that models are useful because they guide experimental design: our agent uses his model to design more valuable samples.

\paragraph{Cognition}

A third literature studies the cognitive foundations of human learning.
\cite{Johnson-Laird-1983-} argues that mental models are core inferential tools.
\cite{Murphy-2002-} explains how humans build models from concepts.
\cite{Tenenbaum-etal-2011-Science} and \cite{Mitchell-2021-NYAS} discuss how concepts help humans generalize.
\cite{Felin-Holweg-2024-StratSci} discuss how humans generate novel insights via directed experimentation.

We present a formal economic model that embeds these ideas.
Our agent has a mental model of states as combinations of concepts.
These concepts allow him to generalize: he can use signals of one state to make inferences about another.
They also facilitate directed experimentation: knowing more concepts allows the agent to design more informative samples (see Theorems~\ref{thm:deeper-vck} and~\ref{thm:versus-size}).

\cite{Ilut-Valchev-2025-} present a related but distinct model.
They isolate two learning modes---``abstract reasoning'' and ``integrating experience''---that correspond to our notions of conceptual and statistical knowledge.
\citeauthor{Ilut-Valchev-2025-} study a dynamic setting, and focus on the ``learning traps'' that arise from reasoning too little or having the wrong data.
In contrast, we study a static setting, and focus on the benefits of reasoning correctly and having the ``right'' data.

For our agent, the ``right'' data focus on an endogenous number of high-variance dimensions.
This form of selective attention is optimal given his decision problem and learning environment.
In contrast, \cite{Gabaix-2014-QJE} and \cite{Schwartzstein-2014-JEEA} take selective attention as given, and study its consequences for decision-making and belief updating.
\cite{Koszegi-Szeidl-2013-QJE}, like us, study an agent who focuses on high-variance dimensions.
But \citeauthor{Koszegi-Szeidl-2013-QJE} assume such behavior as part of their model, whereas we derive it as a consequence of our model.

\paragraph{Human-machine comparisons}

Finally, our paper connects to the literature comparing humans and machines.
\cite{Tenenbaum-etal-2011-Science} discuss how humans use concepts to learn from limited data, whereas machines rely on recognizing patterns in large sets of data \citep{Goodfellow-etal-2016-,Halevy-etal-2009-IEEE}.
Our framework formalizes this distinction: conceptual knowledge compensates for data scarcity (see Theorem~\ref{thm:versus-size}) but loses its value when data are abundant (see Theorem~\ref{thm:vck-tau}).
Moreover, we identify a mechanism through which humans learn faster than machines: humans have conceptual knowledge that empowers them to ``ask the right questions.''

Our findings parallel \citeapos{Iakovlev-Liang-2025-}.
They study the ``value of context'': the gain in predictive performance from choosing the ``right'' model, rather than relying on pattern recognition.
\citeauthor{Iakovlev-Liang-2025-} show that the value of context vanishes with abundant data.
Likewise, the value of conceptual knowledge vanishes with abundant data.
However, our framework offers additional insight: the value of conceptual knowledge rises before it falls, changing direction at the threshold~$\tau'$ identified in Theorem~\ref{thm:vck-tau}.

Empirical work compares human and machine predictions in many domains, such as crime \citep{Berk-2017-JExpCriminol,Kleinberg-etal-2018-QJE}, finance \citep{Jansen-etal-2025-ManSci}, and healthcare \citep{Agarwal-etal-2023-,Mullainathan-Obermeyer-2022-QJE}.
A common finding is that human predictions are less accurate because they are based on intuitive, potentially biased judgments, whereas machine predictions are based on statistical models \citep{Angelova-etal-2025-REStud,Hoffman-etal-2018-QJE}.
Our agent's predictions are also based on a model: his prior on the state vector.
We study the gain from using a restrictive model (i.e., imposing the true prior~$\prior$) rather than a diffuse model (i.e., imposing the na\"ive prior~$\prior\z$).

Finally, \cite{Andrews-etal-2025-} study model-based predictions with those made by ``black box'' algorithms, finding that models dominate at out-of-domain transfer.
This empirical result supports one of our theoretical claims: models embed conceptual knowledge that improves out-of-sample predictive performance (see Appendix Section~\ref{asec:oos}).

\section{Conclusion}
\label{sec:conclusion}

This paper introduces a simple idea: whereas information is valuable because it leads to better decisions, conceptual knowledge is valuable because it leads to better information.
We formalize this idea and study its consequences.
Conceptual knowledge is more valuable when payoff-relevant unknowns are more ``reducible'': when they can be explained with fewer common concepts.
Its value is non-monotone in the quantity and quality of available data, and vanishes with infinite data.
Deeper conceptual knowledge is (weakly) more valuable and compensates for having less data.
So we can improve interventions that give people data by giving them conceptual tools to interpret data.
We can also improve human-AI interactions by recognizing humans' comparative advantage: knowing how to ``ask the right questions.''

Stemming from this paper are several avenues for future research.
One is to analyze the trade-off between concepts and data in a consumer choice setting.
This would require specifying the ``price'' of acquiring concepts vis-\`a-vis observations of data.
Given a price schedule, one could ask many questions about concepts and data:
Are they complements or substitutes?
Are they normal or inferior?
How do these statuses depend on the states' reducibility?

Another avenue is to make our framework dynamic.
For example, the agent could take actions that generate outcomes observed by future agents.
This would allow us to study how conceptual knowledge is ``discovered'' and passed down to new generations.
This discovery process has been studied by other authors \citep{Carnehl-Schneider-2025-ECTA,Gans-2025-}; blending their models and insights with ours may bear fruit.

A third avenue is to study the role of conceptual knowledge in social learning and technology adoption.
This paper and its empirical sibling \citep{Sankar-etal-2025-} came from studying agricultural settings, where heterogeneity in outcomes and the inability to generalize slows learning \citep{Conley-Udry-2010-AER,Munshi-2004-JDevEcon,Tjernstrom-2017-} and technology adoption \citep{Alidaee-2023-,BenYishay-Mobarak-2019-REStud,Laajaj-Macours-2025-ECTA}.
These frictions could be removed by giving people conceptual tools that enable them to generalize.
This paper offers a framework for studying such tools and their value.

{%
\raggedright
\bibliographystyle{apalike}
\bibliography{references}
}

\appendix

\makeatletter
\renewcommand\theequation{\thesection\@arabic\c@equation}
\renewcommand\thefigure{\thesection\@arabic\c@figure}
\renewcommand\thelemma{\thesection\@arabic\c@lemma}
\renewcommand\theproposition{\thesection\@arabic\c@proposition}
\renewcommand\thesubsection{\thesection\@arabic\c@subsection}
\makeatother

\clearpage
\section{Additional material}
\label{app:additional}

\setcounter{equation}{0}
\setcounter{figure}{0}
\setcounter{lemma}{0}
\setcounter{proposition}{0}

\subsection{Connection to statistical learning}
\label{asec:statistical-learning}

This section connects our paper to the literatures on machine and statistical learning, which study how to derive predictive functions from data.%
\footnote{
See \cite{Bishop-2006-} or \cite{Hastie-etal-2009-} for textbook treatments.
}
First, we show that our framework (described in Section~\ref{sec:framework}) can be used to study Bayesian learning about real-valued functions.
Second, we show how the agent's prior derives from his approximating model of an unknown function.

\subsubsection{Function-state equivalence}

Suppose there is a finite set~$\Xcal$ of ``inputs'' and a square-summable function~$f:\Xcal\to\R$ belonging to the set
\[ \Fcal\equiv\left\{g\in\R^\Xcal:\sum_{x\in\Xcal}(g(x))^2<\infty\right\} \]
of such functions.
Endow~$\Fcal$ with the inner product defined by
\[ \inner{g}{g'}=\sum_{x\in\Xcal}g(x)g'(x) \]
for all pairs~$(g,g')\in\Fcal\times\Fcal$.
Let~$\abs{\Xcal}=\dimF$, let~$\ell:\Xcal\to\{1,\ldots,\dimF\}$ be a bijection, and define
\[ \phi_k(x)\equiv\begin{cases}
    1 & \text{if}\ \ell(x)=k \\
    0 & \text{otherwise}
\end{cases} \]
for each~$x\in\Xcal$ and~$k\in\{1,\ldots,\dimF\}$.
Then the indicator functions~$\phi_1,\ldots,\phi_\dimF$ form an orthonormal basis~$\Bcal\equiv\{\phi_k\}_{k=1}^\dimF$ for the inner product space~$(\Fcal,\inner{.}{.})$.
Now let~$\theta_1,\ldots,\theta_\dimF$ be the coordinates of~$f$ over~$\{\phi_k\}_{k=1}^\dimF$:
\[ f=\sum_{k=1}^\dimF\theta_k\phi_k. \]
The agent knows~$\phi_1,\ldots,\phi_\dimF$ but not~$\theta\equiv(\theta_1,\ldots,\theta_\dimF)$, so learning about~$f$ is equivalent to learning about~$\theta$.%
\footnote{
Assuming~$\theta$ is normally distributed is equivalent to assuming~$\{f(x)\}_{x\in\Xcal}$ follows a Gaussian process.
Such processes arise in the economic literature on learning and information acquisition---see, e.g., \cite{Bardhi-2024-ECTA}, \cite{Davies-2024-}, \cite{Ilut-Valchev-2025-}, or \cite{Laajaj-Macours-2025-ECTA}.
See also \citet[Section~6.4]{Bishop-2006-} or \cite{Rasmussen-Williams-2006-} for more information about Gaussian processes and their applications.
}
Moreover, suppose the agent draws an input~$x\in\Xcal$ uniformly at random, and predicts the ``output''~$y\in\R$ with conditional distribution
\[ y\mid x,f\sim\Ncal\left(f(x),\vu\right) \]
given~$x$ and~$f$.
His ``prediction rule''~$\hat{f}\in\Fcal$ maps each realization of~$x$ to a prediction~$\hat{f}(x)$ of~$y$.
This prediction induces a posterior mean squared error (MSE)
\[ \E\left[(y-\hat{f}(x))^2\mid x,\samp\right]=\E\left[(f(x)-\hat{f}(x))^2\mid x,\samp\right]+\vu \]
given~$x$ and the sample~$\samp$, where~$\E$ takes expectations with respect to the joint prior distribution of input-output pairs.
The agent chooses~$\hat{f}$ to minimize the mean posterior MSE across realizations of~$x$:
\begin{equation}
    \label{eq:prediction-rule}
    \hat{f}\in\argmin_{g\in\Fcal}\frac{1}{\abs{\Xcal}}\sum_{x\in\Xcal}\E\left[\left(y-g(x)\right)^2\mid x,\samp\right].
\end{equation}
The optimal actions~$a_1,\ldots,a_\dimF$ defined by~\eqref{eq:action-vector} are precisely the coordinates of~$\hat{f}$ over the basis~$\Bcal$.
By Lemma~\ref{eq:action-vector}, these coordinates equal the posterior mean coordinates of~$f$.
The minimized mean posterior MSE
\[ \min_{g\in\Fcal}\frac{1}{\abs{\Xcal}}\sum_{x\in\Xcal}\E\left[(y-\hat{f}(x))^2\mid x,\samp\right]=\E[\loss(\theta,a)\mid\samp]+\vu \]
equals the expected loss~\eqref{eq:expected-loss} plus a constant~$\vu$ that arises due to the irreducible randomness in the outcome~$y$.
Thus, the prediction problem~\eqref{eq:prediction-rule} is equivalent to the choice problem~\eqref{eq:action-vector}.

\subsubsection{Approximating models}
\label{asec:models}

Suppose the agent knows about a collection~$\psi_1,\ldots,\psi_\dimM\in\Fcal$ of ``features'' that (partially) mediate the relationship between inputs and outputs.%
\footnote{
This~$\dimM$ is the same as the depth parameter defined in Section~\ref{sec:deeper}.
}
These features are linearly independent (but not necessary orthonormal) elements of the function space~$\Fcal$.
They map inputs to known, measurable quantities.
For example, if the inputs are fertilizers, then the features could map fertilizers to quantities of different nutrients.

The agent uses~$\psi_1,\ldots,\psi_\dimM$ to build a ``model''~$m\in\Fcal$ approximating the unknown function~$f$.
This model is a linear combination of features: there is a(n unknown) vector~$\beta\equiv(\beta_1,\ldots,\beta_\dimM)\in\R^\dimM$ such that
\[ m(x)=\sum_{k=1}^\dimM\beta_k\psi_k(x) \]
for each~$x\in\Xcal$.
Then the derivative
\[ \parfrac{m(x)}{\psi_k(x)}=\beta_k \]
of~$m(x)$ with respect to~$\psi_k(x)$ does not depend on the input~$x$.
In this way, the model~$m$ captures the generalizable structure of~$f$ that is common to all inputs.
In contrast, the model's approximation error
\[ \epsilon\equiv f-m \]
captures the idiosyncrasies specific to each input.

The agent uses his knowledge of~$\psi_1,\ldots,\psi_\dimM$ to construct his prior~$\prior$ on~$\theta$.
First, he identifies the subspace
\[ \Fcal^m\equiv\mathrm{span}\{\psi_1,\ldots,\psi_\dimM\} \]
of~$\Fcal$ spanned by the features.
It corresponds to a subspace
\[ \Theta^m\equiv\left\{\vartheta\in\Theta:\sum_{k=1}^\dimF\vartheta_k\phi_k\in\Fcal^m\right\} \]
of the Euclidean space~$\Theta\equiv\R^\dimF$ containing the unknown coordinate vector~$\theta$.
Concretely, if~$\theta^j\equiv(\theta_1^j,\ldots,\theta_\dimF^j)$ contains the (known) coordinates of the~$j^\text{th}$ feature~$\psi_j$ over the orthonormal basis~$\Bcal$, then
\[ \Theta^m=\mathrm{span}\{\theta^1,\ldots,\theta^\dimM\} \]
is the subspace of~$\Theta$ spanned by the vectors~$\theta^1,\ldots,\theta^\dimM$.

Next, the agent constructs an orthonormal basis~$\{\evec_k\}_{k=1}^\dimM$ for~$\Theta^m$ (e.g., by applying the Gram-Schmidt process to~$\theta^1,\ldots,\theta^\dimM$).
If~$\dimM<\dimF$, then he also constructs an orthonormal basis~$\{\evec_k\}_{k=\dimM+1}^\dimF$ for the orthogonal complement
\[ \Theta^\epsilon\equiv\left\{\vartheta\in\Theta:\vartheta^T\vartheta'=0\ \text{for all}\ \vartheta'\in\Theta^m\right\} \]
of~$\Theta^m$.
Then~$\{\evec_k\}_{k=1}^\dimF$ is an orthonormal basis for~$\Theta$.
Letting~$\gamma\equiv(\gamma_1,\ldots,\gamma_\dimF)$ contain the coordinates of~$\theta$ over~$\{\evec_k\}_{k=1}^\dimF$ yields~\eqref{eq:theta-decomposition}; the eigendecomposition~\eqref{eq:Sigma-eigendecomposition} of~$\Sigma$ follows.
Thus, the agent's prior on~$\theta$ derives from his prior on~$\gamma$, which derives from his knowledge of~$\psi_1,\ldots,\psi_\dimM$ (which define the model~$m$).

\subsection{Value of~\texorpdfstring{$\samp$}{S}}
\label{asec:voi}

Consider the sample~$\samp$.
Proposition~\ref{aprop:voi} says that the value~$\voi(\samp)$ of~$\samp$ is non-negative, grows as~$\samp$ grows, and shrinks as~$\vu$ grows.

\begin{proposition}
    \label{aprop:voi}
    The value~$\voi(\samp)$ of the sample~$\samp$
    \begin{enumerate}

        \item[(i)]
        is non-negative,

        \item[(ii)]
        does not fall when~$\samp$ gains observations, and

        \item[(iii)]
        falls when~$\vu$ rises.

    \end{enumerate}
\end{proposition}
\ifbodyproofs\input{proofs/voi}\fi

Sections~\ref{asec:singleton}--\ref{asec:oos} discuss the values of samples with specific structures.

\subsubsection{Singleton samples}
\label{asec:singleton}

Suppose~$\samp=\{(w^{(1)},y^{(1)})\}$ contains a single observation.
Then the Gram matrix~$\gram=w^{(1)}(w^{(1)})^T$ has eigenvalues~$\delta_1=1$ and~$\delta_2=\cdots=\delta_\dimF=0$.
Substituting them into~\eqref{eq:voi-bounds} gives us bounds on the value of~$\samp$:

\begin{proposition}
    \label{aprop:singleton-voi}
    Suppose~$\samp=\{(w^{(1)},y^{(1)})\}$ contains a single observation.
    Then its value
    \begin{equation}
        \label{eq:singleton-voi}
        \voi(\samp)=\frac{(w^{(1)})^T\Sigma^2 w^{(1)}}{\dimF\left((w^{(1)})^T\Sigma w^{(1)}+\vu\right)}
    \end{equation}
    satisfies
    \begin{equation}
        \label{eq:singleton-voi-bounds}
        \frac{\lambda_\dimF^2}{\dimF(\lambda_\dimF+\vu)}
        \overset{\star}{\le}
        \voi(\samp)
        \overset{\star\star}{\le}
        \frac{\lambda_1^2}{\dimF(\lambda_1+\vu)},
    \end{equation}
    where~$\star$ holds with equality if~$\Sigma w^{(1)}=\lambda_\dimF w^{(1)}$ and~$\star\star$ holds with equality if~$\Sigma w^{(1)}=\lambda_1w^{(1)}$.
\end{proposition}
\ifbodyproofs\input{proofs/singleton-voi}\fi

The value of~$\{(w^{(1)},y^{(1)})\}$ is largest when~$w^{(1)}$ is an eigenvector of~$\Sigma$ with corresponding eigenvalue~$\lambda_1=\max\{\lambda_1,\ldots,\lambda_\dimF\}$.
It is smallest when~$w^{(1)}$ is an eigenvector of~$\Sigma$ with corresponding eigenvalue~$\lambda_\dimF=\min\{\lambda_1,\ldots,\lambda_\dimF\}$.
Intuitively, the more ``weight''~$w^{(1)}$ puts on directions in which the prior variance of~$\theta$ is large, the more valuable it is to observe~$(w^{(1)},y^{(1)})$ because the larger is the variance reduction it delivers.
This is especially true when there are few dimensions (i.e., $\dimF$ is small) and when the signal~$y^{(1)}$ is precise (i.e., $\vu$ is small).

\begin{figure}[!t]
    \centering
    \includegraphics[width=0.8\linewidth]{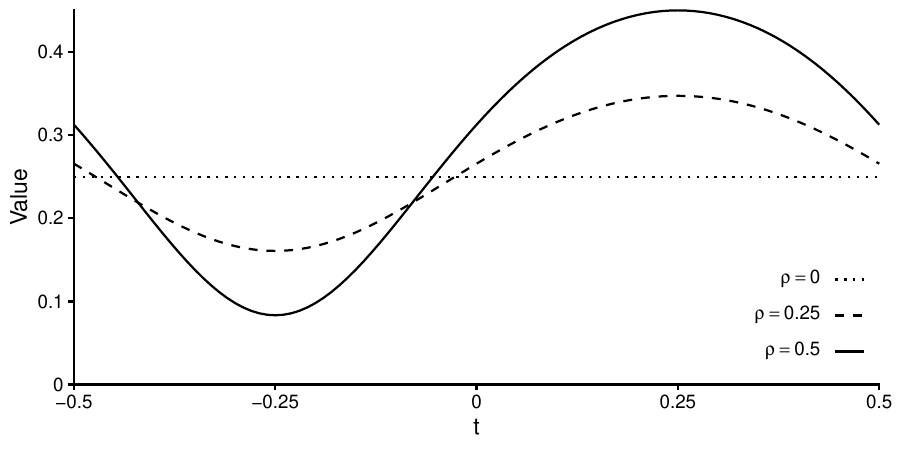}
    \caption{Value~\eqref{eq:singleton-voi-pairwise} of observing~$\samp=\{((\sin(\pi t),\cos(\pi t)),y^{(1)})\}$ when~$\theta$ has prior variance~\eqref{eq:pairwise-Sigma} and~$(\dimF,\vm,\vu)=(2,1,1)$}
    \label{afig:singleton}
\end{figure}

For example, suppose~$\Sigma$ is the matrix~\eqref{eq:pairwise-Sigma} constructed in Example~\ref{eg:pairwise}.
Let~$\dimF=2$ and suppose~$\samp=\{(w^{(1)},y^{(1)})\}$ contains a single observation with
\[ w^{(1)}=(\sin(\pi t),\cos(\pi t)) \]
and~$-1/2\le t\le1/2$.
Increasing~$t$ from~$-1/2$ to~$1/2$ rotates~$w^{(1)}$ clockwise from~$(-1,0)$ to~$(1,0)$.
The value%
\footnote{
We obtain~\eqref{eq:singleton-voi-pairwise} by substituting~$\dimF=2$, the prior variance matrix~\eqref{eq:pairwise-Sigma}, the covariate~$w^{(1)}=(\sin(\pi t),\cos(\pi t))$, and the sample size~$n=1$ into~\eqref{eq:singleton-voi}.
}
%
\begin{equation}
    \label{eq:singleton-voi-pairwise}
    \voi(\samp)=\frac{\left(1+2\rho\sin(2\pi t)+\rho^2\right)\sm^4}{2\left(\left(1+\rho\sin(2\pi t)\right)\vm+\vu\right)}
\end{equation}
of~$\samp$ attains its minimum
when~$t=-1/4$, in which case~$w^{(1)}=(-1/\sqrt{2},1/\sqrt{2})$ equals the unit eigenvector~$\evec_2$ of~$\Sigma$ with the smallest corresponding eigenvalue.
In contrast, the value of~$\samp$ attains its maximum
when~$t=1/4$, in which case~$w^{(1)}=(1/\sqrt{2},1/\sqrt{2})$ equals the unit eigenvector~$\evec_1$ of~$\Sigma$ with the largest corresponding eigenvalue.
Figure~\ref{afig:singleton} shows that~$\voi(\samp)$ rises monotonically as~$t$ rises from~$-1/4$ to~$1/4$, which lowers the angle between~$w^{(1)}$ and~$\evec_1$ from~$90^\circ$ to~$0^\circ$.

\subsubsection{Representative samples}
\label{asec:representative}

Suppose the covariates~$w^{(1)},\ldots,w^{(n)}$ in~$\samp$ are binary vectors: for each~$i\in\{1,\ldots,n\}$, there is an index~$k_i\in\{1,\ldots,\dimF\}$ such that~$w^{(i)}$ has~$k^\text{th}$ component
\[ w_k^{(i)}=\begin{cases}
    1 & \text{if}\ k=k_i \\
    0 & \text{otherwise}.
\end{cases} \]
Then each outcome
\[ y^{(i)}=\theta_{k_i}+u^{(i)} \]
is a ``pure signal'' of the state~$\theta_{k_i}$.
Moreover, the Gram matrix~$\gram$ is diagonal: its~${kk}^\text{th}$ entry
\[ \gram_{kk}=\Abs{\left\{i\in\{1,\ldots,\dimF\}:k_i=k\right\}} \]
counts the outcomes in~$\samp$ that are pure signals of~$\theta_k$.
If~$\gram_{11},\ldots,\gram_{\dimF\dimF}$ are equal (to~$n/\dimF$), then the eigenvalues of~$\gram$ are also equal (to~$n/\dimF$).
These eigenvalues characterize a sample that contains equal information about each state.

Accordingly, we say~$\samp$ is ``representative'' if it induces a Gram matrix with equal eigenvalues.
Then the lower and upper bounds in~\eqref{eq:voi-bounds} are equal, and so~$\samp$ has value
\begin{align*}
    \voi(\samp)
    &= \frac{1}{\dimF}\sum_{k=1}^\dimF\left(\lambda_k-\left(\frac{1}{\lambda_k}+\frac{n}{\dimF\vu}\right)^{-1}\right) \\
    &= \frac{1}{\dimF}\sum_{k=1}^\dimF\frac{n\lambda_k^2}{n\lambda_k+\dimF\vu}. \notag
\end{align*}
This value is larger when the eigenvalues~$\lambda_1,\ldots,\lambda_\dimF$ of~$\Sigma$ are more spread out:
\begin{proposition}
    \label{aprop:representative-mps}
    If~$\samp$ is representative, then its value~$\voi(\samp)$ does not fall when~$\lambda_1,\ldots,\lambda_\dimF$ undergo a mean-preserving spread.
\end{proposition}
\ifbodyproofs\input{proofs/representative-mps}\fi

If~$\samp$ is representative, then it contains equal information about each component of~$\theta$.
But there are diminishing returns to having more information about a given component.
So if the prior variances of~$\gamma_1,\ldots,\gamma_\dimF$ change in a mean-preserving way, then the increased reduction of the higher variances more than offsets the decreased reduction of the lower variances, thereby raising~$\voi(\samp)$.

For example, suppose~$\Sigma$ is the matrix~\eqref{eq:pairwise-Sigma} constructed in Example~\ref{eg:pairwise}.
Then the eigenvalues~$\lambda_1=\left(1+\rho(\dimF-1)\right)\vm$ and~$\lambda_2=\cdots=\lambda_\dimF=(1-\rho)\vm$ of~$\Sigma$ undergo a MPS when~$\rho$ rises.
So if~$\samp$ is representative, then its value
\begin{align*}
    \voi(\samp)
    &= \vm-\frac{1}{\dimF}\left(\frac{1}{\lambda_1}+\frac{n}{\dimF\vu}\right)^{-1}-\left(1-\frac{1}{\dimF}\right)\left(\frac{1}{\lambda_\dimF}+\frac{n}{\dimF\vu}\right)^{-1}
\end{align*}
must be non-decreasing in~$\rho$.
Indeed, the derivative
\begin{align*}
    \parfrac{\voi(\samp)}{\rho}
    &= \left(1-\frac{1}{\dimF}\right)\vm\left(\frac{1}{\lambda_\dimF^2}-\frac{1}{\lambda_1^2}\right)\left(1+\frac{n}{\dimF\vu}\right)^{-2}
\end{align*}
of~$\voi(\samp)$ with respect to~$\rho$ is non-negative because~$\lambda_\dimF\ge\lambda_1$.

\subsubsection{Non-spanning samples}
\label{asec:oos}

If the sample~$\samp$ is \emph{not} representative, then its value can fall when the eigenvalues of~$\Sigma$ undergo a MPS.
This happens, for example, when~$\samp=\{(w^{(1)},y^{(1)})\}$ is a singleton and~$w^{(1)}$ is an eigenvector of~$\Sigma$ corresponding to an eigenvalue that falls under the MPS.%
\footnote{
For example, if~$\theta$ has prior variance~\eqref{eq:pairwise-Sigma}, then the value of~$\samp=\{(w^{(1)},y^{(1)})\}$ is decreasing in~$\rho$ when~$w^{(1)}=\evec_2$.
}
Such a sample is ``non-spanning'': the rank
\[ \dimS\equiv\max\{k\in\{1,\ldots,\dimF\}:\delta_k>0\} \]
of the Gram matrix~$\gram$ is strictly less than~$\dimF$, so there are components of~$\theta$ about which~$\samp$ contains no information because they are outside the column space
\begin{align*}
    \mathrm{col}(G)
    &\equiv \mathrm{span}\{\omega_1,\ldots,\omega_\dimF\} \\
    &= \mathrm{span}\{w^{(1)},\ldots,w^{(n)}\}
\end{align*}
of~$\gram$.
The agent cannot learn about these components from~$\samp$ directly.
But he can learn about them indirectly if he knows how they covary with the components that belong to~$\mathrm{col}(G)$.

For example, suppose the observations in~$\samp$ are pure signals of~$\theta_1,\ldots,\theta_\dimS$.
Then
\[ \mathrm{col}(G)=\left\{v\in\R^\dimF:v_k=0\ \text{for each}\ k>\dimS\right\} \]
is the subspace of~$\R^\dimF$ spanned by the first~$\dimS$ standard basis vectors.
So the first~$\dimS$ components of~$\theta$ are ``on-support'' but the last~$(\dimF-\dimS)$ components are ``off-support.''
Let~$\theta_\samp\equiv(\theta_1,\ldots,\theta_\dimS)$ contain the first~$\dimS$ components of~$\theta$ and define
\[ \xi_k\equiv\Var(\theta_\samp)^{-1}\begin{bmatrix} \Sigma_{11} \\ \vdots \\ \Sigma_{\dimS k} \end{bmatrix} \]
for each~$k\in\{1,\ldots,\dimF\}$.%
\footnote{
The matrix~$\Var(\theta_\samp)$ is invertible because it is a leading principal submatrix of an invertible matrix.
}
Then the posterior variance matrix has trace
\begin{equation}
    \label{eq:oos-trace}
    \trace(\Var(\theta\mid\samp))=\underbrace{\sum_{k=1}^\dimS\Var(\theta_k\mid\samp)+\sum_{k>\dimS}\xi_k^T\Var(\theta_\samp\mid\samp)\xi_k}_{\text{Sampling error}}+\underbrace{\sum_{k>\dimS}\Var(\theta_k\mid\theta_\samp)}_{\text{Extrapolation error}},
\end{equation}
where
\begin{equation}
    \label{eq:oos-conditional-variance}
    \Var(\theta_k\mid\theta_\samp)=\Var(\theta_k)-\xi_k^T\Var(\theta_\samp)\xi_k
\end{equation}
is the prior variance of~$\theta_k$ left unexplained~$\theta_\samp$.%
\footnote{
We derive~\eqref{eq:oos-trace}--\eqref{eq:oos-conditional-variance-pairwise} in Appendix~\ref{app:proofs}.
}
The first two terms on the RHS of~\eqref{eq:oos-trace} are ``sampling errors'' that depend on how much information~$\samp$ contains about the on-support components~$\theta_1,\ldots,\theta_\dimS$.
The third term is an ``extrapolation error'' that depends on how much information these components contain about the off-support components~$\theta_{\dimS+1},\ldots,\theta_\dimF$.
Whereas the sampling error can be reduced by collecting more (or less noisy) data on~$\theta_1,\ldots,\theta_\dimS$, the extrapolation error cannot.
It can only be reduced by making the on-support components of~$\theta$ contain more informatio about the off-support components.

For example, suppose~$\Sigma$ is the matrix~\eqref{eq:pairwise-Sigma} constructed in Example~\ref{eg:pairwise}.
Then
\begin{equation}
    \label{eq:oos-conditional-variance-pairwise}
    \Var(\theta_k\mid\theta_\samp)=\frac{(1-\rho)(1+\rho\dimS)\vm}{1+\rho(\dimS-1)}
\end{equation}
for each~$k>\dimS$ and so the extrapolation error
\[ \sum_{k>\dimS}\Var(\theta_k\mid\theta_\samp)=\frac{(1-\rho)(1+\rho\dimS)(\dimF-\dimS)\vm}{1+\rho(\dimS-1)} \]
falls as the correlation~$\rho$ rises.
It equals~$(\dimF-\dimS)\vm$ when~$\rho=0$, in which case~$\theta_1,\ldots,\theta_\dimS$ provide no information about~$\theta_{\dimS+1},\ldots,\theta_\dimF$ and so~$\Var(\theta_k\mid\theta_\samp)=\Var(\theta_k)=\vm$ for each~$k>\dimS$.
It equals zero in the limit as~$\rho\to1$, in which case~$\theta_1,\ldots,\theta_\dimF$ are fully determined by the coefficient~$\gamma_1$ on their common component and so~$\Var(\theta_k\mid\theta_\samp)=\Var(\theta_k\mid\gamma_1)=0$ for each~$k\in\{1,\ldots,\dimF\}$.

If, in addition, the observations in~$\samp$ have no noise, then~$\Var(\theta_k\mid\samp)=0$ for each~$k\le\dimS$ and so~$\samp$ has value
\begin{align*}
    \lim_{\vu\to0}\voi(\samp)
    &= \frac{1}{\dimF}\left(\sum_{k=1}^\dimS\left(\Var(\theta_k)-0\right)+\sum_{k>\dimS}\left(\Var(\theta_k)-\Var(\theta_k\mid\theta_\samp)\right)\right) \\
    &= \left(1-\frac{(1-\rho)(1+\rho\dimS)(\dimF-\dimS)}{\left(1+\rho(\dimS-1)\right)\dimF}\right)\vm.
\end{align*}
This value rises as~$\rho$ rises, equals~$\dimS\vm/\dimF$ when~$\rho=0$, and equals~$\vm$ in the limit as~$\rho\to1$.
Taking the limit as~$\dimF\to\infty$ gives
\[ \lim_{\dimF\to\infty}\lim_{\vu\to0}\voi(\samp)=\frac{\rho^2\dimS\vm}{\left(1+\rho(\dimS-1)\right)}, \]
which is bounded away from zero if and only if~$\rho>0$.
So if there are many states, and the agent has noise-free data but a limited sampling frame, then his sample has value if and only if he knows the states have a common component that explains some of their prior variances.

This example highlights the importance of conceptual knowledge when making out-of-sample predictions.
If the agent had no conceptual knowledge (and so assumed~$\rho=0$), then his extrapolation error could be arbitrarily large and the value of his sample could be arbitrarily small.
Having conceptual knowledge allows him to use data on the on-support components of~$\theta$ to learn about the off-support components.
This lowers his extrapolation error and ensures his sample has \emph{some} value, even if the sampling frame is limited.

\subsection{KL divergences}
\label{asec:divergence}

In Section~\ref{sec:knowledge}, we claim the KL divergence~\eqref{eq:divergence} quantifies how much the agent's conceptual knowledge allows him to reduce~$\theta$.
Proposition~\ref{aprop:divergence} justifies this claim.
It says the KL divergence from the true prior~$\prior$ to the na\"ive prior~$\prior\z$ is (weakly) larger when states are more reducible (i.e., when the eigenvalues~$\lambda_1,\ldots,\lambda_\dimF$ of~$\Sigma$ are more spread out).

\begin{proposition}
    \label{aprop:divergence}
    The KL divergence from~$\prior$ to~$\prior\z$
    \begin{itemize}

        \item[(i)]
        is non-negative,

        \item[(ii)]
        equals zero when~$\lambda_1,\ldots,\lambda_\dimF$ are equal, and

        \item[(iii)]
        does not fall when~$\lambda_1,\ldots,\lambda_\dimF$ undergo a mean-preserving spread.

    \end{itemize}
\end{proposition}
\ifbodyproofs\input{proofs/divergence}\fi

For example, suppose~$\Sigma$ is the matrix~\eqref{eq:pairwise-Sigma} constructed in Example~\ref{eg:pairwise}.
This matrix has eigenvalues~$\lambda_1=(1+\rho(\dimF-1))\vm$ and~$\lambda_2=\cdots=\lambda_\dimF=(1-\rho)\vm$, which equal~$\overline\lambda=\vm$ when~$\rho=0$ and undergo a MPS when~$\rho$ rises (see Section~\ref{sec:mps}).
So, by Proposition~\ref{aprop:divergence}, the KL divergence
\[ \KLdiv=-\frac{\ln\left(1+\rho(\dimF-1)\right)+(\dimF-1)\ln\left(1-\rho\right)}{2} \]
from the true prior~$\prior$ to the na\"ive prior~$\prior\z$ must equal zero when~$\rho=0$ and be non-decreasing in~$\rho$.
Indeed
\begin{align*}
    \KLdiv\bigg\rvert_{\rho=0}
    &= -\frac{\ln(1)+(\dimF-1)\ln(1)}{2} \\
    &= 0,
\end{align*}
and
\begin{align*}
    \parfrac{}{\rho}\KLdiv
    &= \frac{\dimF-1}{2}\left(\frac{1}{1-\rho}-\frac{1}{1+\rho(\dimF-1)}\right) \\
    &\ge 0
\end{align*}
with equality if and only if~$\rho=0$.

\clearpage
\section{Proofs}
\label{app:proofs}

\addtocontents{toc}{\setcounter{tocdepth}{2}}
\setcounter{equation}{0}
\setcounter{figure}{0}
\setcounter{lemma}{0}
\setcounter{proposition}{0}

\subsection{Claims in Section~\ref{sec:preliminaries}}

\ifbodyproofs\else\input{proofs/expected-loss}\fi

\ifbodyproofs\else\input{proofs/posterior-variance}\fi

\ifbodyproofs\else\input{proofs/voi-bounds}\fi

\ifbodyproofs\else\input{proofs/optimal}\fi

\ifbodyproofs\else\input{proofs/mps}\fi

\subsection{Claims in Sections~\ref{sec:vck}--\ref{sec:versus}}

\ifbodyproofs\else\input{proofs/vck-eigenvalues}\fi

\ifbodyproofs\else\input{proofs/vck-tau}\fi

\ifbodyproofs\else\input{proofs/vck-pairwise}\fi

\ifbodyproofs\else\input{proofs/deeper-dimS}\fi

\ifbodyproofs\else\input{proofs/deeper-vck}\fi

\ifbodyproofs\else\input{proofs/versus-voi}\fi

\ifbodyproofs\else\input{proofs/versus-size}\fi

\subsection{Claims in Appendix Sections~\ref{asec:voi} and~\ref{asec:divergence}}

\ifbodyproofs\else\input{proofs/voi}\fi

\ifbodyproofs\else\input{proofs/singleton-voi}\fi

\ifbodyproofs\else\input{proofs/representative-mps}\fi

\ifbodyproofs\else\input{proofs/oos}\fi

\ifbodyproofs\else\input{proofs/divergence}\fi

\end{document}

%% file: preamble.tex
\usepackage[T1]{fontenc}
\usepackage[margin=1in,includefoot]{geometry}
\usepackage{mathpazo}
\usepackage{xcolor}

\usepackage[hidelinks]{hyperref}

\newcommand{\red}[1]{\textcolor{red}{#1}}

\newif\ifshowtodos
\showtodostrue
\newcommand{\TODO}[1]{\ifshowtodos\par{\color{red}\begin{quote}TODO:#1\end{quote}}\par\fi}

\usepackage{booktabs}
\usepackage{caption}
\usepackage{graphicx}

\let\oldincludegraphics\includegraphics%
\renewcommand{\includegraphics}[2][]{\IfFileExists{#2}{\oldincludegraphics[#1]{#2}}{\red{[FILE NOT FOUND]}}}

\usepackage{amsmath}
\usepackage{amssymb}
\usepackage{amsthm}

\DeclareMathOperator*{\argmax}{\arg\max}
\DeclareMathOperator*{\argmin}{\arg\min}

\DeclareMathOperator{\E}{\mathbb{E}}
\DeclareMathOperator{\Var}{\mathbb{V}}
\DeclareMathOperator{\trace}{tr}
\newcommand{\Abs}[1]{\left\lvert#1\right\rvert}
\newcommand{\abs}[1]{\lvert#1\rvert}
\newcommand{\Bcal}{\mathcal{B}}
\newcommand{\decay}{\alpha}
\newcommand{\der}{\mathrm{d}}
\newcommand{\ecdf}{F}
\newcommand{\emat}{V}
\newcommand{\evec}{v}
\newcommand{\dimF}{K}
\newcommand{\dimM}{J}
\newcommand{\dimS}{R}
\newcommand{\dimSopt}{\bgroup\dimS^*\egroup}
\newcommand{\dimSM}{\bgroup\dimS\M\egroup}
\newcommand{\Fcal}{\mathcal{F}}
\newcommand{\gram}{G}
\newcommand{\inner}[2]{\langle #1,#2\rangle}
\newcommand{\KLdiv}{\mathcal{D}_\text{KL}(\prior\parallel\prior\z)}
\newcommand{\Lcal}{\mathcal{L}}
\newcommand{\loss}{L}
\newcommand{\M}{^{(\dimM)}}
\newcommand{\Mp}{^{(\dimM+1)}}
\newcommand{\Ncal}{\mathcal{N}}
\newcommand{\norm}[1]{\lVert#1\rVert}
\newcommand{\ones}[1]{\mathbf{1}_{#1}}
\newcommand{\parfrac}[2]{\frac{\partial #1}{\partial #2}}
\newcommand{\prior}{\mathbb{P}}
\newcommand{\R}{\mathbb{R}}
\newcommand{\samp}{\bgroup\mathcal{S}\egroup}
\newcommand{\sm}{\sigma}
\newcommand{\su}{\sigma_u}

\newcommand{\vck}{\Pi}
\newcommand{\voi}{\pi}
\newcommand{\vm}{\sm^2}
\newcommand{\vu}{\su^2}

\newcommand{\Xcal}{\mathcal{X}}
\newcommand{\z}{^{(0)}}

\let\oldleft\left
\let\oldright\right
\renewcommand{\left}{\mathopen{}\mathclose\bgroup\oldleft}
\renewcommand{\right}{\aftergroup\egroup\oldright}

\newtheorem{lemma}{Lemma}
\newtheorem{proposition}{Proposition}
\newtheorem{theorem}{Theorem}
\theoremstyle{definition}
\newtheorem{example}{Example}

\newif\ifbodyproofs
\bodyproofstrue

\usepackage{needspace}
\AtBeginEnvironment{lemma}{\Needspace*{8\baselineskip}}
\AtBeginEnvironment{proposition}{\Needspace*{8\baselineskip}}
\AtBeginEnvironment{theorem}{\Needspace*{8\baselineskip}}

\usepackage{natbib}
\newcommand\citeapos[1]{\citeauthor{#1}'s (\citeyear{#1})}

%% file: proofs/expected-loss.tex
\subsubsection{Proof of Lemma~\ref{lem:expected-loss}}

\begin{proof}[\unskip\nopunct]
    We have
    \begin{align*}
        \E[L(\theta,a')\mid\samp]
        &= \frac{1}{\dimF}\sum_{k=1}^\dimF\E\left[(\theta_k-a_k')^2\mid\samp\right] \\
        &= \frac{1}{\dimF}\sum_{k=1}^\dimF\left(\left(\E[\theta_k\mid\samp]-a_k'\right)^2+\Var(\theta_k\mid\samp)\right)
    \end{align*}
    for all~$a'\in\R^\dimF$.
    So~$\E[L(\theta,a')\mid\samp]$ attains its minimum value
    \[ \min_{a'\in\R^\dimF}\,\E[L(\theta,a')\mid\samp]=\frac{1}{\dimF}\sum_{k=1}^\dimF\Var(\theta_k\mid\samp) \]
    when~$a_k'=\E[\theta_k\mid\samp]$ for each~$k\in\{1,\ldots,\dimF\}$.
\end{proof}

%% file: proofs/posterior-variance.tex
\subsubsection{Proof of Lemma~\ref{lem:posterior-variance}}

Our proof of Lemma~\ref{lem:posterior-variance} uses a well-known property of normally distributed random variables:

\begin{lemma}
    \label{lem:normal-conditional}
    Let~$n_1\ge1$ and~$n_2\ge1$ be integers, and let~$z\in\R^{n_1+n_2}$ be normally distributed with mean~$\mu$ and variance~$\Sigma$.
    Partition~$z=(z_1,z_2)$ into vectors~$z_1\in\R^{n_1}$ and~$z_2\in\R^{n_2}$, and let~$\mu=(\mu_1,\mu_2)$ and
    \[ \Sigma=\begin{bmatrix}
        \Sigma_{11} & \Sigma_{12} \\
        \Sigma_{21} & \Sigma_{22}
    \end{bmatrix} \]
    be the corresponding partitions of~$\mu$ and~$\Sigma$.
    If~$\Sigma_{22}$ is invertible, then
    \[ z_1\mid z_2\sim\Ncal\left(\mu_1+\Sigma_{12}\Sigma_{22}^{-1}(z_2-\mu_2),\ \Sigma_{11}-\Sigma_{12}\Sigma_{22}^{-1}\Sigma_{21}\right). \]
\end{lemma}
\begin{proof}\let\qed\relax
    See \citet[p.\! 87]{Bishop-2006-} or \citet[p.\! 55]{DeGroot-2004-}.
\end{proof}

\begin{proof}[Proof of Lemma~\ref{lem:posterior-variance}]
    Let~$y\equiv(y^{(1)},\ldots,y^{(n)})$ and~$u\equiv(u^{(1)},\ldots,u^{(n)})$ be the~$n$-vectors of outcomes and errors, and let
    \[ W\equiv\begin{bmatrix} w^{(1)} & \cdots & w^{(n)} \end{bmatrix}^T \]
    be the~$n\times\dimF$ design matrix.
    Then we can write~\eqref{eq:outcome} in vector form as
    \[ y\equiv W\theta+u. \]
    Consider the concatenation of~$\theta$ and~$y$.
    It is normally distributed with variance
    \[ \Var\left(\begin{bmatrix} \theta \\ y \end{bmatrix}\mid W\right)=\begin{bmatrix} \Sigma & \Sigma W^T \\ W\Sigma & W\Sigma W^T+\vu I_n \end{bmatrix} \]
    under the agent's prior.
    Since observing~$\samp$ is equivalent to observing~$W$ and~$y$, Lemma~\ref{lem:normal-conditional} implies
    \begin{align*}
        \Var(\theta\mid\samp)
        &= \Var(\theta\mid W,y) \\
        &= \Sigma-\Sigma W^T\left(W\Sigma W^T+\vu I_n\right)^{-1}W\Sigma \\
        &= \left(\Sigma^{-1}+\frac{1}{\vu}\gram\right)^{-1}
    \end{align*}
    because~$\gram=W^TW$.
\end{proof}

%% file: proofs/voi-bounds.tex
\subsubsection{Proof of Proposition~\ref{prop:voi-bounds}}

Our proof of Proposition~\ref{prop:voi-bounds} uses the following fact about sums of real, symmetric matrices.

\begin{lemma}
    \label{lem:fan-linskii}
    Let~$n\ge1$ be an integer, let~$A\in\R^{n\times n}$ and~$B\in\R^{n\times n}$ be symmetric matrices with eigenvalues~$a_1\ge\cdots\ge a_n$ and~$b_1\ge\cdots\ge b_n$, and let~$C=A+B$ have eigenvalues~$c_1\ge\cdots\ge c_n$.
    Then
    \[ \sum_{j=1}^k(a_j+b_{n-j+1})\le\sum_{j=1}^kc_j\le\sum_{j=1}^k(a_j+b_j) \]
    for each~$k\in\{1,\ldots,n\}$, with equality when~$k=n$.
\end{lemma}
\begin{proof}\let\qed\relax
    See \citet[Theorem~4.3.47]{Horn-Johnson-2012-}.
\end{proof}

\begin{proof}[Proof of Proposition~\ref{prop:voi-bounds}]
    Now
    \[ \voi(\samp)=\frac{1}{\dimF}\left(\trace(\Sigma)-\trace\left(\left(\Sigma^{-1}+\frac{1}{\vu}\gram\right)^{-1}\right)\right) \]
    by Lemma~\ref{lem:posterior-variance}.
    Moreover, defining~$Z\equiv\emat^T\Omega$ gives
    \begin{align*}
        \left(\Sigma^{-1}+\frac{1}{\vu}\gram\right)^{-1}
        &= \left(\emat\Lambda^{-1}\emat^T+\frac{1}{\vu}\emat\emat^T\Omega\Delta\Omega^T\emat\emat^T\right)^{-1} \\
        &= \emat\left(\Lambda^{-1}+\frac{1}{\vu}Z\Delta Z^T\right)^{-1}\emat^T
    \end{align*}
    and hence
    \begin{align*}
        \trace\left(\left(\Sigma^{-1}+\frac{1}{\vu}\gram\right)^{-1}\right)
        &= \trace\left(\left(\Lambda^{-1}+\frac{1}{\vu}Z\Delta Z^T\right)^{-1}\right)
    \end{align*}
    by the orthogonality of~$\emat$ and the cyclic property of matrix traces.
    So~\eqref{eq:voi-bounds} is equivalent to
    \begin{equation}
        \label{eq:voi-bounds-traces}
        \sum_{k=1}^\dimF\left(\frac{1}{\lambda_k}+\frac{\delta_k}{\vu}\right)^{-1}
        \,\overset{\star\star}{\le}\,
        \trace\left(\left(\Lambda^{-1}+\frac{1}{\vu}Z\Delta Z^T\right)^{-1}\right)
        \,\overset{\star}{\le}\,
        \sum_{k=1}^\dimF\left(\frac{1}{\lambda_k}+\frac{\delta_{\dimF-k+1}}{\vu}\right)^{-1}.
    \end{equation}
    Now~$\Lambda^{-1}$ is real, symmetric, and positive definite.
    It has~$k^\text{th}$ largest eigenvalue~$a_k\equiv1/\lambda_{\dimF-k+1}>0$.
    Moreover, since~$Z$ is orthogonal, the matrix
    \[ B\equiv\frac{1}{\vu}Z\Delta Z^T \]
    is real, symmetric, and positive semi-definite.
    It has~$k^\text{th}$ largest eigenvalue~$b_k\equiv\delta_k/\vu\ge0$.
    Define
    \begin{align*}
        c_k^{**}
        &\equiv a_k+b_{\dimF-k+1} \\
        &= \frac{1}{\lambda_{\dimF-k+1}}+\frac{\delta_{\dimF-k+1}}{\vu} \\
        &> 0
    \end{align*}
    and
    \begin{align*}
        c_k^*
        &\equiv a_k+b_k \\
        &= \frac{1}{\lambda_{\dimF-k+1}}+\frac{\delta_k}{\vu} \\
        &> 0
    \end{align*}
    for each~$k\in\{1,\ldots,\dimF\}$, and consider the matrix~$C\equiv\Lambda^{-1}+B$ with~$k^\text{th}$ largest eigenvalue~$c_k$.
    This matrix is positive definite and so~$c_k>0$ for each~$k$.
    Moreover, by Lemma~\ref{lem:fan-linskii}, we have
    \[ \sum_{j=1}^kc_j^{**}\le\sum_{j=1}^kc_j\le\sum_{j=1}^kc_j^* \]
    for each~$k\in\{1,\ldots,\dimF\}$, with equality when~$k=\dimF$.

    Now define~$g(z)\equiv1/z$ for all~$z>0$.
    Then~$g:(0,\infty)\to\R$ is convex.
    So, by Lemma~\ref{lem:mps}, we have
    \begin{equation}
        \label{eq:voi-bounds-reciprocals}
        \sum_{k=1}^\dimF\frac{1}{c_k^{**}}\le\sum_{k=1}^\dimF\frac{1}{c_k}\le\sum_{k=1}^\dimF\frac{1}{c_k^*}.
    \end{equation}
    But
    \[ \sum_{k=1}^\dimF\frac{1}{c_k^{**}}=\sum_{k=1}^\dimF\left(\frac{1}{\lambda_k}+\frac{\delta_k}{\vu}\right)^{-1} \]
    and
    \[ \sum_{k=1}^\dimF\frac{1}{c_k^*}=\sum_{k=1}^\dimF\left(\frac{1}{\lambda_k}+\frac{\delta_{\dimF-k+1}}{\vu}\right)^{-1} \]
    by the definitions of~$c_1^{**},\ldots,c_\dimF^{**}$ and~$c_1^*,\ldots,c_\dimF^*$, and
    \begin{align*}
        \sum_{k=1}^\dimF\frac{1}{c_k}
        &= \trace\left(C^{-1}\right) \\
        &= \trace\left(\left(\Lambda^{-1}+\frac{1}{\vu}Z\Delta Z^T\right)^{-1}\right)
    \end{align*}
    by the definition of~$C$.
    Substituting these expressions into~\eqref{eq:voi-bounds-reciprocals} yields~\eqref{eq:voi-bounds-traces}, from which~\eqref{eq:voi-bounds} follows.

    It remains to show when the bounds~$\star$ and~$\star\star$ hold with equality.

    Suppose~$\omega_k=\evec_{\dimF-k+1}$ for each~$k\in\{1,\ldots,\dimF\}$.
    Then
    \begin{align*}
        Z
        &= \begin{bmatrix} \evec_1 & \cdots & \evec_\dimF \end{bmatrix}^T
            \begin{bmatrix} \evec_\dimF & \cdots & \evec_1 \end{bmatrix} \\
        &= \begin{bmatrix} & & 1 \\ & \reflectbox{$\ddots$} \\ 1 & \end{bmatrix}
    \end{align*}
    is the~$\dimF\times\dimF$ anti-diagonal matrix with~${jk}^\text{th}$ entry
    \[ Z_{jk}=\begin{cases}
        1 & \text{if}\ j+k=\dimF+1 \\
        0 & \text{if}\ j+k\not=\dimF+1.
    \end{cases} \]
    So the inverse of
    \begin{align*}
        \Lambda^{-1}+\frac{1}{\vu}Z\Delta Z^T
        &= \Lambda^{-1}+\frac{1}{\vu}\begin{bmatrix} \delta_\dimF \\ & \ddots \\ & & \delta_\dimF \end{bmatrix}
    \end{align*}
    has trace
    \[ \trace\left(\left(\Lambda^{-1}+\frac{1}{\vu}Z\Delta Z^T\right)^{-1}\right)=\sum_{k=1}^\dimF\left(\frac{1}{\lambda_k}+\frac{\delta_{\dimF-k+1}}{\vu}\right)^{-1} \]
    and thus~$\star$ holds with equality.

    Now suppose~$\omega_k=\evec_k$ for each~$k\in\{1,\ldots,\dimF\}$.
    Then~$Z$ equals the~$\dimF\times\dimF$ identity matrix.
    So the inverse of
    \[ \Lambda^{-1}+\frac{1}{\vu}Z\Delta Z^T=\Lambda^{-1}+\frac{1}{\vu}\Delta \]
    has trace
    \[ \trace\left(\left(\Lambda^{-1}+\frac{1}{\vu}Z\Delta Z^T\right)^{-1}\right)=\sum_{k=1}^\dimF\left(\frac{1}{\lambda_k}+\frac{\delta_k}{\vu}\right)^{-1} \]
    and thus~$\star\star$ holds with equality.
\end{proof}

%% file: proofs/optimal.tex
\subsubsection{Proof of Proposition~\ref{prop:optimal}}

\begin{proof}[\unskip\nopunct]
    Consider the constrained minimization problem~\eqref{eq:optimal-problem}.
    We can ignore the constraint that~$\delta_k$ is non-increasing in~$k$ because it does not bind (see below).
    So the problem has Lagrangian
    \[ \Lcal\equiv\sum_{k=1}^\dimF\left(\frac{1}{\lambda_k}+\frac{\delta_k}{\vu}\right)^{-1}-\sum_{k=1}^\dimF\eta_k\delta_k-\eta\left(n-\sum_{k=1}^\dimF\delta_k\right), \]
    where~$\eta_k\ge0$ is the Lagrange multiplier on the non-negativity constraint~$\delta_k\ge0$ and~$\eta\in\R$ is the multiplier on the sum constraint.
    Now
    \[ \parfrac{^2\Lcal}{\delta_j\,\partial \delta_k}=\begin{cases}
        \frac{2}{\su^4}\left(\frac{1}{\lambda_k}+\frac{\delta_k}{\vu}\right)^{-3} & \text{if}\ j=k \\
        0 & \text{if}\ j\not=k
    \end{cases} \]
    for each pair~$(j,k)\in\{1,\ldots,\dimF\}^2$, from which it follows that~$\mathcal{L}$ is convex in the vector~$(\delta_1,\ldots,\delta_\dimF)$ whenever it has non-negative components.
    So if~$\delta_1^*,\ldots,\delta_\dimF^*$ solve~\eqref{eq:optimal-problem}, then they satisfy the first-order conditions (FOCs)
    \begin{align*}
        0
        &= \parfrac{\Lcal}{\delta_k} \\
        &= -\frac{1}{\vu}\left(\frac{1}{\lambda_k}+\frac{\delta_k^*}{\vu}\right)^{-2}-\eta_k+\eta,
    \end{align*}
    the complementary slackness conditions~$0=\eta_k\delta_k^*$, and the sum constraint~$\delta_1^*+\cdots+\delta_\dimF^*=n$.

    Suppose the non-negativity constraint on~$\delta_k$ binds.
    Then the FOCs and complementary slackness conditions imply
    \begin{align*}
        0
        &< \eta_k \\
        &= \eta-\frac{\lambda_k^2}{\vu},
    \end{align*}
    which holds if and only if~$\lambda_k<\su\sqrt{\eta}$.
    But~$\lambda_k$ is non-increasing in~$k$ and the FOCs imply that~$\eta$ is strictly positive.
    So there is an integer~$k_0\in\{1,\ldots,\dimF\}$ such that~$\delta_k^*>0$ if and only if~$k\le k_0$.

    Suppose~$k\le k_0$.
    Then~$\eta_k=0$ and so the FOCs imply
    \[ \frac{\vu}{\sqrt{\eta}}=\frac{\vu}{\lambda_k}+\delta_k^*. \]
    The left-hand side is constant in~$k$, from which it follows that
    \[ \frac{\vu}{\lambda_1}+\delta_1^*=\frac{\vu}{\lambda_k}+\delta_k^* \]
    and therefore
    \[ \delta_k^*=\delta_1^*+\vu\left(\frac{1}{\lambda_1}-\frac{1}{\lambda_k}\right). \]
    Then the sum constraint implies
    \begin{align*}
        n
        &= \sum_{k=1}^{k_0}\left(\delta_1^*+\vu\left(\frac{1}{\lambda_1}-\frac{1}{\lambda_k}\right)\right) \\
        &= k_0\delta_1^*+\vu\sum_{k=1}^{k_0}\left(\frac{1}{\lambda_1}-\frac{1}{\lambda_k}\right).
    \end{align*}
    Thus
    \begin{align*}
        \delta_k^*
        &= \frac{1}{k_0}\left(n-\vu\sum_{j=1}^{k_0}\left(\frac{1}{\lambda_1}-\frac{1}{\lambda_j}\right)\right)+\vu\left(\frac{1}{\lambda_1}-\frac{1}{\lambda_k}\right) \\
        &= \frac{n}{k_0}+\vu\left(\frac{1}{k_0}\sum_{j=1}^{k_0}\frac{1}{\lambda_j}-\frac{1}{\lambda_k}\right)
    \end{align*}
    for each~$k\le k_0$ and~$\delta_k^*=0$ for each~$k>k_0$.
    Then
    \begin{align*}
        \sum_{k=1}^\dimF\left(\frac{1}{\lambda_k}+\frac{\delta_k^*}{\vu}\right)^{-1}
        &= \sum_{k=1}^{k_0}\left(\frac{1}{\lambda_k}+\frac{1}{\vu}\left(\frac{n}{k_0}+\vu\left(\frac{1}{k_0}\sum_{j=1}^{k_0}\frac{1}{\lambda_j}-\frac{1}{\lambda_k}\right)\right)\right)^{-1}+\sum_{k>k_0}\left(\frac{1}{\lambda_k}+0\right)^{-1} \\
        &= k_0^2\left(\sum_{j=1}^{k_0}\frac{1}{\lambda_j}+\frac{n}{\vu}\right)^{-1}+\sum_{k>k_0}\lambda_k
    \end{align*}
    is non-increasing in~$k_0$ when~$k_0\le\dimSopt$.
    Thus, the eigenvalues~$\delta_1^*,\ldots,\delta_\dimF^*$ defined by~\eqref{eq:optimal-delta} solve~\eqref{eq:optimal-problem}.
    They are non-increasing because~$\lambda_1,\ldots,\lambda_\dimF$ are non-increasing.
    Moreover, Proposition~\ref{prop:voi-bounds} implies
    \begin{align*}
        \voi(\samp)
        &\le \frac{1}{\dimF}\sum_{k=1}^\dimF\left(\lambda_k-\left(\frac{1}{\lambda_k}+\frac{\delta_k^*}{\vu}\right)^{-1}\right) \\
        &= \frac{1}{\dimF}\left(\sum_{k=1}^\dimF\lambda_k-\left((\dimSopt)^2\left(\sum_{j=1}^\dimSopt\frac{1}{\lambda_j}+\frac{n}{\vu}\right)^{-1}+\sum_{k>\dimSopt}\lambda_k\right)\right) \\
        &= \voi^*,
    \end{align*}
    with equality if~\eqref{eq:optimal-gram} holds.
\end{proof}

%% file: proofs/mps.tex
\subsubsection{Proof of Lemma~\ref{lem:mps}}

\begin{proof}[\unskip\nopunct]
    The result follows from establishing three equivalences:
    \begin{enumerate}
    
        \item
        \emph{(i)$\iff$(ii)}.
        \citet[Theorem~2]{Rothschild-Stiglitz-1970-JET} show that~(i) is equivalent to
        \begin{enumerate}

            \item[(ii')]
            $\int_0^\infty g(z)\,\der\ecdf'(z)\ge\int_0^\infty g(z)\,\der\ecdf(z)$ for all convex functions~$g:(0,\infty)\to\R$, 

        \end{enumerate}
        which is equivalent to~(ii) by the definitions of~$\ecdf$ and~$\ecdf'$.

        \item
        \emph{(ii)$\iff$(iii)}.
        Consider the~$\dimF$-vectors~$\lambda'\equiv(\lambda_1',\ldots,\lambda_\dimF')$ and~$\lambda\equiv(\lambda_1,\ldots,\lambda_\dimF)$.
        \citet[Theorem 2.9]{Arnold-1987-} shows that~(ii) holds precisely when~$\lambda'$ majorizes~$\lambda$.
        But the components of~$\lambda'$ and~$\lambda$ are non-increasing, and so~$\lambda'$ majorizes~$\lambda$ if and only if~(iii) holds.

        \item
        \emph{(iii)$\iff$(iv)}.
        For each~$k\in\{1,\ldots,\dimF\}$ we have
        \begin{align*}
            \sum_{j=1}^k\lambda_j'-\sum_{j=1}^k\lambda_j
            &= \left(\sum_{j=1}^\dimF\lambda_j'-\sum_{j>k}\lambda_j'\right)-\left(\sum_{j=1}^\dimF\lambda_j-\sum_{j>k}\lambda_j\right) \\
            &= \left(\sum_{j=1}^\dimF\lambda_j'-\sum_{j=1}^\dimF\lambda_j\right)-\left(\sum_{j>k}\lambda_j'-\sum_{j>k}\lambda_j\right),
        \end{align*}
        from which it follows that~(iii) and~(iv) are equivalent.\qedhere

    \end{enumerate}
\end{proof}

%% file: proofs/vck-eigenvalues.tex
\subsubsection{Proof of Theorem~\ref{thm:vck-eigenvalues}}

Our proof of Theorem~\ref{thm:vck-eigenvalues} invokes the following lemma.

\begin{lemma}
    \label{lem:optimal-mps}
    The value~$\voi^*$ of an optimal sample does not fall when~$\lambda_1,\ldots,\lambda_\dimF$ undergo a MPS.
\end{lemma}
\begin{proof}
    Now
    \begin{align*}
        \dimSopt
        &\in \argmin_{k_0\in\{1,\ldots,\dimF\}}\left(k_0\left(\frac{1}{k_0}\left(\sum_{k=1}^{k_0}\frac{1}{\lambda_k}+\frac{n}{\vu}\right)\right)^{-1}+\sum_{k>k_0}\lambda_k\right) \\
        &= \argmax_{k_0\in\{1,\ldots,\dimF\}}\left(\sum_{k=1}^{k_0}\lambda_k-k_0^2\left(\sum_{k=1}^{k_0}\frac{1}{\lambda_k}+\frac{n}{\vu}\right)^{-1}\right)
    \end{align*}
    from the proof of Proposition~\ref{prop:optimal}.
    So if~$\lambda_1,\ldots,\lambda_\dimF$ undergo a MPS, then~$\dimSopt$ changes only if doing so makes~$\samp$ more valuable.
    So it suffices to show that for fixed~$\dimSopt$, the MPS does not lower the RHS of~\eqref{eq:optimal-voi}.

    Let~$\lambda_1'\ge\cdots\ge\lambda_\dimF'>0$ be the eigenvalues after the MPS.
    By Lemma~\ref{lem:mps}, the difference
    \begin{equation}
        \label{eq:optimal-mps-eta}
        \eta\equiv\sum_{k=1}^\dimSopt\lambda_k'-\sum_{k=1}^\dimSopt\lambda_k
    \end{equation}
    is non-negative.
    The MPS raises the first bracketed term on the RHS of~\eqref{eq:optimal-voi} by~$\eta$.
    So it suffices to show that the MPS lowers the second bracketed term by at most~$\eta$:
    \begin{equation}
        \label{eq:optimal-mps-bound}
        \underbrace{(\dimSopt)^2\left(\sum_{k=1}^\dimSopt\frac{1}{\lambda_k'}+\frac{n}{\vu}\right)^{-1}}_{S'}-\underbrace{(\dimSopt)^2\left(\sum_{k=1}^\dimSopt\frac{1}{\lambda_k}+\frac{n}{\vu}\right)^{-1}}_{S}\le\eta.
    \end{equation}
    Consider the first term~$S'$ on the LHS of~\eqref{eq:optimal-mps-bound}.
    This term is largest when the harmonic sum
    \[ H'\equiv\sum_{k=1}^\dimSopt\frac{1}{\lambda_k'} \]
    is smallest.
    Defining~$\eta_k\equiv\lambda_k'-\lambda_k$ for each~$k\in\{1,\ldots,\dimF\}$ gives
    \[ H'\equiv\sum_{k=1}^\dimSopt\frac{1}{\lambda_k+\eta_k} \]
    and~$\eta_1+\cdots+\eta_\dimSopt=\eta$.
    Lemma~\ref{lem:mps} implies
    \[ \sum_{j=1}^k\eta_j\ge0 \]
    for each~$k\in\{1,\ldots,\dimSopt\}$.
    Thus
    \[ H'\ge H^*\equiv\sum_{k=1}^\dimSopt\frac{1}{\lambda_k+\eta_k^*}, \]
    where~$\eta_1^*,\ldots,\eta_\dimSopt^*$ solve the constrained minimization problem
    \begin{equation}
        \label{eq:optimal--mps-problem}
        \begin{split}
            \min_{\eta_1,\ldots,\eta_\dimSopt\in\R}\ 
            &\sum_{k=1}^\dimSopt\frac{1}{\lambda_k+\eta_k} \\
            \quad\text{subject to}\quad
            \lambda_k+\eta_k&>0\ \text{for each}\ k\in\{1,\ldots,\dimSopt\}, \\
            \sum_{j=1}^k\eta_j&\ge0\ \text{for each}\ k\in\{1,\ldots,\dimSopt\}, \\
            \ \ \text{and}\ \ 
            \sum_{k=1}^\dimSopt\eta_k&=\eta.
        \end{split}
    \end{equation}
    Setting~$\lambda_k'=\lambda_k+\eta_k^*$ for each~$k\in\{1,\ldots,\dimSopt\}$ yields the ``worst-case'' MPS that maximizes the first term~$S'$ on the LHS of~\eqref{eq:optimal-mps-bound} given the difference~\eqref{eq:optimal-mps-eta}.

    The differences~$\eta_1^*,\ldots,\eta_\dimSopt^*$ that solve~\eqref{eq:optimal--mps-problem} are non-negative.
    To see why, notice that~$\eta_1<0$ is infeasible and assume towards a contradiction that~$\eta_\ell^*<0\le\min\{\eta_1^*,\ldots,\eta_{\ell-1}^*\}$ for some~$\ell>1$.
    Then
    \[ \ell'\equiv\max\{k\in\{1,\ldots,\ell-1\}:\eta_k^*>0\} \]
    must exist, for otherwise~$\eta_1^*,\ldots,\eta_\dimSopt^*$ would violate the constraint
    \[ \sum_{j=1}^\ell\eta_k^*\ge0. \]
    Defining
    \[ \eta_k^\dag\equiv\begin{cases}
        \eta_{\ell'}^*+\eta_\ell^* & \text{if}\ k=\ell' \\
        0 & \text{if}\ \ell'<k=\ell \\
        \eta_k^* & \text{otherwise}
    \end{cases} \]
    gives
    \[ \sum_{j=1}^k\eta_j^\dag=\begin{cases}
        \sum_{j=1}^\ell\eta_j^* & \text{if}\ \ell'\le k\le \ell \\
        \sum_{j=1}^k\eta_j^* & \text{otherwise}
    \end{cases} \]
    for each~$k\in\{1,\ldots,\dimSopt\}$, from which it follows that~$\eta_1^\dag,\ldots,\eta_\dimSopt^\dag$ are feasible.
    But~$\lambda_{\ell'}\ge\lambda_\ell$ and~$\eta_{\ell'}^*>0$, and so~$\lambda_{\ell'}+\eta_{\ell'}^*>\lambda_\ell>0$.
    Thus
    \[ \frac{1}{\lambda_{\ell'}+\eta_{\ell'}^*+\eta_\ell^*}+\frac{1}{\lambda_\ell}<\frac{1}{\lambda_{\ell'}+\eta_{\ell'}^*}+\frac{1}{\lambda_\ell+\eta_\ell^*} \]
    because~$g(z)\equiv1/z$ is a strictly decreasing and convex function of~$z>0$.
    But then
    \begin{align*}
        \sum_{k=1}^\dimSopt\frac{1}{\lambda_k+\eta_k^\dag}
        &= \sum_{k<\ell'}\frac{1}{\lambda_k+\eta_k^*}+\frac{1}{\lambda_{\ell'}+\eta_{\ell'}^*+\eta_\ell^*}+\frac{1}{\lambda_\ell}+\sum_{k>\ell}\frac{1}{\lambda_k+\eta_k^*} \\
        &< \frac{1}{\lambda_{\ell'}+\eta_{\ell'}^*}+\frac{1}{\lambda_\ell+\eta_\ell^*}+\sum_{k\notin\{\ell',\ell\}}\frac{1}{\lambda_k+\eta_k^*} \\
        &= \sum_{k=1}^\dimSopt\frac{1}{\lambda_k+\eta_k^*},
    \end{align*}
    contradicting the optimality of~$\eta_1^*,\ldots,\eta_\dimSopt^*$.
    So they must be non-negative because~$\ell$ cannot exist.

    Finally, we use the non-negativity of~$\eta_1^*,\ldots,\eta_\dimSopt^*$ to establish the upper bound~\eqref{eq:optimal-mps-bound} on~$(S'-S)$.
    Let~$k\in\{1,\ldots,\dimSopt\}$ and consider the derivative
    \[ \parfrac{S}{\lambda_k}=\left(\frac{\dimSopt}{\lambda_k}\left(\sum_{k=1}^\dimSopt\frac{1}{\lambda_k}+\frac{n}{\vu}\right)^{-1}\right)^2 \]
    of~$S$ with respect to~$\lambda_k$.
    This derivative is non-negative.
    It is also bounded above by one, since
    \[ \sum_{k=1}^\dimSopt\frac{1}{\lambda_k}+\frac{n}{\vu}\ge\frac{\dimSopt}{\lambda_k} \]
    by the definition of~$\dimSopt$.
    So~$S$ is a~1-Lipschitz function of~$\lambda_1,\ldots,\lambda_\dimSopt$: changing~$\lambda_k$ by~$\eta_k$ changes~$S$ by at most~$\abs{\eta_k}$.
    Letting~$S^*$ be the value of~$S$ that obtains from changing~$\lambda_k$ by~$\eta_k^*$ gives
    \begin{align*}
        S'-S
        &\overset{\star}{\le} S^*-S \\
        &\le \abs{S^*-S} \\
        &\overset{\star\star}{\le} \sum_{k=1}^\dimF\abs{\eta_k^*},
    \end{align*}
    where~$\star$ uses the maximality of~$S^*$ (induced by the minimality of~$H^*$) and~$\star\star$ uses the Lipschitz property.
    But~$\eta_1^*,\ldots,\eta_\dimSopt^*$ are non-negative and sum to~$\eta$, from which the bound~\eqref{eq:optimal-mps-bound} follows:
    \begin{align*}
        S'-S
        &\le \sum_{k=1}^\dimF\eta_k^* \\
        &= \eta.\qedhere
    \end{align*}
\end{proof}

\begin{proof}[Proof of Theorem~\ref{thm:vck-eigenvalues}]
    It suffices to prove~(ii) and~(iii), which together imply~(i).
    This is because every distribution of~$\lambda_1,\ldots,\lambda_\dimF$ is a MPS of the degenerate distribution under which they are equal (to their mean~$\overline\lambda$).

    Consider~(ii).
    If~$\lambda_1,\ldots,\lambda_\dimF$ are equal, then~$\lambda_k=\overline\lambda$ for each~$k\in\{1,\ldots,\dimF\}$, and so~$\dimSopt=\dimS\z$ and~$\voi^*=\voi\z$ by definition.
    Thus~$\vck\equiv\voi^*-\voi\z=0$.

    Now consider~(iii).
    The value~$\voi\z$ of the na\"ive agent's optimal sample depends on~$\lambda_1,\ldots,\lambda_\dimF$ via their mean~$\overline\lambda$ only.
    It does not change when~$\lambda_1,\ldots,\lambda_\dimF$ undergo a MPS.
    Since~$\voi^*$ does not fall under the MPS (by Lemma~\ref{lem:optimal-mps}), neither does~$\vck\equiv\voi^*-\voi\z$.
\end{proof}

%% file: proofs/vck-tau.tex
\subsubsection{Proof of Theorem~\ref{thm:vck-tau}}

\begin{proof}[\unskip\nopunct]
    Define
    \[ \tau_k\equiv\overline\lambda\left(\frac{k}{\lambda_k}-\sum_{j=1}^k\frac{1}{\lambda_j}\right) \]
    for each~$k\in\{1,\ldots,\dimF\}$.
    Then~$\tau_1=0$, and for each~$k<\dimF$ the difference
    \[ \tau_{k+1}-\tau_k=k\overline\lambda\left(\frac{1}{\lambda_{k+1}}-\frac{1}{\lambda_k}\right) \]
    is non-negative because~$\overline\lambda>0$ and~$\lambda_{k+1}\le\lambda_k$.
    So~$\tau_k$ is non-decreasing in~$k$ and hence
    \begin{align*}
        \dimSopt
        &= \max\left\{k\in\{1,\ldots,\dimF\}:\tau_k\le\tau\right\}
    \end{align*}
    is non-decreasing in~$\tau$.
    Now define~$\tau_{\dimF+1}\equiv\infty$ and suppose~$\tau\in[\tau_k,\tau_{k+1})$ for some~$k\in\{1,\ldots,\dimF\}$.
    Then~$\dimSopt=k$ and so
    \begin{align*}
        \vck
        &= \vck_k \\
        &\equiv \frac{\overline\lambda}{\dimF}\left(\sum_{j=1}^k\frac{\lambda_j}{\overline\lambda}-k^2\left(\sum_{j=1}^k\frac{\overline\lambda}{\lambda_j}+\tau\right)^{-1}-\frac{\dimF\tau}{\dimF+\tau}\right).
    \end{align*}
    Each piece~$\vck_k$ is continuous in~$\tau$.
    Moreover, for each~$k<\dimF$ the difference
    \[ \vck_{k+1}-\vck_k=-\frac{\overline\lambda}{\dimF}\left((k+1)^2\left(\sum_{j=1}^{k+1}\frac{\overline\lambda}{\lambda_j}+\tau\right)^{-1}-k^2\left(\sum_{j=1}^k\frac{\overline\lambda}{\lambda_j}+\tau\right)^{-1}-\frac{\lambda_{k+1}}{\overline\lambda}\right) \]
    between consecutive pieces converges to zero as~$\tau\to\tau_{k+1}$.
    It follows that~$\vck$ is continuous in~$\tau$.
    So to determine whether~$\vck$ is increasing or decreasing in~$\tau$, it suffices to analyze its derivative
    \begin{align}
        \parfrac{\vck_k}{\tau}
        &= \frac{\overline\lambda}{\dimF}\left(k^2\left(\sum_{j=1}^k\frac{\overline\lambda}{\lambda_j}+\tau\right)^{-2}-\left(\frac{\dimF}{\dimF+\tau}\right)^2\right) \label{eq:vck-tau-derivative}
    \end{align}
    on each piece~$\vck_k$.

    Consider the final piece
    \begin{align*}
        \vck_\dimF
        &= \frac{\overline\lambda}{\dimF}\left(\sum_{j=1}^\dimF\frac{\lambda_j}{\overline\lambda}-\dimF^2\left(\sum_{j=1}^\dimF\frac{\overline\lambda}{\lambda_j}+\tau\right)^{-1}-\frac{\dimF\tau}{\dimF+\tau}\right) \\
        &= \dimF\overline\lambda\left(\frac{1}{\dimF+\tau}-\left(\sum_{j=1}^\dimF\frac{\overline\lambda}{\lambda_j}+\tau\right)^{-1}\right).
    \end{align*}
    If~$\lambda_1,\ldots,\lambda_\dimF$ are equal (i.e., if~$\lambda_1=\lambda_\dimF$), then~$\lambda_k=\overline\lambda$ and~$\tau_k=0$ for each~$k\in\{1,\ldots,\dimF\}$, and so
    \begin{align*}
        \vck\bigg\rvert_{\lambda_1=\lambda_\dimF}
        &= \vck_\dimF\bigg\rvert_{\lambda_1=\lambda_\dimF} \\
        &= \dimF\overline\lambda\left(\frac{1}{\dimF+\tau}-\left(\sum_{j=1}^\dimF\frac{\overline\lambda}{\overline\lambda}+\tau\right)^{-1}\right) \\
        &= 0
    \end{align*}
    for all~$\tau\ge0$.
    Whereas if~$\lambda_1,\ldots,\lambda_\dimF$ are not equal (i.e., if~$\lambda_1>\lambda_\dimF$), then
    \begin{align*}
        \sum_{j=1}^\dimF\frac{\overline\lambda}{\lambda_j}
        &> \frac{\dimF\overline\lambda}{\frac{1}{\dimF}\sum_{j=1}^\dimF\lambda_j} \\
        &= \dimF
    \end{align*}
    by Jensen's inequality and the definition of~$\overline\lambda$, from which it follows that
    \[ \parfrac{\vck_\dimF}{\tau}\bigg\rvert_{\lambda_1>\lambda_\dimF}=\dimF\overline\lambda\left(\left(\sum_{j=1}^\dimF\frac{\overline\lambda}{\lambda_j}+\tau\right)^{-2}-\left(\frac{1}{\dimF+\tau}\right)^2\right) \]
    is strictly negative.
    Thus~$\vck$ is non-increasing in~$\tau$ whenever~$\tau\ge\tau_\dimF$.
    Moreover,
    \begin{align*}
        \lim_{\tau\to\infty}\vck
        &= \lim_{\tau\to\infty}\vck_\dimF \\
        &= \dimF\overline\lambda\lim_{\tau\to\infty}\left(\frac{1}{\dimF+\tau}-\left(\sum_{j=1}^\dimF\frac{\overline\lambda}{\lambda_j}+\tau\right)^{-1}\right) \\
        &= 0.
    \end{align*}
    So if~$\lambda_1,\ldots,\lambda_\dimF$ are equal, then~$\tau_\dimF=0$ and the result follows from letting~$\tau'=0$.

    It remains to show that if~$\lambda_1,\ldots,\lambda_\dimF$ are \emph{not} equal, then there exists~$\tau'\in(0,\tau_\dimF)$ such that~$\vck$ is increasing in~$\tau$ if and only if~$\tau<\tau'$.

    Suppose~$\tau\in[\tau_k,\tau_{k+1})$ for some~$k<\dimF$.
    Then~$\vck$ is increasing in~$\tau$ if and only if~\eqref{eq:vck-tau-derivative} exceeds zero, which happens precisely when
    \begin{align*}
        \tau
        &< \tau_k' \\
        &\equiv \frac{\dimF}{\dimF-k}\left(k\left(1-\frac{\overline\lambda}{\lambda_k}\right)+\tau_k\right).
    \end{align*}
    So~$\vck_k$ is decreasing in~$\tau\in[\tau_k,\tau_{k+1})$ if~$\tau_k'<\tau_k$, increasing if~$\tau_k'\ge\tau_{k+1}$, and increasing-and-then-decreasing if~$\tau_k\le\tau_k'<\tau_{k+1}$.
    Now~$\tau_k'\ge\tau_k$ if and only if
    \[ \frac{\dimF-k}{\lambda_k}+\sum_{j=1}^k\frac{1}{\lambda_j}\le\frac{\dimF}{\overline\lambda}, \]
    whereas~$\tau_k'<\tau_{k+1}$ if and only if
    \[ \frac{\dimF}{\overline\lambda}<\frac{\dimF-(k+1)}{\lambda_{k+1}}+\sum_{j=1}^{k+1}\frac{1}{\lambda_j}. \]
    So defining
    \[ \eta_k\equiv\frac{\dimF-k}{\lambda_k}+\sum_{j=1}^k\frac{1}{\lambda_j} \]
    for each~$k\in\{1,\ldots,\dimF\}$ gives~$\tau_k'\in[\tau_k,\tau_{k+1})$ if and only if~$K/\overline\lambda\in[\eta_k,\eta_{k+1})$.
    But~$\eta_k$ is non-decreasing in~$k$ because~$\lambda_{k+1}\le\lambda_k$ and therefore
    \begin{align*}
        \eta_{k+1}-\eta_k
        &= (\dimF-k)\left(\frac{1}{\lambda_{k+1}}-\frac{1}{\lambda_k}\right) \\
        &\ge 0.
    \end{align*}
    It follows that~$\tau_k'\in[\tau_k,\tau_{k+1})$ for at most one~$k<\dimF$.
    But there is at least one such~$k$ when~$\lambda_1,\ldots,\lambda_\dimF$ are not equal.
    To see why, notice that
    \begin{align*}
        \lim_{\tau\to0}\parfrac{\vck}{\tau}
        &= \lim_{\tau\to0}\parfrac{\vck_1}{\tau} \\
        &= \frac{\overline\lambda}{\dimF}\lim_{\tau\to0}\left(\left(\frac{\overline\lambda}{\lambda_1}+\tau\right)^{-2}-\left(\frac{\dimF}{\dimF+\tau}\right)^2\right) \\
        &= \frac{\overline\lambda}{\dimF}\left(\left(\frac{\lambda_1}{\overline\lambda}\right)^2-1\right)
    \end{align*}
    is strictly positive when~$\lambda_1>\overline\lambda$, which holds precisely when~$\lambda_1,\ldots,\lambda_\dimF$ are not equal, in which case the value~$\vck$ is decreasing in~$\tau$ whenever~$\tau>\tau_\dimF$.
    So~$\vck$ is initially increasing in~$\tau$ and eventually decreasing in~$\tau$, which, by continuity, means its derivative with respect to~$\tau$ changes sign at least once.
    Therefore, if~$\lambda_1,\ldots,\lambda_\dimF$ are not equal, then there is a unique~$k<\dimF$ such that~$\tau_k'\in[\tau_k,\tau_{k+1})$.
    Letting~$\tau'=\tau_k'>0$ completes the proof.
\end{proof}

%% file: proofs/vck-pairwise.tex
\subsubsection{Proof of Proposition~\ref{prop:vck-pairwise}}

Our proof of Proposition~\ref{prop:vck-pairwise} invokes the following lemma.

\begin{lemma}
    \label{lem:optimal-pairwise}
    Suppose~$\theta$ has prior variance~\eqref{eq:pairwise-Sigma} with~$\vm>0$ and~$\rho\in[0,1)$.
    \begin{enumerate}

        \item[(i)]
        There is a threshold~$\rho'\in(0,1)$ such that
        \begin{equation}
            \label{eq:dimSopt-pairwise}
            \dimSopt=\begin{cases}
                \dimF & \text{if}\ \rho\le\rho' \\
                1 & \text{if}\ \rho>\rho'.
            \end{cases}
        \end{equation}
        %

        \item[(ii)]
        The value~$\voi^*$ of an optimal sample rises when~$\rho$ rises.

    \end{enumerate}
\end{lemma}
\begin{proof}
    Consider~(i).
    If~$\lambda_1\ge\lambda_2=\cdots=\lambda_\dimF$, then
    \[ \dimSopt=\begin{cases}
        1 & \text{if}\ \frac{1}{\lambda_1}+\frac{n}{\vu}<\frac{1}{\lambda_2} \\
        \dimF & \text{otherwise}.
    \end{cases} \]
    Now~\eqref{eq:pairwise-Sigma} has eigenvalues~$\lambda_1=\left(1+\rho(\dimF-1)\right)\vm$ and~$\lambda_2=\cdots=\lambda_\dimF=(1-\rho)\vm$.
    So~$\dimSopt=\dimF$ if and only if
    \begin{align*}
        0
        &\le \frac{1}{\left(1+\rho(\dimF-1)\right)\vm}-\frac{1}{(1-\rho)\vm}+\frac{n}{\vu} \\
        &= \frac{1}{\vm}\left(\frac{1}{1+\rho(\dimF-1)}-\frac{1}{1-\rho}+\tau\right).
    \end{align*}
    The bracketed term on the RHS is continuous and decreasing in~$\rho$, strictly positive when~$\rho=0$, and unbounded below as~$\rho\to1$.
    So, by the intermediate value theorem, there exists~$\rho'\in(0,1)$ such that~\eqref{eq:dimSopt-pairwise} holds.
    
    Now consider~(ii).
    Substituting~\eqref{eq:dimSopt-pairwise} into~\eqref{eq:optimal-voi} gives
    \begin{align}
        \voi^*
        &= \frac{1}{\dimF}\begin{cases}
            \sum_{k=1}^\dimF\lambda_k-\dimF^2\left(\sum_{k=1}^\dimF\frac{1}{\lambda_k}+\frac{n}{\vu}\right)^{-1} & \text{if}\ \rho\le\rho' \\
            \lambda_1-\left(\frac{1}{\lambda_1}+\frac{n}{\vu}\right)^{-1} & \text{if}\ \rho>\rho'
        \end{cases} \notag \\
        &= \frac{\vm}{\dimF}\begin{cases}
            \dimF-\dimF^2\left(\frac{1}{1+\rho(\dimF-1)}+\frac{\dimF-1}{1-\rho}+\tau\right)^{-1} & \text{if}\ \rho\le\rho' \\
            1+\rho(\dimF-1)-\left(\frac{1}{1+\rho(\dimF-1)}+\tau\right)^{-1} & \text{if}\ \rho>\rho',
        \end{cases} \label{eq:optimal-pairwise-voi}
    \end{align}
    which is piecewise increasing in~$\rho$:
    \begin{align*}
        \parfrac{}{\rho}\left[\voi^*\big\rvert_{\rho\le\rho'}\right]
        &= \dimF(\dimF-1)\left(\frac{1}{1+\rho(\dimF-1)}+\frac{\dimF-1}{1-\rho}+\tau\right)^{-2}\left(\frac{1}{(1-\rho)^2}-\frac{1}{\left(1+\rho(\dimF-1)\right)^2}\right) \\
        &\ge 0
    \end{align*}
    with equality if and only if~$\rho=0$, and
    \begin{align*}
        \parfrac{}{\rho}\left[\voi^*\big\rvert_{\rho>\rho'}\right]
        &= \frac{(\dimF-1)\vm}{\dimF}\left(1+\left(\frac{1}{1+\tau\left(1+\rho(\dimF-1)\right)}\right)^2\right) \\
        &> 0.\qedhere
    \end{align*}
\end{proof}

\begin{proof}[Proof of Proposition~\ref{prop:vck-pairwise}]
    Suppose the sample~$\samp$ is optimal.
    Then its value~$\voi^*$ equals
    \[ \voi\z=\frac{\vm\tau}{\dimF+\tau} \]
    when~$\rho=0$.
    Now~$\voi^*$ is increasing in~$\rho$ (by Lemma~\ref{lem:optimal-pairwise}), whereas~$\voi\z$ is constant in~$\rho$.
    So~$\vck=\voi^*-\voi\z$ equals zero when~$\rho=0$ and is increasing in~$\rho$.
    
    It remains to prove~(iii).
    Now~\eqref{eq:pairwise-Sigma} has eigenvalues~$\lambda_1=\left(1+\rho(\dimF-1)\right)\vm$ and~$\lambda_2=\cdots=\lambda_\dimF=(1-\rho)\vm$, which have mean~$\overline\lambda=\vm$.
    Defining
    \begin{align*}
        \tau_\dimF
        &\equiv \overline\lambda\left(\frac{1}{\lambda_2}-\frac{1}{\lambda_1}\right) \\
        &= \frac{\rho\dimF}{(1-\rho)\left(1+\rho(\dimF-1)\right)}
    \end{align*}
    gives
    \begin{align*}
        \dimSopt
        &= \begin{cases}
            1 & \text{if}\ \frac{1}{\lambda_1}+\frac{n}{\vu}<\frac{1}{\lambda_\dimF} \\
            \dimF & \text{if}\ \frac{1}{\lambda_1}+\frac{n}{\vu}\ge\frac{1}{\lambda_\dimF}
        \end{cases} \\
        &= \begin{cases}
            1 & \text{if}\ \tau<\tau_\dimF \\
            \dimF & \text{if}\ \tau\ge\tau_\dimF,
        \end{cases}
    \end{align*}
    which when substituted into~\eqref{eq:optimal-voi} gives
    \begin{align*}
        \vck
        &= \frac{\vm}{\dimF}\begin{cases}
            1+\rho(\dimF-1)-\left(\frac{1}{1+\rho(\dimF-1)}+\tau\right)^{-1}-\frac{\dimF\tau}{\dimF+\tau} & \text{if}\ \tau<\tau_\dimF \\
            \dimF^2\left((\dimF+\tau)^{-1}-\left(\sum_{k=1}^\dimF\frac{\vm}{\lambda_k}+\tau\right)^{-1}\right) & \text{if}\ \tau\ge\tau_\dimF.
        \end{cases}
    \end{align*}
    The first piece is (weakly) concave in~$\tau$: differentiating it with respect to~$\tau$ gives
    \[ \parfrac{}{\tau}\left[\vck\big\rvert_{\tau<\tau_\dimF}\right]=\frac{\vm}{\dimF}\left(\left(\frac{1}{1+\rho(\dimF-1)}+\tau\right)^{-2}-\left(\frac{\dimF}{\dimF+\tau}\right)^2\right), \]
    which is strictly positive if and only if
    \[ \tau<\tau'\equiv\frac{\rho\dimF}{1+\rho(\dimF-1)}. \]
    In contrast, our proof of Theorem~\ref{thm:vck-tau} shows that the second piece (with~$\tau\ge\tau_\dimF$) is non-increasing in~$\tau$.
    But~$\tau'\le\tau_\dimF$, from which~(iii) follows.
\end{proof}

%% file: proofs/deeper-dimS.tex
\subsubsection{Proof of Lemma~\ref{lem:deeper-dimS}}

\begin{proof}[\unskip\nopunct]
    Define
    \[ \tau_k\equiv\overline\lambda\left(\frac{k}{\lambda_k}-\sum_{j=1}^k\frac{1}{\lambda_j}\right) \]
    for each~$k\in\{1,\ldots,\dimF\}$ so that
    \[ \dimSopt=\max\{k\in\{1,\ldots,\dimF\}:\tau_k\le\tau\} \]
    as in the proof of Theorem~\ref{thm:vck-tau}.
    Fix~$\dimM\in\{0,\ldots,\dimF\}$ and define
    \begin{align*}
        \tau_k\M
        &\equiv \overline\lambda\left(\frac{k}{\lambda_k\M}-\sum_{j=1}^k\frac{1}{\lambda_j\M}\right) \\
        &= \overline\lambda\begin{cases}
            0 & \text{if}\ \dimM=0 \\
            \frac{k}{\lambda_k}-\sum_{j=1}^k\frac{1}{\lambda_j} & \text{if}\ \dimM>0\ \text{and}\ k\le\dimM \\
            \frac{\dimM}{\lambda_\dimF\M}-\sum_{j=1}^\dimM\frac{1}{\lambda_j} & \text{if}\ \dimM>0\ \text{and}\ k>\dimM.
        \end{cases}
    \end{align*}
    for each~$k\in\{1,\ldots,\dimF\}$.
    Then~$\tau_1\M=0$, and for each~$k<\dimF$ the difference
    \begin{align*}
        \tau_{k+1}\M-\tau_k\M
        &= \overline\lambda\begin{cases}
            0 & \text{if}\ \dimM=0 \\
            k\left(\frac{1}{\lambda_{k+1}}-\frac{1}{\lambda_k}\right) & \text{if}\ \dimM>0\ \text{and}\ k\le\dimM-1 \\
            \dimM\left(\frac{1}{\lambda_\dimF\M}-\frac{1}{\lambda_\dimM}\right) & \text{if}\ \dimM>0\ \text{and}\ k=\dimM \\
            0 & \text{if}\ \dimM>0\ \text{and}\ k>\dimM.
        \end{cases}
    \end{align*}
    is non-negative because~$\lambda_{k+1}\le\lambda_k$ and~$\lambda_\dimF\M\le\lambda_\dimM$.
    So~$\tau_k\M$ is non-decreasing in~$k$ and
    \[ \dimSM=\max\left\{k\in\{1,\ldots,\dimF\}:\tau_k\M\le\tau\right\}. \]
    Define~$\tau_0\z\equiv0$ and notice~$\tau_\dimM\M=\cdots=\tau_\dimF\M$.
    So if~$\tau\ge\tau_\dimM\M$, then~$\dimSM=\dimF$; if~$\tau<\tau_\dimM\M$, then
    \begin{align*}
        \dimSM
        &= \max\left\{k\in\{1,\ldots,\dimM\}:\tau_k\M\le\tau\right\} \\
        &= \min\left\{\dimM,\max\left\{k\in\{1,\ldots,\dimF\}:\tau_k\le\tau\right\}\right\} \\
        &= \min\left\{\dimM,\dimSopt\right\}.
    \end{align*}
    But if~$\dimM<\dimF$, then
    \begin{align*}
        \tau_{\dimM+1}\Mp-\tau_\dimM\M
        &= \overline\lambda\left(\left(\frac{\dimM+1}{\lambda_{\dimM+1}}-\sum_{j=1}^{\dimM+1}\frac{1}{\lambda_j}\right)-\left(\frac{\dimM}{\lambda_\dimM}-\sum_{j=1}^\dimM\frac{1}{\lambda_j}\right)\right) \\
        &= \dimM\overline\lambda\left(\frac{1}{\lambda_{\dimM+1}}-\frac{1}{\lambda_\dimM}\right)
    \end{align*}
    is non-negative because~$\lambda_{\dimM+1}\le\lambda_\dimM$.
    So~$\tau_\dimM\M$ is non-decreasing in~$\dimM$, from which it follows that
    \[ \dimM'\equiv\max\left\{k\in\{0,\ldots,\dimF\}:\tau_j^{(j)}\le\tau\ \text{for each}\ j\in\{0,\ldots,k\}\right\} \]
    exists and
    \[ \dimSM=\begin{cases}
        \dimF & \text{if}\ \dimM\le\dimM' \\
        \min\{\dimM,\dimSopt\} & \text{if}\ \dimM>\dimM'.
    \end{cases} \]
    Clearly~$\dimM'$ is non-decreasing in~$\tau$.
\end{proof}

%% file: proofs/deeper-vck.tex
\subsubsection{Proof of Theorem~\ref{thm:deeper-vck}}

\begin{proof}[\unskip\nopunct]
    Now~(i) follows from~(ii)--(iv), while~(ii) follows from the definition of~$\vck\M=\voi\M-\voi\z$ and~(iii) follows from Lemma~\ref{lem:versus-voi}(i).
    For~(iv), suppose~$\dimM\ge\dimSopt$.
    Then~$\dimSM=\dimSopt$ by Lemma~\ref{lem:deeper-dimS}, so
    \begin{align*}
        \voi\M
        &= \frac{1}{\dimF}\left(\sum_{k=1}^\dimSopt\lambda_k\M-(\dimSopt)^2\left(\sum_{k=1}^\dimSopt\frac{1}{\lambda_k\M}+\frac{n}{\vu}\right)^{-1}\right) \\
        &= \frac{1}{\dimF}\left(\sum_{k=1}^\dimSopt\lambda_k-(\dimSopt)^2\left(\sum_{k=1}^\dimSopt\frac{1}{\lambda_k}+\frac{n}{\vu}\right)^{-1}\right) \\
        &= \voi^*
    \end{align*}
    and hence~$\vck\M=\vck$ by definition.
\end{proof}

%% file: proofs/versus-voi.tex
\subsubsection{Proof of Lemma~\ref{lem:versus-voi}}

\begin{proof}[\unskip\nopunct]
    We prove~(i) and~(ii) separately:
    \begin{enumerate}
    
        \item[(i)]
        It suffices to show that~$\lambda_1\M,\ldots,\lambda_\dimF\M$ undergo a MPS when~$\dimM$ rises.
        Then~(i) follows from an argument similar to that used to prove Lemma~\ref{lem:optimal-mps}.

        Fix~$\dimM<\dimF$.
        For each~$k\in\{1,\ldots,\dimF\}$ we have
        \[ \lambda_k\Mp-\lambda_k\M=\begin{cases}
            0 & \text{if}\ k\le\dimM \\
            \lambda_{\dimM+1}-\lambda_\dimF\M & \text{if}\ k=\dimM+1 \\
            \lambda_\dimF\Mp-\lambda_\dimF\M & \text{if}\ k>\dimM+1
        \end{cases} \]
        and hence
        \begin{align}
            \sum_{j=1}^k\lambda_j\Mp-\sum_{j=1}^k\lambda_j\M
            &= \begin{cases}
                0 & \text{if}\ k\le\dimM \\
                \lambda_{\dimM+1}-\lambda_\dimF\M & \text{if}\ k=\dimM+1 \\
                \lambda_{\dimM+1}-\lambda_\dimF\M+\left(k-(\dimM+1)\right)\left(\lambda_\dimF\Mp-\lambda_\dimF\M\right) & \text{if}\ k>\dimM+1
            \end{cases} \notag \\
            &= \begin{cases}
                0 & \text{if}\ k\le\dimM \\
                \lambda_{\dimM+1}-\lambda_\dimF\M & \text{if}\ k=\dimM+1 \\
                (\dimF-k)\left(\lambda_\dimF\M-\lambda_\dimF\Mp\right) & \text{if}\ k>\dimM+1.
            \end{cases} \label{eq:deeper-cumsum}
        \end{align}
        because~$\lambda_{\dimM+1}=(\dimF-\dimM)\lambda_\dimF\M-\left(\dimF-(\dimM+1)\right)\lambda_\dimF\Mp$.
        We also have
        \begin{align*}
            \lambda_{\dimM+1}
            &= \frac{1}{\dimF-\dimM}\sum_{k>\dimM}\lambda_{\dimM+1} \\
            &\ge \frac{1}{\dimF-\dimM}\sum_{k>\dimM}\lambda_k \\
            &= \lambda_\dimF\M
        \end{align*}
        because~$\lambda_{\dimM+1}\ge\cdots\ge\lambda_\dimF$.
        Likewise~$\lambda_{\dimM+1}\ge\lambda_\dimF\Mp$ and so
        \begin{align*}
            \lambda_\dimF\M-\lambda_\dimF\Mp
            &= \frac{1}{\dimF-\dimM}\sum_{k>\dimM}\lambda_k-\frac{1}{\dimF-(\dimM+1)}\sum_{k>\dimM+1}\lambda_k \\
            &= \frac{1}{\dimF-\dimM}\lambda_{\dimM+1}+\left(\frac{1}{\dimF-\dimM}-\frac{1}{\dimF-(\dimM+1)}\right)\sum_{k>\dimM+1}\lambda_k \\
            &= \frac{1}{\dimF-\dimM}\lambda_{\dimM+1}-\frac{1}{\dimF-\dimM}\lambda_\dimF\Mp \\
            &\ge 0.
        \end{align*}
        So the difference~\eqref{eq:deeper-cumsum} is non-negative and equals zero when~$k=\dimF$.
        Thus, by Lemma~\ref{lem:mps}, the eigenvalues~$\lambda_1\M,\ldots,\lambda_\dimF\M$ undergo a MPS when~$\dimM$ rises.

        \item[(ii)]
        Fix~$\dimM\in\{0,\ldots,\dimF\}$ and define
        \[ t_k\M\equiv k^2\left(\sum_{j=1}^k\frac{1}{\lambda_j\M}+\frac{n}{\vu}\right)^{-1}+\sum_{j>k}\lambda_j\M\]
        for each~$k\in\{1,\ldots,\dimF\}$.
        Then
        \[ \voi\M=\overline\lambda-\frac{1}{\dimF}\min\left\{t_k\M:k\in\{1,\ldots,\dimF\}\right\} \]
        from the proof of Proposition~\ref{prop:optimal}.
        But
        \[ \parfrac{t_k\M}{n}=-\frac{k^2}{\vu}\left(\sum_{j=1}^k\frac{1}{\lambda_j\M}+\frac{n}{\vu}\right)^{-2} \]
        is strictly negative, from which it follows that~$\voi\M$ is increasing in~$n$.\qedhere

    \end{enumerate}
\end{proof}

%% file: proofs/versus-size.tex
\subsubsection{Proof of Theorem~\ref{thm:versus-size}}

Our proof of Theorem~\ref{thm:versus-size} invokes the following Lemma:

\begin{lemma}
    \label{lem:versus-mps}
    Fix~$\dimM\in\{0,\ldots,\dimF\}$.
    Then~$\lambda_1\M,\ldots,\lambda_\dimF\M$ undergo a MPS when~$\lambda_1,\ldots,\lambda_\dimF$ undergo a MPS.
\end{lemma}
\begin{proof}
    Fix~$k\in\{1,\ldots,\dimF\}$.
    By Lemma~\ref{lem:mps}, the cumulative sum
    \[ \sum_{j=1}^{\min\{k,\dimM\}}\lambda_j \]
    does not fall when~$\lambda_1,\ldots,\lambda_\dimF$ undergo a MPS, while the tail sum
    \[ \sum_{j>\dimM}\lambda_j \]
    does not rise under the MPS.
    So the MPS does not lower
    \begin{align*}
        \sum_{j=1}^k\lambda_j\M
        &= \begin{cases}
            \sum_{j=1}^k\lambda_j & \text{if}\ k\le\dimM \\
            \sum_{j=1}^\dimM\lambda_j+(k-\dimM)\lambda_\dimF\M & \text{if}\ k>\dimM
        \end{cases} \\
        &= \begin{cases}
            \sum_{j=1}^{\min\{k,\dimM\}}\lambda_j & \text{if}\ k\le\dimM \\
            \sum_{j=1}^\dimF\lambda_j-\frac{\dimF-k}{\dimF-\dimM}\sum_{j>\dimM}\lambda_j & \text{if}\ k>\dimM
        \end{cases}
    \end{align*}
    and leaves it unchanged when~$k=\dimF$.
    The result follows from Lemma~\ref{lem:mps}.
\end{proof}
\begin{proof}[Proof of Theorem~\ref{thm:versus-size}]
    We prove~(i) and~(ii) separately:

    \begin{enumerate}

        \item[(i)]
        Fix~$n\ge0$.
        Now~$\voi\M$ is non-decreasing in~$\dimM$ (by Lemma~\ref{lem:versus-voi}), so if~$\voi\M\ge\voi_0$ then~$\voi\Mp\ge\voi_0$.
        Thus
        \[ \{n\ge0:\voi\M\ge\voi_0\}\subseteq\{n\ge0:\voi\Mp\ge\voi_0\} \]
        and therefore~$n_{\pi_0}\M\ge n_{\pi_0}\Mp$.

        \item[(ii)]
        It suffices to show that~$\voi\M$ does not fall when~$\lambda_1,\ldots,\lambda_\dimF$ undergo a MPS.
        Then, since~$\voi\M$ is increasing in~$n$ (by Lemma~\ref{lem:versus-voi}), the MPS expands~$\{n\ge0:\voi\M\ge\voi_0\}$ and so cannot raise~$n_{\pi_0}\M$.
        But the argument used to prove Lemma~\ref{lem:optimal-mps} implies that~$\voi\M$ does not fall when~$\lambda_1\M,\ldots,\lambda_\dimF\M$ undergo a MPS, which, by Lemma~\ref{lem:versus-mps}, happens when~$\lambda_1,\ldots,\lambda_\dimF$ undergo a MPS.\qedhere

    \end{enumerate}
\end{proof}

%% file: proofs/voi.tex
\subsubsection{Proof of Proposition~\ref{aprop:voi}}

\begin{proof}[\unskip\nopunct]
    Let~$\samp'$ be a superset of~$\samp$.
    Then
    \begin{align*}
        \Var(\theta_k)
        &= \E[\Var(\theta_k\mid\samp)]+\Var(\E[\theta_k\mid\samp]) \\
        &\ge \Var(\theta_k\mid\samp) \\
        &= \E[\Var(\theta_k\mid\samp,\samp')\mid\samp]+\Var(\E[\theta_k\mid\samp,\samp']\mid\samp) \\
        &\ge \Var(\theta_k\mid\samp,\samp') \\
        &= \Var(\theta_k\mid\samp')
    \end{align*}
    for each~$k\in\{1,\ldots,\dimF\}$, where the first two equalities hold by the law of total variance, the inequalities hold because the posterior variance of~$\theta_k$ is non-negative and non-random (by Lemma~\ref{lem:posterior-variance}), and the last equality holds because~$\samp'$ is a superset of~$\samp$.
    It follows that~$0\le\voi(\samp)\le\voi(\samp')$, thereby establishing~(i) and~(ii).

    Now consider~(iii).
    Differentiating the posterior variance matrix~\eqref{eq:posterior-variance} with respect to~$\vu$ gives
    \begin{align*}
        \parfrac{}{\vu}\Var(\theta\mid\samp)
        &= -\left(\Sigma^{-1}+\frac{1}{\vu}\gram\right)\left(\parfrac{}{\vu}\left[\Sigma^{-1}+\frac{1}{\vu}\gram\right]\right)\left(\Sigma^{-1}+\frac{1}{\vu}\gram\right) \\
        &= \frac{1}{\su^4}\left(\Sigma^{-1}\gram\Sigma^{-1}+\frac{1}{\vu}\gram^2\Sigma^{-1}+\frac{1}{\vu}\Sigma^{-1}\gram^2+\frac{1}{\su^4}\gram^3\right),
    \end{align*}
    which is the sum of four matrices with strictly positive traces.
    Thus
    \[ \trace\left(\parfrac{}{\vu}\Var(\theta\mid\samp)\right)>0 \]
    and therefore
    \begin{align*}
        \parfrac{\voi(\samp)}{\vu}
        &= -\frac{1}{\dimF}\trace\left(\parfrac{}{\vu}\Var(\theta\mid\samp)\right) \\
        &< 0
    \end{align*}
    because traces are linear operators.
\end{proof}

%% file: proofs/singleton-voi.tex
\subsubsection{Proof of Proposition~\ref{aprop:singleton-voi}}

Our proof of Proposition~\ref{aprop:singleton-voi} uses the following fact about rank-one updates of invertible matrices.

\begin{lemma}[Sherman-Morrison formula]
    \label{lem:sherman-morrison}
    Let~$n\ge1$ be an integer, let~$A\in\R^{n\times n}$ be invertible, and let~$u\in\R^n$ and~$v\in\R^n$.
    If~$v^TA^{-1}u\not=-1$, then
    \[ \left(A+uv^T\right)^{-1}=A^{-1}-\frac{A^{-1}uv^TA^{-1}}{1+v^TA^{-1}u}. \]
\end{lemma}
\begin{proof}\let\qed\relax
    See \cite{Bartlett-1951-AoMS}.
\end{proof}

\begin{proof}[Proof of Proposition~\ref{aprop:singleton-voi}]
    Suppose~$\samp$ contains a single observation and let~$w\equiv w^{(1)}$ for convenience.
    Then, by Lemmas~\ref{lem:posterior-variance} and~\ref{lem:sherman-morrison}, we have
    \begin{align*}
        \Var(\theta\mid\samp)
        &= \left(\Sigma^{-1}+\frac{1}{\vu}ww^T\right)^{-1} \\
        &= \Sigma-\frac{\Sigma ww^T\Sigma}{w^T\Sigma w+\vu}.
    \end{align*}
    Thus
    \begin{align*}
        \dimF\voi(\samp)
        &= \trace(\Sigma-\Var(\theta\mid\samp)) \notag \\
        &= \trace\left(\frac{\Sigma ww^T\Sigma}{w^T\Sigma w+\vu}\right) \notag \\
        &= \frac{w^T\Sigma^2w}{w^T\Sigma w+\vu},
    \end{align*}
    where the last equality holds by the linearity and cyclic property of matrix traces.
    Equation~\eqref{eq:singleton-voi} follows.
    The inequalities~\eqref{eq:singleton-voi-bounds} follow from Proposition~\ref{prop:voi-bounds}, as do the choices of~$w^{(1)}$ that make~$\star$ and~$\star\star$ hold with equality.
\end{proof}

%% file: proofs/representative-mps.tex
\subsubsection{Proof of Proposition~\ref{aprop:representative-mps}}

\begin{proof}[\unskip\nopunct]
    Define~$g(z)\equiv nz^2/\dimF(nz+\dimF\vu)$ for all~$z>0$.
    Then~$g:(0,\infty)\to\R$ is convex.
    So if~$\samp$ is representative, then, by Lemma~\ref{lem:mps}, its value
    \[ \voi(\samp)=\sum_{k=1}^\dimF g(\lambda_k) \]
    does not fall when~$\lambda_1,\ldots,\lambda_\dimF$ undergo a MPS.
\end{proof}

%% file: proofs/divergence.tex
\subsubsection{Proof of Proposition~\ref{aprop:divergence}}

\begin{proof}[\unskip\nopunct]
    It suffices to prove~(ii) and~(iii), which together imply~(i).
    
    If~$\lambda_1,\ldots,\lambda_\dimF$ are equal, then~$\lambda_k=\overline\lambda$ for each~$k\in\{1,\ldots,\dimF\}$ and so
    \begin{align*}
        \KLdiv
        &= -\frac{1}{2}\sum_{k=1}^\dimF\ln(1) \\
        &= 0,
    \end{align*}
    thus establishing~(ii).
    For~(iii), consider the function~$g:(0,\infty)\to\R$ defined by
    \[ g(z)\equiv\frac{\ln(\overline\lambda)-\ln(z)}{2}. \]
    This function is convex on its domain.
    So, by Lemma~\ref{lem:mps}, the KL divergence
    \[ \KLdiv=\sum_{k=1}^\dimF g(\lambda_k) \]
    from~$\prior$ to~$\prior\z$ does not fall when~$\lambda_1,\ldots,\lambda_\dimF$ undergo a MPS.
\end{proof}

%% file: proofs/oos.tex
\subsubsection{Derivation of~\eqref{eq:oos-trace}--\eqref{eq:oos-conditional-variance-pairwise}}

\begin{proof}[Derivation of~\eqref{eq:oos-trace} and~\eqref{eq:oos-conditional-variance}]
    Let~$k>\dimS$.
    Now~$\theta_1,\ldots,\theta_\dimS,\theta_k$ are jointly normally distributed under the agent's prior, and so Lemma~\ref{lem:normal-conditional} implies that~$\theta_k$ is conditionally normally distributed with mean
    \begin{equation}
        \label{eq:oos-conditional-mean}
        \E[\theta_k\mid\theta_\samp]=\E[\theta_k]+\xi_k^T(\theta_\samp-\E[\theta_\samp])
    \end{equation}
    and variance~\eqref{eq:oos-conditional-variance} given~$\theta_\samp$.
    So
    \begin{align*}
        \Var(\theta_k\mid\samp)
        &= \Var(\E[\theta_k\mid\samp,\theta_\samp]\mid\samp)+\E[\Var(\theta_k\mid\samp,\theta_\samp)\mid\samp] \\
        &= \Var(\E[\theta_k\mid\theta_\samp]\mid\samp)+\E[\Var(\theta_k\mid\theta_\samp)\mid\samp] \\
        &= \xi_k^T\Var(\theta_\samp\mid\samp)\xi_k+\Var(\theta_k\mid\theta_\samp),
    \end{align*}
    where the first equality holds by the law of total variance, the second holds because~$\theta_k$ is conditionally independent of~$\samp$ given~$\theta_\samp$, and the third uses~\eqref{eq:oos-conditional-mean} and the non-randomness of~\eqref{eq:oos-conditional-variance}.
    So
    \begin{align*}
        \trace(\Var(\theta\mid\samp))
        &= \sum_{k=1}^\dimF\Var(\theta_k\mid\samp) \\
        &= \sum_{k=1}^\dimS\Var(\theta_k\mid\samp)+\sum_{k>\dimS}\left(\xi_k^T\Var(\theta_\samp\mid\samp)\xi_k+\Var(\theta_k\mid\theta_\samp)\right).\qedhere
    \end{align*}
\end{proof}

\begin{proof}[Derivation of~\eqref{eq:oos-conditional-variance-pairwise}]
    Suppose~$\theta$ has prior variance~\eqref{eq:pairwise-Sigma}.
    Then
    \[ \Var(\theta_\samp)=\left(\rho\ones{\dimS}\ones{\dimS}+(1-\rho)I_\dimS\right)\vm \]
    has inverse
    \begin{align*}
        \Var(\theta_\samp)^{-1}
        &= \frac{1}{(1-\rho)\vm}\left(I_\dimS+\frac{\rho}{1-\rho}\ones{\dimS}\ones{\dimS}^T\right)^{-1} \\
        &= \frac{1}{(1-\rho)\vm}\left(I_\dimS-\frac{\rho}{1-\rho+\rho\ones{\dimS}^T\ones{\dimS}}\ones{\dimS}\ones{\dimS}^T\right)
    \end{align*}
    by Lemma~\ref{lem:sherman-morrison}.
    Now if~$k>\dimS$, then
    \[ \begin{bmatrix} \Sigma_{11} \\ \vdots \\ \Sigma_{\dimS k} \end{bmatrix}=\rho\vm\ones{\dimS} \]
    and so
    \begin{align*}
        \xi_k^T\Var(\theta_\samp)\xi_k
        &= \begin{bmatrix} \Sigma_{11} \\ \vdots \\ \Sigma_{\dimS k} \end{bmatrix}^T\Var(\theta_\samp)^{-1}\begin{bmatrix} \Sigma_{11} \\ \vdots \\ \Sigma_{\dimS k} \end{bmatrix} \\
        &= (\rho\vm\ones{\dimS})^T\left(\frac{1}{(1-\rho)\vm}\left(I_\dimS-\frac{\rho}{1-\rho+\rho\ones{\dimS}^T\ones{\dimS}}\ones{\dimS}\ones{\dimS}^T\right)\right)(\rho\vm\ones{\dimS}) \\
        &= \frac{\rho^2\dimS\vm}{1+\rho(\dimS-1)}
    \end{align*}
    because~$\ones{\dimS}^T\ones{\dimS}=\dimS$.
    Substituting this expression and~$\Var(\theta_k)=\vm$ into~\eqref{eq:oos-conditional-variance} yields~\eqref{eq:oos-conditional-variance-pairwise}.
\end{proof}